\documentclass[a4paper,12pt]{elsarticle}

\usepackage[T1]{fontenc}
\usepackage{natbib}
\usepackage{float} % 表格紧跟文字下方
\usepackage{siunitx}
\usepackage{geometry}
\usepackage{tikz}
\usepackage{fontawesome}
\usepackage{tabularx}
\usetikzlibrary{patterns}
\usepackage{amsmath, graphicx,amssymb,epsfig,psfrag,subfigure,bm,bbm,booktabs}
\usepackage{mathrsfs}
\usepackage{cases}
\usepackage{multirow} % for cmd 'multirow', 'multicolumn'
\usepackage{color}
\usepackage{hyperref}
\hypersetup{hypertex=true,
	colorlinks=true,
	linkcolor=blue,
	anchorcolor=blue,
	citecolor=blue}

\newcommand{\bpm}{\begin{pmatrix}}
\newcommand{\epm}{\end{pmatrix}}
\newcommand{\beqa}{\begin{eqnarray}}
\newcommand{\eeqa}{\end{eqnarray}}
\newcommand{\beqas}{\begin{eqnarray*}}
\newcommand{\eeqas}{\end{eqnarray*}}
\let\sss = \scriptscriptstyle

\newcommand{\rU}{\mathrm{U}}

\newcommand{\rQ}{\mathrm{Q}}

\newcommand{\rT}{\mathrm{T}}
\newcommand{\re}{\mathrm{e}}
\newcommand{\rn}{\mathrm{n}}

\newcommand{\mC}{\mathbb{C}}

\newcommand{\CL}{\mathcal{L}}
\newcommand{\CS}{\mathcal{S}}
\newcommand{\CA}{\mathcal{A}}
\newcommand{\CB}{\mathcal{B}}
\newcommand{\CD}{\mathcal{D}}

\newcommand{\CM}{\mathcal{M}}
\newcommand{\CP}{\mathcal{P}}
\newcommand{\CT}{\mathcal{T}}

\newcommand{\pdhfrac}[2]{\mathchoice{\frac{#1}{#2}}{#1/#2}{#1/#2}{#1/#2}}

\newcommand{\pd}[2]{\pdhfrac{{\partial}#1}{{\partial}#2}}
\newcommand{\spd}[2]{\pdhfrac{\partial^2#1}{{\partial}#2^2}}
\newcommand{\mpd}[3]{\pdhfrac{\partial^2#1}{{\partial}#2{\partial}#3}}

\newcommand{\ve}{\textbf{e}}

\newcommand{\vQ}{\textbf{Q}}
\newcommand{\vR}{\textbf{R}}

\newcommand{\vA}{\textbf{A}}
\newcommand{\vB}{\textbf{B}}
\newcommand{\vF}{\textbf{F}}
\newcommand{\vK}{\textbf{K}}
\newcommand{\vN}{\textbf{N}}
\newcommand{\vU}{\mathbf{U}}
\newcommand{\vr}{\mathbf{r}}

\newcommand{\vt}{\mathbf{t}}
\newcommand{\vu}{\mathbf{u}}

\newcommand{\vf}{\boldsymbol{f}}
\newcommand{\vp}{\boldsymbol{p}}
\newcommand{\vT}{\boldsymbol{T}}
\newcommand{\vM}{\boldsymbol{M}}
\newcommand{\vP}{\boldsymbol{P}}
\newcommand{\vD}{\boldsymbol{D}}

\newcommand{\vsi}{\boldsymbol{\sigma}}
\newcommand{\veps}{\boldsymbol{\varepsilon}}

\newcommand{\vphi}{\boldsymbol{\varphi}}
\newcommand{\vCA}{\boldsymbol{\CA}}
\newcommand{\vCB}{\boldsymbol{\CB}}
\newcommand{\vCD}{\boldsymbol{\CD}}
\newcommand{\vCT}{\boldsymbol{\CT}}
\newcommand{\vCM}{\boldsymbol{\CM}}

\newcommand{\bsigma}{\bar{\sigma}}

\newcommand{\bmC}{\bar{\mC}}

\newcommand{\tve}{\widetilde{\ve}}

\newcommand{\wvu}{\widetilde{\vu}}

\newcommand{\wa}{\widehat{a}}
\newcommand{\wh}{\widehat{h}}

\newcommand{\wt}{\widehat{t}}
\newcommand{\wf}{\widehat{f}}

\newcommand{\wE}{\widehat{E}}
\newcommand{\wT}{\widehat{T}}
\newcommand{\wN}{\widehat{N}}
\newcommand{\wM}{\widehat{M}}
\newcommand{\wu}{\widehat{u}}
\newcommand{\wP}{\widehat{P}}
\newcommand{\wQ}{\widehat{Q}}
\newcommand{\wCP}{\widehat{\CP}}
\newcommand{\wCT}{\widehat{\CT}}
\newcommand{\wCM}{\widehat{\CM}}
\newcommand{\wload}{\widehat{p}}

\newcommand{\wrho}{\widehat{\rho}}
\newcommand{\wxi}{\widehat{\xi}}
\newcommand{\wsigma}{\widehat{\sigma}}
\newcommand{\wk}{\widehat{\kappa}}

\newcommand{\wphi}{\widehat{\varphi}}

\newcommand{\mO}{\mathcal{O}}

\newcommand{\intd}{\mathrm{d}}
\newcommand{\df}{\displaystyle\frac}

\newcolumntype{Z}{>{\centering\let\newline\\\arraybackslash\hspace{0pt}}X}

\makeatletter
\newcommand*\bigcdot{\mathpalette\bigcdot@{.5}}
\newcommand*\bigcdot@[2]{\mathbin{\vcenter{\hbox{\scalebox{#2}{$\m@th#1\bullet$}}}}}
\makeatother

\journal{International Journal of Engineering Science}

\begin{document}

\begin{frontmatter}

\title{Asymptotic Formulation of the Role of Shear Loads on Multi-Layered Thin Shells and Classification of Their Deformation Modes}

%% Group authors per affiliation:
\author[DUT1]{Xiwei Pan}

\author[DUT1,DUT2]{Yichao Zhu\corref{mycorrespondingauthor}}
\cortext[mycorrespondingauthor]{Corresponding authors}
\ead{yichaozhu@dlut.edu.cn}

\address[DUT1]{Department of Engineering Mechanics, Dalian University of Technology, Dalian, 116023, P. R. China}
\address[DUT2]{State Key Laboratory of Structural Analysis, Optimization and CAE Software for Industrial Equipment, Dalian University of Technology}

\begin{abstract}
	Shell structures are generally modeled based on kinematic hypotheses, where some of the parameters are preferentially evaluated in a phenomenological manner. In this article, asymptotic analysis against the underlying three-dimensional equation system is considered so as to provide a rational framework for modeling and interpreting the deformation behavior of multi-layered thin shells (MTSs). Capable of accurately predicting both overall stiffness and detailed stress distribution, the proposed shell theory shows its distinguishing features at least in the following aspects. Firstly, it naturally introduces a rule for classifying the deformation modes of MTSs based on the magnitude of the maximum dimensionless principal curvature. Secondly, for each class, the hierarchy in the order of the involved field quantities is examined, and it is shown that when the product of the maximum principal curvature and the characteristic shell size reaches the magnitude of unity or larger, the resulting shell theory cannot be treated by natural extension of plate theories. Lastly, it is demonstrated that, for moderate shear forces and comparable material properties, a leading-order multi-layered shell theory derived from asymptotic analysis should suffice to output satisfactory predictions over the shell stiffness, as well as its internal stress distribution. Numerical examples of the deformation and strength analysis for MTSs are also presented to show the reliability of the leading-order model.
\end{abstract}

\begin{keyword}
	\texttt{Multi-layered thin shell, Surface shear force, Transverse shear stress, Shell strength, Shell classification, Asymptotic analysis}
\end{keyword}

\end{frontmatter}

\section{Introduction}
Shell structures, owing to their excellence in load-bearing behavior and containment of space, have seen extensive applications across various engineering disciplines including civil, aerospace, biomechanics, etc. Moreover, they offer a visually striking and aesthetically pleasing solution for the realm of architectural design.

Multi-layered thin shells (MTSs), as a crucial component thereof, further enhance the versatility of material distribution to accommodate various external loading conditions and deformation patterns. Due to the geometric characteristics of such structures, where the dimension along shell thickness direction is significantly smaller than the other two surface dimensions, conducting three-dimensional (3D) solid-based analyses using direct brick elements or solid shell elements (\cite{klinkel1999continuum}) entails substantial computational overhead in order to accurately capture the stress variation along that direction, which is a cumbersome yet unnecessary task. A prevalent approach involves introducing assumptions regarding structural responses implied by the geometry, thereby associating internal field variables with the mid-surface of the shell to derive a reduced two-dimensional (2D) model. This facilitates problem analysis while allowing the outcomes to be readily extrapolated to the entire configuration based on the presumed response patterns.

In the majority of prior endeavors to achieve the aforementioned 3D-to-2D transformation, pioneered by \cite{love1892treatise}, \cite{reissner1941new}, \cite{reissner1952stress}, \cite{naghdi1957theory}, \cite{naghdi1957effect}, \cite{sanders1960improved}, \cite{koiter1960consistent}, considerable attention has been devoted to establishing appropriate stress-strain relations that are compatible with the original 3D constitutive equations of the structure. In general, there are two primary strategies: the axiomatic approach and the asymptotic approach. The former requires a ``priori guess'' regarding the deformation modes of the shell under study, typically guided by experimental or empirical insights. According to this assumption, shell deformation can be approximately described by the stretching and bending of its mid-surface. From the earliest days of Love’s introduction of Kirchhoff's assumptions on plate bending to the solution for isotropic shells (\cite{love1892treatise}), axiom-based shell theories have come a long way, but they have basically revolved around the quest for more accurate modeling of transverse stresses. In classical shell theory (CST), the well-known Kirchhoff-Love hypothesis postulates a state of plane strain while simultaneously neglecting the effect of transverse stresses in the stress-strain relationships. This yields a set of governing equations for three displacements on the shell mid-surface, effectively characterizing the mechanical response of isotropic homogeneous thin shells, and the corresponding error introduced by the simplification is roughly of $\mO\left(h/R\right)$ (\cite{koiter1960consistent}), where $h$ is the overall thickness and $R$ denotes the minimum principal radius of curvature of the shell mid-surface. However, due to trivial transverse strains, CST is limited in its applicability to more complex scenarios, such as shear deformation.

In response to the problems mentioned above, the subsequent improved theories partially or completely discard the Kirchhoff-Love hypothesis including assumptions regarding the smallness of $h/R$ and the non-deformability of the normal. \cite{byrne1944theory}, \cite{flugge2013stresses} and \cite{vlasov1951basic} were more careful in their treatment of $h/R$, retaining terms that were initially considered negligible in the strain-displacement relationships as well as in the stress resultant equations. In contrast to the results obtained directly from the Kirchhoff-Love hypothesis, which are called ``the first-order approximations'', theories derived along this line are referred to as ``the second-order approximations''. Another class of second-order approximations enjoy a solid foundation rooted in the variational principle proposed by \cite{reissner1950variational} and take into account the effects of transverse shear and normal stresses by either choosing more flexible expressions for the normal displacement (\cite{naghdi1957theory}), directly incorporating effects of thickness variation into the transverse stress components (\cite{naghdi1957effect}), or a combination of both (\cite{reissner1952stress}). Similar to the aforementioned methods, none of these approaches provide solid, explicit reasons for these choices or a discussion on the need for modifications.

Reissner-Mindlin theory, also known as the first-order shear deformation theory (FSDT), was first proposed by \cite{reissner1945effect} and \cite{1951Influence} for elastic plates and then extended to laminated shells (\cite{dong1972laminated}). This hypothesis relaxes the normality constraint while preserving the original inextensibility condition. The resulting theory introduces two additional variables in the in-plane displacements to account for rotations of the normal and incorporates constitutive relations for transverse shear stresses. The shear correction factor (SCF) $K$ is adopted to partially address inaccuracies arising from the uniform distribution of shear strains along the thickness direction, thereby providing accurate strain energy contributions. For isotropic, homogeneous elastic problems, a commonly used value for $K$, as suggested by \cite{reissner1945effect}, is $\frac{5}{6}$, while specific values catering for more complicated cases such as orthotropic laminates (\cite{10.1115/1.3422950}), sandwich structures (\cite{birman2002choice}), and layered plates and shells (\cite{gruttmann2017shear}) are also available, albeit requiring much more effort. The diversity in the value of $K$ reflects, to some extent, the ongoing need for improvement in the FSDT.

For multi-layered or laminated shells, analyses based on the FSDT encounter challenges. Firstly, the uniformly distributed transverse shear strain fails to accurately capture the true distribution in multi-layered structures. Secondly, determining the SCF for shells with complex cross-sections poses difficulties. In this context, higher order theories (HOTs) turn out to be more appropriate for considering shear effects, where more enriched kinematic assumptions for the displacement field are generally adopted. When the assumed displacement patterns are applied to all layers simultaneously, it leads to the equivalent single layer (ESL) method (\cite{abrate2017equivalent}, \cite{TORNABENE2022109119}), which is computationally efficient as the unknowns are independent of the number of material layers. However, it struggles to satisfy interlaminar stress continuities and surface boundary conditions. In contrast, layerwise theories (LWTs) apply the assumed patterns independently to each ply (\cite{reddy1987generalization}, \cite{barbero1990general}), ensuring perfect bonding between plies. However, LWTs require a large number of degrees of freedom to model each stack layer individually. The advantages of the above two can be integrated within the framework of zigzag theories (refer to the review work of \cite{carrera2003historical} for more details). \cite{carrera2003theories} later proposed a unified method, i.e., the Carrera unified formulation (CUF), by choosing more general expansions (with prescribed orders of expansion and types of base function) for the displacement field. And the axiomatic/asymptotic technique has been subsequently developed under the CUF framework, which presents systematic asymptotic analyses based on initially-assumed expressions. The error caused by the deactivation of each term included in the expressions is examined, so as to maintain all terms having significant contributions to the displacement pattern. These models tend to address shear deformation by employing more complex expressions, such as higher-order or piecewise functions, a path that seems endless. Furthermore, they rarely examine the specific orders of field variables in shells with varying curvature, nor do they assess the influence of shear deformation on the internal stress distribution to determine whether such complexity is warranted.

The aforementioned axiomatic approaches pursue a more accurate modeling of shells by either pre-assuming the displacement expressions or adding terms without clear physical backgrounds. These operations introduce many ambiguities to the analytical process. For instance, while dropping, retaining and adding certain terms in various literatures do improve the accuracy of the original theory to some extent, the exact order of their effects is difficult to assess. And most likely, as \cite{koiter1960consistent} pointed out, many apparently different refinements of the first approximation are essentially of the same order of magnitude. To address such an issue, asymptotic approaches (\cite{gol1963derivation}, \cite{berdichevskii1979variational}, \cite{goldenveizer1966principles}, \cite{yu2002asymptotic}, \cite{ciarlet2019asymptotic} \cite{zhao2022revisiting}, \cite{zhao2023general}) have been proposed to provide approximate shell formulations with known accuracy from a mathematically rigorous perspective. Theories of this type typically feature a small parameter implied by physical characteristics, with the corresponding variables and equations expressed as series expansions. This allows the shell equations to be truncated to the desired order. \cite{gol1963derivation} and \cite{goldenveizer1966principles} were among the first to successfully derive 2D asymptotic forms of the original 3D dynamic equations for isotropic plates and shells. \cite{ciarlet2019asymptotic} demonstrated the possibility of asymptotically justifying the intrinsic equations of Koiter’s model regarding linearly elastic shells from the full 3D equations. Note also that, the classical plate theory based on the Kirchhoff-Love hypothesis has also been shown to be asymptotically justified (\cite{zhao2022revisiting}). In addition, \cite{berdichevskii1979variational} and \cite{yu2002asymptotic} developed the variational asymptotic method (VAM) by performing asymptotic analysis (AA) directly on the full three-dimensional energy formulation. Through this approach, the original 3D problem is decoupled into a 1D, through-the-thickness problem and a 2D, surface problem. Despite having encompassed all terms whose influences are of the same order of magnitude, the development of asymptotic approaches turns out to be more difficult than axiomatic ones, mainly due to the need to address separate analysis for each parameter. So, for such methods, it is crucial to achieve satisfactory accuracy at the first few orders or even at the leading order. In plate structures, especially microstructural plates (\cite{zhao2023general}), a leading-order asymptotic theory has been shown to be sufficient for modeling the main behavior without the need to account for additional transverse shear strain effects.

In this paper, we propose a method based on AA for a more rational modeling of MTSs and a clearer interpretation of their deformation modes. Through AA of the 3D equations, we identify three distinct shell deformation patterns corresponding to weakly, normally, and significantly curved shells, and determine the specific orders of relevant quantities such as the displacement and stress components, external loads, observation time, within each curvature regime. In fact, curvature, as a complex geometric factor, introduces much greater diversity into shell models. For instance, in cases of normal curvature, the normal displacement is of the same order of magnitude as the in-plane components, rather than predominating over them as it does in plates or weakly curved shells. Therefore, unlike previous theories, we believe that the shell model is not a direct extension of thin plate theories. While computations may still be carried out under traditional order-of-magnitude assumptions, it is important to recognize that for shells with larger curvature, the scalings of field variables can fundamentally differ. According to AA, surface shear forces primarily affect the redistribution of in-plane stress components at the leading order, while the transverse shear effects enter the equations at higher-order equilibrium. The leading-order transverse shear stresses can be obtained directly from equilibrium equations without the need to incorporate shear effects into the stiffness matrix. Furthermore, since our formulations for each order are derived directly from the full 3D equations, quantities typically neglected in classical theories are inherently accounted for in an asymptotic manner---they vanish at lower orders and appear explicitly at higher orders. Thus, this proposed theory remains fully consistent with the three-dimensional equations. For clarity, we compare the proposed model with several representative classical theories in Table~\ref{Table1_theories_comparison}, focusing on key aspects such as the necessity of kinematic assumptions, the number of unknowns, compatibility with three-dimensional model, and whether the discussion includes the order of magnitude of displacements. In the table below, $N$ represents the expansion order with respect to thickness coordinate in the CUF, and $L$ denotes the number of layers in the shell under investigation.
\begin{table}
	\centering
	\small
	\setlength{\abovecaptionskip}{-0.cm}
	\setlength{\belowcaptionskip}{0.35cm}
	\caption{Comparison of the proposed model with representative classical shell theories.\label{Table1_theories_comparison}}
	\renewcommand{\arraystretch}{1.4} % adjusting row spacing
	\begin{tabular}{ccccc}
		\hline
		Shell & Kinematic & Number of & Compatibility with & Discussion on \\
		Theories & Assumptions & Unknowns & 3D Elasticity & $u_i$ Orders\\
		\hline
		CST & $\gamma_{\alpha3}=\varepsilon_{33}=0$ & 3 & \faTimes & \faTimes \\
		FSDT & $\gamma_{\alpha3}\neq0,\ \varepsilon_{33}=0$ & 5 & \faTimes & \faTimes \\
		CUF & $\vu=F_\tau\vu_\tau,\ \tau=0,\,\cdots,N$ & $3\left(N+1\right)$ & \faCheck & \faTimes \\
		 & The form of $F_\tau(\wxi_3)$ & & & \\
		Reddy's & $\vu=\sum_{k=1}^{L+1}\vu_k(\wxi_1,\wxi_2)\varphi_k(\wxi_3)$ & $3\left(L+1\right)$ & \faCheck & \faTimes \\
		Layerwise & The form of $\varphi_k(\wxi_3)$ & & & \\
		Present & \faTimes & 3 & \faCheck & \faCheck \\
		\hline
	\end{tabular}
\end{table}

The remainder of this article is organized as follows. Sec.~\ref{Sec_problem_settings} introduces basic problem settings, including the representation of shell geometry, governing equations formulated in the orthogonal curvilinear coordinate system, and the associated boundary conditions. Following this, AA is performed on these equations based on a small parameter implied by geometric thinness. In Sec.~\ref{Sec_asymptotic_analysis}, the cases of weakly, normally, and significantly curved shells are discussed, culminating in a unified leading-order shell formulations. Corresponding isogeometric discretizations of the involved equations are also provided. Specific numerical examples are presented in Sec.~\ref{Sec_numerical_example} All examined cases are benchmarked against results from finite element analysis (FEA) with extremely fine mesh to validate the proposed model and to clarify the role of surface shear forces. Finally, Sec.~\ref{Sec_conclusions} concludes the article.

\section{Problem Settings}\label{Sec_problem_settings}
\subsection{Shell Geometry}
An MTS consisting of $N$ layers, where the thickness ratio of each layer from bottom to top with respect to the overall thickness $h$ is denoted by $\lambda_{\sss\CL},\,\CL = 1\,\cdots,N$, is considered here first, with a detailed illustration shown in Fig.~\ref{Fig_shell_geometry}. For the referenced geometry of the shell mid-surface, it can be generally parametrized by
\begin{equation}\label{representation_midsurface}
	\CS:\,\vr = \vr\left(\wxi_1,\wxi_2\right),
\end{equation}
where $\wxi_\alpha,\,\alpha = 1,2$ are two selected parameters, and the symbol ``$\wedge$'' attached to a variable represents its non-dimensional counterpart. Thus the length scale is actually stored in $\vr$. To facilitate further analysis, unless specified, Greek letters are adopted throughout the article to indicate components associated with the tangent plane at each point, e.g., $\alpha = 1,2$, and Latin letters are used to represent quantities that take values in the whole three-dimensional space, e.g., $i = 1,2,3$.

Based on the mid-surface representation given by Eq.~\eqref{representation_midsurface}, any point inside an MTS of thickness $h$ can be obtained by offsetting a specific point on $\CS$ by $h\,\wxi_3$ along the normal direction $\ve_3$, that is,
\begin{equation}\label{shell_3d}
	\Omega:\,\vr_\mathrm{s} = \vr\left(\wxi_1,\wxi_2\right)+h\,\wxi_3\,\ve_3,
\end{equation}
where $\ve_3$, as illustrated in Fig.~\ref{Fig_shell_geometry}, is the unit normal to the shell mid-surface $\CS$, and $\wxi_3$ takes its values within the range $\left[-0.5,0.5\right]$.

\begin{figure}[!ht]
	\centering
	% Pattern Info
	\tikzset{
		pattern size/.store in=\mcSize, 
		pattern size = 5pt,
		pattern thickness/.store in=\mcThickness, 
		pattern thickness = 0.3pt,
		pattern radius/.store in=\mcRadius, 
		pattern radius = 1pt}
	\makeatletter
	\pgfutil@ifundefined{pgf@pattern@name@_qen3rkook}{
		\pgfdeclarepatternformonly[\mcThickness,\mcSize]{_qen3rkook}
		{\pgfqpoint{0pt}{0pt}}
		{\pgfpoint{\mcSize+\mcThickness}{\mcSize+\mcThickness}}
		{\pgfpoint{\mcSize}{\mcSize}}
		{
			\pgfsetcolor{\tikz@pattern@color}
			\pgfsetlinewidth{\mcThickness}
			\pgfpathmoveto{\pgfqpoint{0pt}{0pt}}
			\pgfpathlineto{\pgfpoint{\mcSize+\mcThickness}{\mcSize+\mcThickness}}
			\pgfusepath{stroke}
	}}
	\makeatother
	\tikzset{every picture/.style={line width=0.75pt}} %set default line width to 0.75pt
	\begin{tikzpicture}[x=0.75pt,y=0.75pt,yscale=-1,xscale=1]
		%uncomment if require: \path (0,310); %set diagram left start at 0, and has height of 310
		%Straight Lines [id:da36918328584380244] 
		\draw    (176.5,163.92) -- (210.33,164.11) ;
		\draw [shift={(210.33,164.11)}, rotate = 0.33] [color={rgb, 255:red, 0; green, 0; blue, 0 }  ][fill={rgb, 255:red, 0; green, 0; blue, 0 }  ][line width=0.75]      (0, 0) circle [x radius= 1.34, y radius= 1.34]   ;
		%Straight Lines [id:da9053174159274353] 
		\draw    (256.33,62.44) -- (265.83,82.44) ;
		%Shape: Brace [id:dp6311501171804732] 
		\draw  [color={rgb, 255:red, 74; green, 74; blue, 74 }  ,draw opacity=1 ] (433.57,97.05) .. controls (436.4,95.72) and (437.14,93.64) .. (435.81,90.81) -- (435.81,90.81) .. controls (433.9,86.78) and (434.36,84.09) .. (437.18,82.76) .. controls (434.36,84.09) and (432,82.74) .. (430.09,78.71)(430.95,80.53) -- (430.09,78.71) .. controls (428.76,75.89) and (426.68,75.15) .. (423.86,76.48) ;
		%Straight Lines [id:da35523365447229294] 
		\draw [line width=0.75]  [dash pattern={on 4.5pt off 4.5pt}]  (405.29,129.23) -- (474.56,105.28) ;
		%Straight Lines [id:da07233732541508364] 
		\draw [line width=0.75]  [dash pattern={on 4.5pt off 4.5pt}]  (386.16,161.83) -- (471.56,191.28) ;
		%Curve Lines [id:da9403222831769158] 
		\draw    (470.1,109.05) .. controls (506.71,93.14) and (528.05,110) .. (560.05,100.86) ;
		%Curve Lines [id:da4928935069228082] 
		\draw    (476.01,193.48) .. controls (525,183.71) and (540.37,203.56) .. (566.9,189.43) ;
		%Curve Lines [id:da12959440785184606] 
		\draw    (450.9,134) .. controls (508.05,112) and (548.33,134.29) .. (584.05,121.43) ;
		%Curve Lines [id:da15482765093062012] 
		\draw    (457.29,176.57) .. controls (530.62,156.86) and (532.62,184) .. (586.33,168.29) ;
		%Curve Lines [id:da9373229637198448] 
		\draw    (448.62,157.43) .. controls (524.62,131.14) and (551.76,162.57) .. (592.33,142) ;
		%Straight Lines [id:da537292749553314] 
		\draw    (109.19,210.96) -- (98.15,223.35) ;
		\draw [shift={(96.82,224.85)}, rotate = 311.69] [fill={rgb, 255:red, 0; green, 0; blue, 0 }  ][line width=0.08]  [draw opacity=0] (12,-3) -- (0,0) -- (12,3) -- cycle    ;
		%Straight Lines [id:da6638219130844178] 
		\draw    (109.19,210.96) -- (129.57,210.63) ;
		\draw [shift={(131.57,210.6)}, rotate = 179.09] [fill={rgb, 255:red, 0; green, 0; blue, 0 }  ][line width=0.08]  [draw opacity=0] (12,-3) -- (0,0) -- (12,3) -- cycle    ;
		%Straight Lines [id:da20486534575707882] 
		\draw    (109.19,210.96) -- (109.13,191.49) ;
		\draw [shift={(109.13,189.49)}, rotate = 89.82] [fill={rgb, 255:red, 0; green, 0; blue, 0 }  ][line width=0.08]  [draw opacity=0] (12,-3) -- (0,0) -- (12,3) -- cycle    ;
		%Straight Lines [id:da3721198640803509] 
		\draw    (358.83,224.78) -- (367.89,242.94) ;
		%Straight Lines [id:da5574307354350003] 
		\draw [color={rgb, 255:red, 74; green, 74; blue, 74 }  ,draw opacity=1 ]   (199.18,215.33) -- (364.58,239.98) ;
		\draw [shift={(366.56,240.28)}, rotate = 188.48] [fill={rgb, 255:red, 74; green, 74; blue, 74 }  ,fill opacity=1 ][line width=0.08]  [draw opacity=0] (12,-3) -- (0,0) -- (12,3) -- cycle    ;
		\draw [shift={(197.2,215.03)}, rotate = 8.48] [fill={rgb, 255:red, 74; green, 74; blue, 74 }  ,fill opacity=1 ][line width=0.08]  [draw opacity=0] (12,-3) -- (0,0) -- (12,3) -- cycle    ;
		%Curve Lines [id:da4934550417558048] 
		\draw [line width=0.75]    (265.83,82.44) .. controls (306.17,76.11) and (424.83,100.44) .. (430.83,97.11) .. controls (436.83,93.78) and (365.5,214.11) .. (358.83,224.78) .. controls (352.17,235.44) and (201.83,202.11) .. (193.3,206.75) .. controls (184.77,211.39) and (226.5,124.44) .. (265.83,82.44) -- cycle ;
		%Curve Lines [id:da27283659724230946] 
		\draw [fill={rgb, 255:red, 74; green, 144; blue, 226 }  ,fill opacity=0.75 ][line width=0.75]  [dash pattern={on 4.5pt off 4.5pt}]  (261.08,72.44) .. controls (301.42,66.11) and (420.08,90.44) .. (426.08,87.11) .. controls (432.08,83.78) and (360.75,204.11) .. (354.08,214.78) .. controls (347.42,225.44) and (197.08,192.11) .. (188.55,196.75) .. controls (180.02,201.39) and (221.75,114.44) .. (261.08,72.44) -- cycle ;
		%Straight Lines [id:da06349636831487304] 
		\draw    (307.17,147.89) -- (350.26,153.55) ;
		\draw [shift={(352.24,153.81)}, rotate = 187.48] [color={rgb, 255:red, 0; green, 0; blue, 0 }  ][line width=0.75]    (10.93,-3.29) .. controls (6.95,-1.4) and (3.31,-0.3) .. (0,0) .. controls (3.31,0.3) and (6.95,1.4) .. (10.93,3.29)   ;
		%Straight Lines [id:da17652518902258585] 
		\draw    (307.17,147.89) -- (329.23,110.11) ;
		\draw [shift={(330.24,108.38)}, rotate = 120.28] [color={rgb, 255:red, 0; green, 0; blue, 0 }  ][line width=0.75]    (10.93,-3.29) .. controls (6.95,-1.4) and (3.31,-0.3) .. (0,0) .. controls (3.31,0.3) and (6.95,1.4) .. (10.93,3.29)   ;
		%Straight Lines [id:da5226747657862965] 
		\draw    (307.17,147.89) -- (286.59,108.38) ;
		\draw [shift={(285.67,106.61)}, rotate = 62.48] [color={rgb, 255:red, 0; green, 0; blue, 0 }  ][line width=0.75]    (10.93,-3.29) .. controls (6.95,-1.4) and (3.31,-0.3) .. (0,0) .. controls (3.31,0.3) and (6.95,1.4) .. (10.93,3.29)   ;
		%Straight Lines [id:da8721607042381831] 
		\draw [color={rgb, 255:red, 208; green, 2; blue, 27 }  ,draw opacity=0.8 ]   (307.17,147.89) -- (327.72,150.59) ;
		\draw [shift={(329.7,150.85)}, rotate = 187.48] [color={rgb, 255:red, 208; green, 2; blue, 27 }  ,draw opacity=0.8 ][line width=0.75]    (10.93,-3.29) .. controls (6.95,-1.4) and (3.31,-0.3) .. (0,0) .. controls (3.31,0.3) and (6.95,1.4) .. (10.93,3.29)   ;
		%Straight Lines [id:da9907434091852221] 
		\draw [color={rgb, 255:red, 208; green, 2; blue, 27 }  ,draw opacity=0.8 ]   (307.17,147.89) -- (317.7,129.86) ;
		\draw [shift={(318.7,128.13)}, rotate = 120.28] [color={rgb, 255:red, 208; green, 2; blue, 27 }  ,draw opacity=0.8 ][line width=0.75]    (10.93,-3.29) .. controls (6.95,-1.4) and (3.31,-0.3) .. (0,0) .. controls (3.31,0.3) and (6.95,1.4) .. (10.93,3.29)   ;
		%Straight Lines [id:da7841138739005244] 
		\draw [color={rgb, 255:red, 208; green, 2; blue, 27 }  ,draw opacity=0.8 ]   (307.17,147.89) -- (297.34,129.02) ;
		\draw [shift={(296.42,127.25)}, rotate = 62.48] [color={rgb, 255:red, 208; green, 2; blue, 27 }  ,draw opacity=0.8 ][line width=0.75]    (10.93,-3.29) .. controls (6.95,-1.4) and (3.31,-0.3) .. (0,0) .. controls (3.31,0.3) and (6.95,1.4) .. (10.93,3.29)   ;
		%Curve Lines [id:da10078135792170562] 
		\draw [line width=0.75]    (256.33,62.44) .. controls (296.67,56.11) and (415.33,80.44) .. (421.33,77.11) .. controls (427.33,73.78) and (356,194.11) .. (349.33,204.78) .. controls (342.67,215.44) and (192.33,182.11) .. (183.8,186.75) .. controls (175.27,191.39) and (217,104.44) .. (256.33,62.44) -- cycle ;
		%Straight Lines [id:da4131884163506119] 
		\draw    (183.38,187.25) -- (192.88,207.25) ;
		%Straight Lines [id:da1437215775124161] 
		\draw    (349.33,204.78) -- (358.83,224.78) ;
		%Straight Lines [id:da1887836262145226] 
		\draw    (421.33,77.11) -- (430.83,97.11) ;
		%Shape: Ellipse [id:dp04916780900684681] 
		\draw  [dash pattern={on 4.5pt off 4.5pt}][line width=0.75]  (386.16,161.83) .. controls (379.19,157.73) and (377.82,147.12) .. (383.1,138.12) .. controls (388.38,129.12) and (398.31,125.14) .. (405.29,129.23) .. controls (412.26,133.32) and (413.63,143.94) .. (408.35,152.94) .. controls (403.07,161.94) and (393.14,165.92) .. (386.16,161.83) -- cycle ;
		%Shape: Ellipse [id:dp4897499962775622] 
		\draw  [dash pattern={on 5.63pt off 4.5pt}][line width=1.5]  (447.61,152.49) .. controls (445.75,121.69) and (476.56,94.78) .. (516.43,92.36) .. controls (556.3,89.95) and (590.13,112.96) .. (591.99,143.76) .. controls (593.85,174.55) and (563.04,201.47) .. (523.17,203.88) .. controls (483.3,206.29) and (449.47,183.28) .. (447.61,152.49) -- cycle ;
		%Straight Lines [id:da6450997999601913] 
		\draw    (192.88,207.25) -- (199.89,220.61) ;
		%Shape: Ellipse [id:dp8195811363142982] 
		\draw  [pattern=_qen3rkook,pattern size=6pt,pattern thickness=0.75pt,pattern radius=0pt, pattern color={rgb, 255:red, 0; green, 0; blue, 0}] (237.47,165.19) .. controls (234.6,159.07) and (239.82,150.56) .. (249.14,146.19) .. controls (258.47,141.81) and (268.35,143.23) .. (271.22,149.35) .. controls (274.09,155.47) and (268.86,163.97) .. (259.54,168.35) .. controls (250.22,172.72) and (240.34,171.31) .. (237.47,165.19) -- cycle ;
		%Straight Lines [id:da12146999407720882] 
		\draw    (259.62,131.52) -- (263.26,150.69) ;
		\draw [shift={(263.82,153.64)}, rotate = 259.25] [fill={rgb, 255:red, 0; green, 0; blue, 0 }  ][line width=0.08]  [draw opacity=0] (8.93,-4.29) -- (0,0) -- (8.93,4.29) -- cycle    ;
		%Straight Lines [id:da2508187066989849] 
		\draw    (239.42,138.6) -- (242.92,159.24) ;
		\draw [shift={(243.42,162.2)}, rotate = 260.38] [fill={rgb, 255:red, 0; green, 0; blue, 0 }  ][line width=0.08]  [draw opacity=0] (8.93,-4.29) -- (0,0) -- (8.93,4.29) -- cycle    ;
		%Straight Lines [id:da32572348664981776] 
		\draw    (250.57,137.62) -- (255.77,162.36) ;
		\draw [shift={(256.39,165.3)}, rotate = 258.13] [fill={rgb, 255:red, 0; green, 0; blue, 0 }  ][line width=0.08]  [draw opacity=0] (8.93,-4.29) -- (0,0) -- (8.93,4.29) -- cycle    ;
		%Straight Lines [id:da5472964485726302] 
		\draw    (188.5,142.08) -- (212.33,142.11) ;
		\draw [shift={(212.33,142.11)}, rotate = 0.07] [color={rgb, 255:red, 0; green, 0; blue, 0 }  ][fill={rgb, 255:red, 0; green, 0; blue, 0 }  ][line width=0.75]      (0, 0) circle [x radius= 1.34, y radius= 1.34]   ;
		%Straight Lines [id:da5079365139183334] 
		\draw    (363,62.75) -- (332.27,73.14) ;
		\draw [shift={(327.5,74.75)}, rotate = 161.32] [color={rgb, 255:red, 0; green, 0; blue, 0 }  ][line width=0.75]      (0, 0) circle [x radius= 6.03, y radius= 6.03]   ;
		%Straight Lines [id:da21436448936053543] 
		\draw    (201,119.08) -- (214.83,119.11) ;
		\draw [shift={(214.83,119.11)}, rotate = 0.12] [color={rgb, 255:red, 0; green, 0; blue, 0 }  ][fill={rgb, 255:red, 0; green, 0; blue, 0 }  ][line width=0.75]      (0, 0) circle [x radius= 1.34, y radius= 1.34]   ;
		
		% Text Node
		\draw (438.45,74.89) node [anchor=north west][inner sep=0.75pt]  [font=\normalsize] [align=left] {\begin{minipage}[lt]{8.67pt}\setlength\topsep{0pt}
				\begin{center}
					{\fontfamily{ptm}\selectfont \textcolor[rgb]{0.29,0.29,0.29}{{\small $h$}}}
				\end{center}
				
		\end{minipage}};
		% Text Node
		\draw (330.2,90.48) node [anchor=north west][inner sep=0.75pt]  [font=\footnotesize]  {$\widehat{\xi _{2}}$};
		% Text Node
		\draw (515.37,145.76) node [anchor=north west][inner sep=0.75pt]  [font=\normalsize,rotate=-271.2]  {$\cdots $};
		% Text Node
		\draw (498.99,174.25) node [anchor=north west][inner sep=0.75pt]  [font=\small,rotate=-1.28] [align=left] {{\fontfamily{ptm}\selectfont Layer 1}};
		% Text Node
		\draw (499.56,150.76) node [anchor=north west][inner sep=0.75pt]  [font=\small,rotate=-1.78] [align=left] {{\fontfamily{ptm}\selectfont Layer 2}};
		% Text Node
		\draw (498.78,105.65) node [anchor=north west][inner sep=0.75pt]  [font=\small,rotate=-2.19] [align=left] {{\fontfamily{ptm}\selectfont Layer N}};
		% Text Node
		\draw (132.83,201.22) node [anchor=north west][inner sep=0.75pt]  [font=\footnotesize]  {$\widetilde{\boldsymbol{e_{2}}}$};
		% Text Node
		\draw (113.37,182.87) node [anchor=north west][inner sep=0.75pt]  [font=\footnotesize]  {$\widetilde{\boldsymbol{e_{3}}}$};
		% Text Node
		\draw (84.61,224) node [anchor=north west][inner sep=0.75pt]  [font=\footnotesize]  {$\widetilde{\boldsymbol{e_{1}}}$};
		% Text Node
		\draw (282.81,131.8) node [anchor=north west][inner sep=0.75pt]  [font=\footnotesize]  {$\boldsymbol{e_{3}}$};
		% Text Node
		\draw (302.84,119.57) node [anchor=north west][inner sep=0.75pt]  [font=\footnotesize]  {$\boldsymbol{e_{2}}$};
		% Text Node
		\draw (319.1,154.1) node [anchor=north west][inner sep=0.75pt]  [font=\footnotesize]  {$\boldsymbol{e_{1}}$};
		% Text Node
		\draw (354.65,147.45) node [anchor=north west][inner sep=0.75pt]  [font=\footnotesize]  {$\widehat{\xi _{1}}$};
		% Text Node
		\draw (270.99,89.47) node [anchor=north west][inner sep=0.75pt]  [font=\footnotesize]  {$\widehat{\xi _{3}}$};
		% Text Node
		\draw (273.27,233.58) node [anchor=north west][inner sep=0.75pt]   [align=left] {\begin{minipage}[lt]{8.95pt}\setlength\topsep{0pt}
				\begin{center}
					{\fontfamily{ptm}\selectfont \textcolor[rgb]{0.29,0.29,0.29}{$L$}}
				\end{center}
				
		\end{minipage}};
		% Text Node
		\draw (333.63,188.55) node [anchor=north west][inner sep=0.75pt]  [font=\small]  {$\si{\ohm}$};
		% Text Node
		\draw (168.99,134.43) node [anchor=north west][inner sep=0.75pt]  [font=\small]  {$\CS$};
		% Text Node
		\draw (297.05,153.16) node [anchor=north west][inner sep=0.75pt]  [font=\small]  {$P$};
		% Text Node
		\draw (369.05,52.73) node [anchor=north west][inner sep=0.75pt]  [font=\small]  {$\partial \mathrm{_{u}} \si{\ohm}$};
		% Text Node
		\draw (244.51,121.54) node [anchor=north west][inner sep=0.75pt]    {$\boldsymbol{p}$};
		% Text Node
		\draw (246.04,175.25) node [anchor=north west][inner sep=0.75pt]  [font=\small]  {$\partial \mathrm{_{t}} \si{\ohm}$};
		% Text Node
		\draw (180.99,110.43) node [anchor=north west][inner sep=0.75pt]  [font=\small]  {$\CS_{\mathrm{t}}$};
		% Text Node
		\draw (154.99,156.43) node [anchor=north west][inner sep=0.75pt]  [font=\small]  {$\CS_{\mathrm{b}}$};
		% Text Node
		\draw (592.33,112.43) node [anchor=north west][inner sep=0.75pt]  [font=\small]  {$\CS_{N-1}$};
		% Text Node
		\draw (595.66,162.43) node [anchor=north west][inner sep=0.75pt]  [font=\small]  {$\CS_{1}$};
		% Text Node
		\draw (601.37,154.43) node [anchor=north west][inner sep=0.75pt]  [font=\normalsize,rotate=-271.2]  {$\cdots $};
	\end{tikzpicture}
	\caption{Representation of the multi-layered thin shell. The overall shell entity is denoted by $\Omega$ and the shell mid-surface (blue part) is denoted by $\CS$, with $\CS_{\mathrm{b}}$ and $\CS_{\mathrm{t}}$ being the bottom and top surfaces of the structure, respectively. $\CS_I\left(I=1,\,\cdots,N-1\right)$ represents the interface between two adjacent layers $I$ and $\left(I+1\right)$. A set of orthogonal curvilinear coordinate systems $\left(\wxi_1,\wxi_2,\wxi_3\right)$ is attached to $\CS$, and the corresponding in-plane orthonormal basis vectors $\ve_\alpha$ are aligned to the local principal directions of the shell mid-surface. Another orthonormal basis $\left\lbrace\tve_1,\tve_2,\tve_3\right\rbrace$ are aligned to directions of the global Cartesian coordinate system.\label{Fig_shell_geometry}}
\end{figure}

Without loss of generalities, one can always choose coordinates $\wxi_\alpha,\,\alpha = 1,2$ that parameterize ``lines of curvature'' except at umbilical points of the regular surface, which then yields an orthogonal curvilinear coordinate system $\left(\wxi_1,\wxi_2,\wxi_3\right)$. Accordingly, the first and second fundamental forms of the mid-surface with respect to two parameters are both diagonal, i.e.,
\begin{equation}\label{condition_curvature_lines}
	\pd{\vr}{\wxi_\alpha}\cdot\pd{\vr}{\wxi_\beta} = 0;\quad\mpd{\vr}{\wxi_\alpha}{\wxi_\beta}\cdot\ve_3 = 0,
\end{equation}
where the subindicies satisfy $\alpha\neq\beta$. Therefore, an orthonormal basis $\left\lbrace\ve_1,\ve_2,\ve_3\right\rbrace$, as shown in Fig.~\ref{Fig_shell_geometry}, is defined on the tangent plane of a given point $P$ on the shell mid-surface by
\begin{equation}\label{basis_midsurface}
	\ve_\alpha = \frac{1}{a_\alpha}\pd{\vr}{\wxi_\alpha};\quad\ve_3 = \ve_1\times\ve_2,
\end{equation}
note that the Einstein summation convention is not adopted here, nor will it be adopted later unless specified. And $a_\alpha$ above take the expressions
\begin{equation}
	a_\alpha\left(\wxi_1,\wxi_2\right) = \left|\pd{\vr}{\wxi_\alpha}\right|.
\end{equation}

Alternatively, the basis vectors $\ve_i$ can be defined, from the perspective of space $\Omega$ which is occupied by the shell entity (Eq.~\eqref{shell_3d}), by
\begin{equation}\label{basis_lame_coefficient}
	\ve_i = \frac{1}{h_i}\pd{\vr_\mathrm{s}}{\wxi_i};\quad h_i\left(\wxi_1,\wxi_2,\wxi_3\right) = \left|\pd{\vr_\mathrm{s}\left(\wxi_1,\wxi_2,\wxi_3\right)}{\wxi_i}\right|,
\end{equation}
where $h_i\left(\wxi_1,\wxi_2,\wxi_3\right)$ are the \textit{Lam\'{e} coefficients}.

Derivatives of the above basis vectors can be expressed based on the linear combination of themselves in general, that is,
\begin{equation}
	\pd{\ve_i}{\wxi_j} = \sum_{k}\Gamma_{ijk}\ve_k,
\end{equation}
the related coefficients $\Gamma_{ijk}$ are called the \textit{Christoffel symbols}. Having determined the 27 constants $\Gamma_{ijk}$, the expressions of the specific derivatives become
\begin{equation}\label{derivative_basis}
	\pd{\ve_i}{\wxi_j} = \frac{\ve_j}{h_i}\pd{h_j}{\wxi_i}-\delta_{ij}\sum_{k}\frac{\ve_k}{h_k}\pd{h_i}{\wxi_k},
\end{equation}
and one can refer to the works of \cite{reissner1941new} and \cite{howell_kozyreff_ockendon_2008} for detailed derivations.

To determine the mathematical expressions of the \textit{Lam\'{e} coefficients} $h_i$, one has to first calculate the terms $\pd{\ve_3}{\wxi_\alpha}$. Since the principal curvatures of a regular parametric surface $\CS$ at a point $P$ are exactly associated with the two eigenvalues of the corresponding \textit{Weingarten map} of that point, which is defined as a tangent map $g_\ast$ from the tangent space $T_PS$ to itself: $W = -g_\ast:\,T_PS\to T_PS$, we have the following relation
\begin{equation}\label{weingarten_relation}
	W\left(\ve_\alpha\right) = -\frac{1}{a_\alpha}g_\ast\left(\pd{\vr}{\wxi_\alpha}\right) = -\frac{1}{a_\alpha}\pd{\ve_3}{\wxi_\alpha} = \kappa_\alpha\ve_\alpha,
\end{equation}
where $\kappa_\alpha\left(\wxi_1,\wxi_2\right)$ are the principal curvatures of the shell mid-surface.

Based on Eqs.~\eqref{basis_midsurface} and \eqref{weingarten_relation}, the derivatives of shell parametrization $\vr_\mathrm{s}$ taken with respect to the curvilinear coordinates are expressed by
\begin{equation}
	\pd{\vr_\mathrm{s}}{\wxi_\alpha} = a_\alpha\left(1-h\,\kappa_\alpha\wxi_3\right)\,\ve_\alpha;\quad\pd{\vr_\mathrm{s}}{\wxi_3} = h\,\ve_3,
\end{equation}
so the \textit{Lam\'{e} coefficients} $h_i$ take the form
\begin{equation}
	h_\alpha = a_\alpha\left(1-h\,\kappa_\alpha\wxi_3\right);\quad h_3 = h.
\end{equation}

\subsection{Basic Equations in Orthogonal Curvilinear Coordinate Systems}
Based on the global-to-local coordinate transformation in shell representation mentioned above, one can write down the three-dimensional governing equations for an MTS in the curvilinear coordinate system, i.e.,
\begin{enumerate}
	\item[(1)] the dynamic equilibrium equations in the curvilinear coordinates read
	\begin{equation}\label{equilibrium_concise}
		\rho\,\ddot{u}_i = \ve_i\cdot\left(\nabla\cdot\vsi\right)+f_i = \sum_{j}\frac{1}{H}\pd{}{\wxi_j}\left(\frac{H\sigma_{ij}}{h_j}\right)+\sum_{k}\frac{1}{h_ih_k}\left(\sigma_{ik}\pd{h_i}{\wxi_k}-\sigma_{kk}\pd{h_k}{\wxi_i}\right)+f_i,
	\end{equation}
	where $\ddot{u}_i = \frac{\partial^2u_i}{\partial t^2}$ and $H = h_1h_2h_3$;
	\item[(2)] the strain components read
	\begin{equation}\label{geometric_concise}
		\varepsilon_{ij} = \frac{1}{2}\left(\frac{1}{h_j}\pd{u_i}{\wxi_j}-\frac{u_i}{h_ih_j}\pd{h_i}{\wxi_j}+\frac{1}{h_i}\pd{u_j}{\wxi_i}-\frac{u_j}{h_ih_j}\pd{h_j}{\wxi_i}\right)+\delta_{ij}\sum_{k}\frac{u_k}{h_jh_k}\pd{h_j}{\wxi_k};
	\end{equation}
\end{enumerate}
one also needs the (3D) constitutive laws to close the system. At each point in $\Omega$, the local curvilinear coordinate basis is found by rotating the usual Cartesian basis $\left\lbrace\tve_1,\tve_2,\tve_3\right\rbrace$, and the rotation is characterized by an orthogonal matrix $\rQ$ whose rows are the basis vectors as defined by Eq.~\eqref{basis_midsurface}, that is, $\rQ = \left(\ve_1,\ve_2,\ve_3\right)^\rT$. Thus a congruent transformation linking two second-order tensors in different orthogonal coordinates, such as $\vA = \left(a_{ij}\right)$ in $\left\lbrace\ve_1,\ve_2,\ve_3\right\rbrace$ and $\vB = \left(b_{ij}\right)$ in $\left\lbrace\tve_1,\tve_2,\tve_3\right\rbrace$, can be expressed based on the above orthogonal matrix $\rQ$
\begin{equation}\label{congruent_trans}
	\vA = \rQ\vB\rQ^T\quad \text{or}\quad a_{ij} = \ve_i^T\vB\ve_j.
\end{equation}

When considered in a fixed Cartesian coordinate system, the constitutive laws read
\begin{equation}\label{constitutive_cartesian}
	\sigma_{ij} = \lambda\left(\varepsilon_{kk}\right)\delta_{ij}+2G\varepsilon_{ij},
\end{equation}
where the two coefficients $\lambda$ and $G$ are called the ``\textit{Lam\'{e} constants}''. Based on the orthonormal basis vectors in such a coordinate system, Eq.~\eqref{constitutive_cartesian} can be re-expressed in the form of tensor
\begin{equation}
	\sigma_{ij}\tve_i\tve_j^T = \lambda\left(\varepsilon_{kk}\right)\delta_{ij}\tve_i\tve_j^T+2G\varepsilon_{ij}\tve_i\tve_j^T,
\end{equation}
it is noted that, the Einstein summation rule is adopted in the above two equations. And the congruent transformation (Eq.~\eqref{congruent_trans}) indicates that the constitutive laws remain unchanged in different orthogonal coordinate systems.

\subsection{Boundary Conditions}
For the investigated MTS, the following boundary conditions are involved for both surfaces (top and bottom) and interfaces (between laminate layers). Firstly, on the shell surfaces, we have
\begin{equation}\label{bc_surface}
	\sigma_{\alpha 3} = p_\alpha\left(\wxi_1,\wxi_2\right),\ \wxi_3 = \frac{1}{2};\quad\sigma_{33} = p_3\left(\wxi_1,\wxi_2\right),\ \wxi_3 = \frac{1}{2};\quad\sigma_{i3} = 0,\ \wxi_3 = -\frac{1}{2},
\end{equation}
where $p_i$ are the external loads per area imposed on the top surface of the shell $\CS_\mathrm{t}$.

Secondly, on the interface of any two adjacent layers, all the displacement components measured in the local curvilinear coordinate system should remain continuous. Thus
\begin{equation}\label{bc_interface_disp}
	\left.u_i\right|_-^+ = 0,\quad\text{on}\ \wxi_3 = \CS_I,
\end{equation}
where $\left.u_i\right|_-^+$ represents the difference in the values of $u_i$ on both sides of the interface, and $\CS_I$ denotes the $\wxi_3$-coordinate of the interface between the $I$-th and $(I+1)$-th layer, $I = 1,2,\,\cdots, N-1$. Moreover, the transverse stress components between layers should also be continuous, that is,
\begin{equation}\label{bc_interface_sigma}
	\left.\sigma_{i3}\right|_-^+ = 0,\quad\text{on}\ \wxi_3 = \CS_I.
\end{equation}

\section{Asymptotic Reduction of the 3D Governing Systems for the MTS}\label{Sec_asymptotic_analysis}
\subsection{Non-dimensionalization}\label{Sec_scaling}
In view of the slender nature of an MTS, and a small parameter $\epsilon$ is thus introduced by
\begin{equation}\label{epsilon}
	\epsilon = \frac{h}{L}\ll 1,
\end{equation}
where $L$ is a parameter characterizing the shell surface dimension, while $h$ is recalled to be the shell thickness.

For AA, the 3D equations listed above are preferably non-dimensionalized, and generally, non-dimensionalization is carried out in three aspects: (1) surface geometry; (2) material properties (3) field variables. Firstly, for quantities that characterize the surface geometry, the typical lenth scale $L$ is selected as the reference, that is,
\begin{equation}\label{length_scaling}
	\vr\left(\wxi_1,\wxi_2\right) = L\,\widehat{\vr}\left(\wxi_1,\wxi_2\right);\quad a_\alpha\left(\wxi_1,\wxi_2\right) = L\,\wa_\alpha\left(\wxi_1,\wxi_2\right);\quad\kappa_\alpha = \frac{1}{L}\,\wk_\alpha,
\end{equation}
here the symbol ``$\wedge$'' is recalled to denote a non-dimensional quantity. One thing worth mentioning is that the dimensionless principal curvatures $\wk_\alpha$ here cannot be hastily assumed to be of $\mO\left(1\right)$, as diverse geometries of shell structures encompasses cases where principal curvatures approach zero, e.g., weakly curved shell ($\wk_\alpha\approx 0,\text{ for }\alpha=1\text{ and }2$) or cylindrical shell ($\wk_2=0$). It is therefore necessary to categorize shells based on the exact magnitudes of principal curvatures, which will be presented in the subsequent discussions. 

Secondly, the material properties involved, such as Young's modulus and density, also need be non-dimensionalized. As we are investigating shells with multiple layers, the material properties of each layer are compared to those of the bottommost layer
\begin{equation}
	E_{\sss\CL} = E_1\cdot\wE_{\sss\CL};\quad\rho_{\sss\CL} = \rho_1\cdot\wrho_{\sss\CL},
\end{equation}
where the two introduced quantities $E_1$ and $\rho_1$ are the representative Young's modulus and density of the material assigned to \textit{Layer 1}, respectively.

Thirdly, the non-dimensionalization of certain involved field variables is directly conducted from the traditional plate theory, either based on experimentally summarized axioms (\cite{reissner1941new},\cite{vlasov1951basic}) or AA (\cite{zhao2022revisiting}). Consequently, it is found that the displacement components are non-dimensionalized by
\begin{equation}\label{disp_scaling}
	u_\alpha = \epsilon \rU\wu_\alpha,\quad u_3 = \rU\wu_3,
\end{equation}
and the stress field is typically non-dimensionalized by
\begin{equation}\label{stress_scaling}
	\sigma_{\alpha\beta} = \sigma^\star\,\wsigma_{\alpha\beta};\quad\sigma_{\alpha3} = \epsilon\,\sigma^\star\,\wsigma_{\alpha3};\quad\sigma_{33} = \epsilon^2\,\sigma^\star\,\wsigma_{33},
\end{equation}
where the introduced parameter is chosen to be $\sigma^\star = \frac{\epsilon^2\rU E_1}{h}$ according to the non-dimensionalization manipulation of the constitutive law related to the in-plane shear stress $\sigma_{12}$. Note that, although the conclusions regarding displacement and stress fields from plate theory are directly applied here, their specific order relationships necessitate further elucidation through comprehensive analyses of shell structures with different magnitudes of principal curvatures.

As partly implied by the dynamic equilibrium equation Eq.~\eqref{equilibrium_concise} and the scaling relations of field quantities Eqs.~\eqref{disp_scaling}-\eqref{stress_scaling}, the appropriate time-scale is chosen to be
\begin{equation}\label{time_scaling}
	t = \epsilon^{m_1}L\sqrt{\frac{\rho_1}{E_1}}\cdot\wt,
\end{equation}
where the exponent $m_1$ is left unspecified until the magnitude of principal curvature against the shell surface dimension $L$ is determined. Additionally, the prescribed body force vector is non-dimensionalized by
\begin{equation}\label{bodyforce_scaling}
	\vf_{\sss\CL} = \epsilon^{m_2}\frac{\rU E_1}{h^2}\cdot\widehat{\vf}_{\sss\CL},
\end{equation}
where similar to Eq.~\eqref{time_scaling}, the exponent $m_2$ remains to be undetermined for the moment. Detailed discussions on the modification of scaling relations of displacement and stress components (Eqs.~\eqref{disp_scaling}-\eqref{stress_scaling}), as well as the determination of specific orders of time and body force (Eq.~\eqref{time_scaling}-\eqref{bodyforce_scaling}), will be covered later when investigating shells of different curvature patterns in Secs.~\ref{Sec_weaklycurved_shell}, \ref{Sec_normal_shell}, and \ref{Sec_significantlycurved_shell}.

\subsection{Asymptotic Expansions}
To capture the shell behavior in different orders, the asymptotic expansions of the (non-dimensional) displacement and stress fields in terms of the small parameter $\epsilon$ are first given below
\begin{subequations}
	\begin{align}
		\wu_i &\sim \wu_i^{(0)}\left(\wxi_1,\wxi_2,\wxi_3\right)+\epsilon\wu_i^{(1)}\left(\wxi_1,\wxi_2,\wxi_3\right)+\,\cdots;\\
		\wsigma_{ij} &\sim \wsigma_{ij}^{(0)}\left(\wxi_1,\wxi_2,\wxi_3\right)+\epsilon\wsigma_{ij}^{(1)}\left(\wxi_1,\wxi_2,\wxi_3\right)+\,\cdots,
	\end{align}
\end{subequations}
where the subindicies $i,j = 1,2,3;\,\left(\wxi_1,\wxi_2\right)\in \CS,\,\wxi_3\in\left[-0.5,0.5\right]$.

Based on the previous scaling rules of field variables, the non-dimensional constitutive laws linking the local stress and displacement components are given in detail by
\begin{subequations}\label{constitutive_dimensionless}
	\begin{align}
		\begin{split}
		\epsilon^2\wsigma_{11} = &\frac{\wE_{\sss\CL}}{(1+\nu_{\sss\CL})(1-2\nu_{\sss\CL})}\left[\epsilon^2\frac{1-\nu_{\sss\CL}}{\wh_1}\left(\pd{\wu_1}{\wxi_1}+\frac{\wu_2}{\wh_2}\pd{\wh_1}{\wxi_2}-\frac{\wk_1}{\epsilon}\wu_3\wa_1\right)+\right.\\
		&\left.\epsilon^2\frac{\nu_{\sss\CL}}{\wh_2}\left(\pd{\wu_2}{\wxi_2}+\frac{\wu_1}{\wh_1}\pd{\wh_2}{\wxi_1}-\frac{\wk_2}{\epsilon}\wu_3\wa_2\right)+\nu_{\sss\CL}\pd{\wu_3}{\wxi_3}\right];\label{constitutive_dimensionless1}\end{split}\\
		\wsigma_{12} = &\frac{\wE_{\sss\CL}}{2(1+\nu_{\sss\CL})}\left[\frac{1}{\wh_1}\left(\pd{\wu_2}{\wxi_1}-\frac{\wu_2}{\wh_2}\pd{\wh_2}{\wxi_1}\right)+\frac{1}{\wh_2}\left(\pd{\wu_1}{\wxi_2}-\frac{\wu_1}{\wh_1}\pd{\wh_1}{\wxi_2}\right)\right];\label{constitutive_dimensionless2}\\
		\begin{split}
		\epsilon^2\wsigma_{22} = &\frac{\wE_{\sss\CL}}{(1+\nu_{\sss\CL})(1-2\nu_{\sss\CL})}\left[\epsilon^2\frac{\nu_{\sss\CL}}{\wh_1}\left(\pd{\wu_1}{\wxi_1}+\frac{\wu_2}{\wh_2}\pd{\wh_1}{\wxi_2}-\frac{\wk_1}{\epsilon}\wu_3\wa_1\right)+\right.\\
		&\left.\epsilon^2\frac{1-\nu_{\sss\CL}}{\wh_2}\left(\pd{\wu_2}{\wxi_2}+\frac{\wu_1}{\wh_1}\pd{\wh_2}{\wxi_1}-\frac{\wk_2}{\epsilon}\wu_3\wa_2\right)+\nu_{\sss\CL}\pd{\wu_3}{\wxi_3}\right];\label{constitutive_dimensionless3}\end{split}\\
		\epsilon^2\wsigma_{13} = &\frac{\wE_{\sss\CL}}{2(1+\nu_{\sss\CL})}\left[\frac{1}{\wh_1}\left(\pd{\wu_3}{\wxi_1}+\epsilon\wu_1\wk_1\wa_1\right)+\pd{\wu_1}{\wxi_3}\right];\label{constitutive_dimensionless4}\\
		\epsilon^2\wsigma_{23} = &\frac{\wE_{\sss\CL}}{2(1+\nu_{\sss\CL})}\left[\frac{1}{\wh_2}\left(\pd{\wu_3}{\wxi_2}+\epsilon\wu_2\wk_2\wa_2\right)+\pd{\wu_2}{\wxi_3}\right];\label{constitutive_dimensionless5}\\
		\begin{split}
		\epsilon^4\wsigma_{33} = &\frac{\wE_{\sss\CL}}{(1+\nu_{\sss\CL})(1-2\nu_{\sss\CL})}\left[\epsilon^2\frac{\nu_{\sss\CL}}{\wh_1}\left(\pd{\wu_1}{\wxi_1}+\frac{\wu_2}{\wh_2}\pd{\wh_1}{\wxi_2}-\frac{\wk_1}{\epsilon}\wu_3\wa_1\right)+\right.\\
		&\left.\epsilon^2\frac{\nu_{\sss\CL}}{\wh_2}\left(\pd{\wu_2}{\wxi_2}+\frac{\wu_1}{\wh_1}\pd{\wh_2}{\wxi_1}-\frac{\wk_2}{\epsilon}\wu_3\wa_2\right)+\left(1-\nu_{\sss\CL}\right)\pd{\wu_3}{\wxi_3}\right],\label{constitutive_dimensionless6}\end{split}
	\end{align}
\end{subequations}
where the subindex $\CL = 1,2,\,\cdots,N$ is recalled to represent the layer index occupied by \textit{Material $\CL$}.

The original governing equation Eq.~\eqref{equilibrium_concise} can be similarly non-dimensionalized as
\begin{subequations}\label{equilibrium_dimensionless}
	\begin{align}
		\begin{split}
			\epsilon^{-2m_1}\wrho_{\sss\CL}\spd{\wu_1}{\wt}-\epsilon^{m_2-3}\wf_{\sss\CL}^1&=\frac{1}{\wh_1\wh_2}\left[\pd{}{\wxi_1}\left(\wh_2\,\wsigma_{11}\right)+\pd{}{\wxi_2}\left(\wh_1\,\wsigma_{12}\right)+\wsigma_{12}\pd{\wh_1}{\wxi_2}-\wsigma_{22}\pd{\wh_2}{\wxi_1}\right]\\
			&+\pd{\wsigma_{13}}{\wxi_3}-\epsilon\frac{2\wk_1\wa_1\wsigma_{13}}{\wh_1}-\epsilon\frac{\wk_2\wa_2\wsigma_{13}}{\wh_2};\label{equilibrium_dimensionless1}
		\end{split}\\
		\begin{split}
			\epsilon^{-2m_1}\wrho_{\sss\CL}\spd{\wu_2}{\wt}-\epsilon^{m_2-3}\wf_{\sss\CL}^2&=\frac{1}{\wh_1\wh_2}\left[\pd{}{\wxi_1}\left(\wh_2\,\wsigma_{12}\right)+\pd{}{\wxi_2}\left(\wh_1\,\wsigma_{22}\right)+\wsigma_{12}\pd{\wh_2}{\wxi_1}-\wsigma_{11}\pd{\wh_1}{\wxi_2}\right]\\
			&+\pd{\wsigma_{23}}{\wxi_3}-\epsilon\frac{\wk_1\wa_1\wsigma_{23}}{\wh_1}-\epsilon\frac{2\wk_2\wa_2\wsigma_{23}}{\wh_2};\label{equilibrium_dimensionless2}
		\end{split}\\
		\begin{split}
			\epsilon^{-2m_1-2}\wrho_{\sss\CL}\spd{\wu_3}{\wt}-\epsilon^{m_2-4}\wf_{\sss\CL}^3&=\frac{1}{\wh_1\wh_2}\left[\pd{}{\wxi_1}\left(\wh_2\wsigma_{13}\right)+\pd{}{\wxi_2}\left(\wh_1\wsigma_{23}\right)\right]\\
			&+\pd{\wsigma_{33}}{\wxi_3}-\epsilon\frac{\wk_1\wa_1}{\wh_1}\left(\wsigma_{33}-\frac{1}{\epsilon^2}\wsigma_{11}\right)-\epsilon\frac{\wk_2\wa_2}{\wh_2}\left(\wsigma_{33}-\frac{1}{\epsilon^2}\wsigma_{22}\right).\label{equilibrium_dimensionless3}
		\end{split}
	\end{align}
\end{subequations}

Note that, the terms $\frac{1}{\wh_\alpha}$ here denote the inverse of the dimensionless \textit{Lam\'{e} coefficients}, which can be further expressed in the form of series expansions, that is,
\begin{equation}\label{lame_coefficient_expansion}
	\frac{1}{\wh_\alpha} = \frac{1}{\wa_\alpha\left(1-\epsilon\wk_\alpha\wxi_3\right)} = \frac{1}{\wa_\alpha}\left[1+\epsilon\wk_\alpha\wxi_3+\left(\epsilon\wk_\alpha\wxi_3\right)^2+\cdots\right],\quad\text{if }|\epsilon\wk_\alpha\wxi_3|<1.
\end{equation}

\subsection{Classification of Shell Models with Curvature Patterns}
Despite the relative insignificance of the terms $\epsilon\wk_\alpha$ in Eqs.~\eqref{constitutive_dimensionless}-\eqref{equilibrium_dimensionless} compared to 1, the exact magnitude of $\wk_\alpha$ after non-dimensionalization about the shell surface scale $L$ varies for cases of weakly curved or normally curved shells. These variations result in shifts in the order relationships of field variables across different shell types, thereby affecting structural properties such as load-bearing capacity and vibration characteristics. Hence, three specific types of shells (see Fig.~\ref{Fig_classification}) corresponding to different curvature patterns are investigated below.
\begin{figure}[!htbp]
\setlength{\abovecaptionskip}{0.2cm}
\setlength{\belowcaptionskip}{-0.3cm}
\centering
\tikzset{every picture/.style={line width=0.75pt}} %set default line width to 0.75pt
\begin{tikzpicture}[x=0.75pt,y=0.75pt,yscale=-1,xscale=1]
%uncomment if require: \path (0,465); %set diagram left start at 0, and has height of 465
%Shape: Block Arc [id:dp45187666435588447] 
\draw  [fill={rgb, 255:red, 74; green, 144; blue, 226 }  ,fill opacity=0.7 ] (202.52,86.54) .. controls (203.21,70.17) and (261.11,56.94) .. (332.43,56.94) .. controls (404.09,56.94) and (462.19,70.29) .. (462.35,86.77) -- (456.13,86.77) .. controls (455.95,73.72) and (400.64,63.16) .. (332.43,63.16) .. controls (264.6,63.16) and (209.53,73.61) .. (208.74,86.56) -- cycle ;
%Down Arrow [id:dp5781367799918236] 
\draw  [color={rgb, 255:red, 0; green, 0; blue, 0 }  ,draw opacity=0.5 ][fill={rgb, 255:red, 155; green, 155; blue, 155 }  ,fill opacity=0.5 ] (312.51,42.68) -- (322.97,42.68) -- (322.97,28) -- (343.89,28) -- (343.89,42.68) -- (354.36,42.68) -- (333.43,52.46) -- cycle ;
%Curve Right Arrow [id:dp6210656809871475] 
\draw  [fill={rgb, 255:red, 255; green, 255; blue, 255 }  ,fill opacity=1 ] (281.39,64.3) .. controls (281.39,55.68) and (274.42,48.69) .. (265.82,48.69) -- (265.82,44.02) .. controls (274.42,44.02) and (281.39,51.01) .. (281.39,59.63) ;\draw  [fill={rgb, 255:red, 255; green, 255; blue, 255 }  ,fill opacity=1 ] (281.39,59.63) .. controls (281.39,66.03) and (277.54,71.53) .. (272.04,73.94) -- (272.04,69.86) -- (265.82,77.58) -- (272.04,82.7) -- (272.04,78.61) .. controls (277.54,76.2) and (281.39,70.7) .. (281.39,64.3)(281.39,59.63) -- (281.39,64.3) ;
%Curve Left Arrow [id:dp967944492373217] 
\draw  [fill={rgb, 255:red, 255; green, 255; blue, 255 }  ,fill opacity=1 ] (383.46,59.7) .. controls (383.46,51.04) and (390.45,44.02) .. (399.08,44.02) -- (399.08,48.71) .. controls (390.45,48.71) and (383.46,55.73) .. (383.46,64.39) ;\draw  [fill={rgb, 255:red, 255; green, 255; blue, 255 }  ,fill opacity=1 ] (383.46,64.39) .. controls (383.46,70.82) and (387.31,76.34) .. (392.83,78.76) -- (392.83,82.7) -- (399.08,77.72) -- (392.83,70.14) -- (392.83,74.08) .. controls (387.31,71.66) and (383.46,66.13) .. (383.46,59.7)(383.46,64.39) -- (383.46,59.7) ;
%Shape: Block Arc [id:dp057791226887446534] 
\draw  [fill={rgb, 255:red, 74; green, 144; blue, 226 }  ,fill opacity=0.7 ] (308.06,395.58) .. controls (307.58,336.99) and (319.02,289.48) .. (333.67,289.36) .. controls (348.36,289.25) and (360.72,336.84) .. (361.28,395.68) .. controls (361.28,396.01) and (361.28,396.35) .. (361.28,396.68) -- (355.43,396.51) .. controls (355.43,396.25) and (355.43,395.99) .. (355.43,395.73) .. controls (354.9,340.12) and (345.18,295.12) .. (333.72,295.21) .. controls (322.29,295.3) and (313.43,340.23) .. (313.91,395.65) -- cycle ;
%Shape: Block Arc [id:dp4630341589414646] 
\draw  [fill={rgb, 255:red, 74; green, 144; blue, 226 }  ,fill opacity=0.7 ] (252.9,239.65) .. controls (252.9,239.6) and (252.9,239.55) .. (252.9,239.5) .. controls (252.9,195.56) and (288.52,159.95) .. (332.46,159.95) .. controls (376.37,159.95) and (411.97,195.53) .. (412.01,239.43) -- (405.8,239.43) .. controls (405.77,198.96) and (372.94,166.16) .. (332.46,166.16) .. controls (291.95,166.16) and (259.11,199) .. (259.11,239.5) .. controls (259.11,239.55) and (259.11,239.59) .. (259.11,239.64) -- cycle ;
%Down Arrow [id:dp777891023308767] 
\draw  [color={rgb, 255:red, 0; green, 0; blue, 0 }  ,draw opacity=0.5 ][fill={rgb, 255:red, 155; green, 155; blue, 155 }  ,fill opacity=0.5 ] (311.5,145.69) -- (321.96,145.69) -- (321.96,131.01) -- (342.88,131.01) -- (342.88,145.69) -- (353.35,145.69) -- (332.42,155.47) -- cycle ;
%Left Arrow [id:dp16097350075030503] 
\draw   (289.62,193.04) -- (294.28,182.61) -- (295.86,185.79) -- (312.47,177.16) -- (315.64,183.52) -- (299.03,192.15) -- (300.62,195.33) -- cycle ;
%Left Arrow [id:dp08945965438208581] 
\draw   (270.59,234.35) -- (266.37,223.69) -- (269.73,224.72) -- (275.04,206.47) -- (281.76,208.53) -- (276.45,226.78) -- (279.81,227.81) -- cycle ;
%Left Arrow [id:dp03328361703828797] 
\draw   (395.67,234.93) -- (386.17,228.77) -- (389.49,227.6) -- (383.4,209.58) -- (390.03,207.26) -- (396.11,225.28) -- (399.43,224.12) -- cycle ;
%Left Arrow [id:dp1539789703832699] 
\draw   (375.98,194.14) -- (364.86,195.86) -- (366.61,192.76) -- (350.46,183.23) -- (353.96,177.03) -- (370.11,186.55) -- (371.86,183.45) -- cycle ;
%Right Arrow [id:dp752331736540351] 
\draw  [color={rgb, 255:red, 74; green, 74; blue, 74 }  ,draw opacity=0.5 ][fill={rgb, 255:red, 155; green, 155; blue, 155 }  ,fill opacity=0.5 ] (371.5,157.68) -- (385.7,167) -- (386.68,165.5) -- (394.18,174.71) -- (382.74,171.5) -- (383.73,170) -- (369.53,160.68) -- cycle ;
%Right Arrow [id:dp062313183513304615] 
\draw  [color={rgb, 255:red, 74; green, 74; blue, 74 }  ,draw opacity=0.5 ][fill={rgb, 255:red, 155; green, 155; blue, 155 }  ,fill opacity=0.5 ] (404.88,183.54) -- (413.12,198.39) -- (414.69,197.51) -- (417.05,209.16) -- (408.41,201) -- (409.98,200.13) -- (401.74,185.28) -- cycle ;
%Right Arrow [id:dp6146846987422527] 
\draw  [color={rgb, 255:red, 74; green, 74; blue, 74 }  ,draw opacity=0.5 ][fill={rgb, 255:red, 155; green, 155; blue, 155 }  ,fill opacity=0.5 ] (296.6,159.23) -- (282.49,168.68) -- (283.49,170.17) -- (272.08,173.49) -- (279.49,164.21) -- (280.49,165.7) -- (294.61,156.25) -- cycle ;
%Right Arrow [id:dp24770257088582937] 
\draw  [color={rgb, 255:red, 74; green, 74; blue, 74 }  ,draw opacity=0.5 ][fill={rgb, 255:red, 155; green, 155; blue, 155 }  ,fill opacity=0.5 ] (263,185.14) -- (254.48,199.83) -- (256.03,200.73) -- (247.24,208.72) -- (249.82,197.12) -- (251.37,198.02) -- (259.9,183.34) -- cycle ;
%Right Arrow [id:dp9537772105117015] 
\draw  [color={rgb, 255:red, 74; green, 74; blue, 74 }  ,draw opacity=0.5 ][fill={rgb, 255:red, 155; green, 155; blue, 155 }  ,fill opacity=0.5 ] (355.78,295) -- (361.05,310.3) -- (363.01,309.72) -- (362.61,321.08) -- (355.18,312.03) -- (357.14,311.46) -- (351.87,296.15) -- cycle ;
%Right Arrow [id:dp686078093233977] 
\draw  [color={rgb, 255:red, 74; green, 74; blue, 74 }  ,draw opacity=0.5 ][fill={rgb, 255:red, 155; green, 155; blue, 155 }  ,fill opacity=0.5 ] (370.84,368.22) -- (371.59,384.27) -- (373.64,384.19) -- (370.04,395.05) -- (365.44,384.52) -- (367.49,384.43) -- (366.73,368.39) -- cycle ;
%Right Arrow [id:dp4588317601943781] 
\draw  [color={rgb, 255:red, 74; green, 74; blue, 74 }  ,draw opacity=0.5 ][fill={rgb, 255:red, 155; green, 155; blue, 155 }  ,fill opacity=0.5 ] (316.04,296.11) -- (311.13,311.52) -- (313.1,312.06) -- (305.88,321.25) -- (305.22,309.9) -- (307.19,310.44) -- (312.11,295.04) -- cycle ;
%Right Arrow [id:dp24839197738007246] 
\draw  [color={rgb, 255:red, 74; green, 74; blue, 74 }  ,draw opacity=0.5 ][fill={rgb, 255:red, 155; green, 155; blue, 155 }  ,fill opacity=0.5 ] (302.49,368.41) -- (301.55,384.45) -- (303.6,384.55) -- (298.88,395.04) -- (295.4,384.15) -- (297.45,384.25) -- (298.39,368.21) -- cycle ;
%Curve Left Arrow [id:dp052847087405133664] 
\draw  [fill={rgb, 255:red, 255; green, 255; blue, 255 }  ,fill opacity=1 ] (314.45,338.39) .. controls (309.61,337.92) and (306.12,333.77) .. (306.65,329.13) -- (309.18,329.38) .. controls (308.65,334.02) and (312.14,338.16) .. (316.98,338.64) ;\draw  [fill={rgb, 255:red, 255; green, 255; blue, 255 }  ,fill opacity=1 ] (316.98,338.64) .. controls (320.57,338.99) and (323.89,337.22) .. (325.58,334.38) -- (327.7,334.59) -- (325.43,330.97) -- (320.94,333.93) -- (323.06,334.13) .. controls (321.37,336.97) and (318.04,338.74) .. (314.45,338.39)(316.98,338.64) -- (314.45,338.39) ;
%Curve Right Arrow [id:dp9536632409034398] 
\draw  [fill={rgb, 255:red, 255; green, 255; blue, 255 }  ,fill opacity=1 ] (351.94,338.29) .. controls (356.72,337.62) and (359.99,333.33) .. (359.23,328.71) -- (361.75,328.36) .. controls (362.51,332.98) and (359.24,337.27) .. (354.46,337.94) ;\draw  [fill={rgb, 255:red, 255; green, 255; blue, 255 }  ,fill opacity=1 ] (354.46,337.94) .. controls (350.91,338.43) and (347.52,336.79) .. (345.71,334.02) -- (347.91,333.71) -- (343.18,330.95) -- (340.98,334.68) -- (343.19,334.37) .. controls (345,337.14) and (348.39,338.78) .. (351.94,338.29)(354.46,337.94) -- (351.94,338.29) ;
%Left Arrow [id:dp294488454838266] 
\draw   (321.62,365.07) -- (317.3,360.06) -- (319.6,360.21) -- (320.22,350.23) -- (324.82,350.54) -- (324.21,360.52) -- (326.51,360.67) -- cycle ;
%Left Arrow [id:dp7060839645268777] 
\draw   (347.32,365.16) -- (342.35,360.7) -- (344.66,360.58) -- (343.89,350.61) -- (348.49,350.37) -- (349.26,360.34) -- (351.56,360.22) -- cycle ;
%Left Arrow [id:dp05246360772303116] 
\draw   (321.33,393.82) -- (316.7,389.07) -- (319.01,389.09) -- (318.97,379.1) -- (323.59,379.14) -- (323.62,389.14) -- (325.93,389.16) -- cycle ;
%Left Arrow [id:dp5594372062006872] 
\draw   (348.09,394.18) -- (343.4,389.47) -- (345.71,389.47) -- (345.54,379.48) -- (350.16,379.47) -- (350.32,389.47) -- (352.63,389.46) -- cycle ;
%Down Arrow [id:dp5678826540598962] 
\draw  [color={rgb, 255:red, 0; green, 0; blue, 0 }  ,draw opacity=0.5 ][fill={rgb, 255:red, 155; green, 155; blue, 155 }  ,fill opacity=0.5 ] (93,448.71) -- (98.25,448.71) -- (98.25,441.51) -- (108.75,441.51) -- (108.75,448.71) -- (114,448.71) -- (103.5,453.51) -- cycle ;
%Right Arrow [id:dp9919949699071104] 
\draw  [color={rgb, 255:red, 74; green, 74; blue, 74 }  ,draw opacity=0.5 ][fill={rgb, 255:red, 155; green, 155; blue, 155 }  ,fill opacity=0.5 ] (245.14,440.39) -- (249.51,448.26) -- (250.82,447.53) -- (251.11,454.24) -- (245.57,450.45) -- (246.88,449.72) -- (242.52,441.85) -- cycle ;
%Right Arrow [id:dp509980265780072] 
\draw  [color={rgb, 255:red, 0; green, 0; blue, 0 }  ,draw opacity=1 ] (364.14,439.89) -- (368.51,447.76) -- (369.82,447.03) -- (370.11,453.74) -- (364.57,449.95) -- (365.88,449.22) -- (361.52,441.35) -- cycle ;
%Curve Left Arrow [id:dp8042088976546449] 
\draw  [fill={rgb, 255:red, 255; green, 255; blue, 255 }  ,fill opacity=1 ] (506.92,452.68) .. controls (502.06,452.76) and (498.12,449.04) .. (498.12,444.37) -- (500.66,444.33) .. controls (500.66,449) and (504.6,452.72) .. (509.46,452.64) ;\draw  [fill={rgb, 255:red, 255; green, 255; blue, 255 }  ,fill opacity=1 ] (509.46,452.64) .. controls (513.06,452.58) and (516.17,450.45) .. (517.52,447.44) -- (519.65,447.4) -- (516.99,444.06) -- (512.86,447.51) -- (514.99,447.48) .. controls (513.63,450.49) and (510.53,452.62) .. (506.92,452.68)(509.46,452.64) -- (506.92,452.68) ;

% Text Node
\draw (288.5,95.64) node [anchor=north west][inner sep=0.75pt]  [font=\LARGE] [align=left] {{\fontfamily{ptm}\selectfont {\small Bending Mode}}};
% Text Node
\draw (281,248.65) node [anchor=north west][inner sep=0.75pt]  [font=\LARGE] [align=left] {{\fontfamily{ptm}\selectfont {\small Membrane Mode}}};
% Text Node
\draw (57,178.15) node [anchor=north west][inner sep=0.75pt]  [font=\small]  {$ \begin{array}{l}
\max([\widehat{\kappa }_{\alpha }])\mathcal{\sim O}( 1)\\
\mathrm{Eq.~\eqref{curv_normal}}
\end{array}$};
% Text Node
\draw (57,62.09) node [anchor=north west][inner sep=0.75pt]  [font=\small]  {$ \begin{array}{l}
\max([\widehat{\kappa }_{\alpha }]) \sim \mathcal{O}( \epsilon )\\
\mathrm{Eq.~\eqref{curv_nearflat}}
\end{array}$};
% Text Node
\draw (481.67,177.64) node [anchor=north west][inner sep=0.75pt]  [font=\small]  {$ \begin{array}{l}
u_{i} \sim \mathcal{O}( \epsilon )\\
\sigma _{\alpha 3} ,\ \sigma _{33} \sim \mathcal{O}( \epsilon \sigma _{\alpha \beta })
\end{array}$};
% Text Node
\draw (487.67,50.47) node [anchor=north west][inner sep=0.75pt]  [font=\small]  {$ \begin{array}{l}
u_{\alpha } \sim \mathcal{O}( \epsilon u_{3})\\
\sigma _{\alpha 3} \sim \mathcal{O}( \epsilon \sigma _{\alpha \beta })\\
\sigma _{33} \sim \mathcal{O}\left( \epsilon ^{2} \sigma _{\alpha \beta }\right)
\end{array}$};
% Text Node
\draw (294.5,407.48) node [anchor=north west][inner sep=0.75pt]  [font=\LARGE] [align=left] {{\fontfamily{ptm}\selectfont {\small Mixed Mode}}};
% Text Node
\draw (57,322.15) node [anchor=north west][inner sep=0.75pt]  [font=\small]  {$ \begin{array}{l}
\max([\widehat{\kappa }_{\alpha }]) \sim \mathcal{O}\left(\frac{1}{\epsilon }\right)\\
\mathrm{Eq.~\eqref{curv_significant}}
\end{array}$};
% Text Node
\draw (478,320.64) node [anchor=north west][inner sep=0.75pt]  [font=\small]  {$ \begin{array}{l}
u_{3} \sim \mathcal{O}( \epsilon u_{\alpha })\\
\sigma _{\alpha 3} ,\ \sigma _{33} \sim \mathcal{O}( \epsilon \sigma _{\alpha \beta })
\end{array}$};
% Text Node
\draw (114.5,439.77) node [anchor=north west][inner sep=0.75pt]  [font=\normalsize] [align=left] {{\fontfamily{ptm}\selectfont {\footnotesize : Transverse Load}}};
% Text Node
\draw (252.5,439.8) node [anchor=north west][inner sep=0.75pt]  [font=\normalsize] [align=left] {{\fontfamily{ptm}\selectfont {\footnotesize : In-plane Load}}};
% Text Node
\draw (371.5,439.8) node [anchor=north west][inner sep=0.75pt]  [font=\normalsize] [align=left] {{\fontfamily{ptm}\selectfont {\footnotesize : In-plane Tension}}};
% Text Node
\draw (521,439.8) node [anchor=north west][inner sep=0.75pt]  [font=\normalsize] [align=left] {{\fontfamily{ptm}\selectfont {\footnotesize : Moment}}};
\end{tikzpicture}
\caption{Classification of shells based on the magnitude of maximum dimensionless principal curvature $\wk_\alpha$. The three panels from top to bottom illustrate the main deformation modes of (a) weakly curved shells, (b) normally curved shells, and (c) significantly curved shells under external loading.\label{Fig_classification}}
\end{figure}

\subsubsection{Weakly Curved Shell}\label{Sec_weaklycurved_shell}
When the MTS under consideration is nearly flat, as shown in the upper panel of Fig.~\ref{Fig_classification}, the maximum curvature of the shell mid-surface is almost imperceptible. Mathematically, we have the following estimation of non-dimensionalized principal curvatures
\begin{equation}\label{curv_nearflat}
	\max\left(\left[\wk_\alpha\right]\right)\sim\mO\left(\epsilon\right),
\end{equation}
the symbol ``$\left[\,\bigcdot\,\right]$'' here denotes the set of principal curvatures corresponding to all points investigated on the shell mid-surface $\CS$.

From the general (non-dimensional) constitutive laws Eq.~\eqref{constitutive_dimensionless}, the local out-of-plane displacement component is found to be independent of the normal variable in the leading order, i.e.,
\begin{equation}\label{u3_midsurface}
	\wu_3^{(0)} = \wu_3^\ast\left(\wxi_1,\wxi_2\right).
\end{equation}
And with Eqs.~\eqref{constitutive_dimensionless4} and \eqref{constitutive_dimensionless5}, the linear relationship between two tangential displacement components at any point inside the investigated MTS and the normal variable $\wxi_3$ is derived
\begin{equation}\label{u_alpha_linearize}
	\wu_\alpha^{(0)}\left(\wxi_1,\wxi_2,\wxi_3\right) = \wu_\alpha^\ast\left(\wxi_1,\wxi_2\right)-\frac{\wxi_3}{\wa_\alpha}\pd{\wu_3^\ast\left(\wxi_1,\wxi_2\right)}{\wxi_\alpha},
\end{equation}
here, the displacement components affiliated with the symbol ``$\ast$'' represent their counterparts related to the shell mid-surface. Eqs.~\eqref{u3_midsurface} and \eqref{u_alpha_linearize} coincide with the displacement patterns given by the classical Kirchhoff-Love hypothesis, but only for the leading order. Specifically, for an MTS whose principal curvatures satisfy Eq.~\eqref{curv_nearflat}, the original conclusions regarding the local displacement components $\wu_i$ can be retained up to $\mO\left(\epsilon\right)$, that is,
\begin{subequations}\label{disp_relation_weakly}
	\begin{align}
		&\wu_\alpha^{(d)}\left(\wxi_1,\wxi_2,\wxi_3\right) = \wu_\alpha^{\ast(d)}\left(\wxi_1,\wxi_2\right)-\frac{\wxi_3}{\wa_\alpha}\pd{\wu_3^{(d)}\left(\wxi_1,\wxi_2\right)}{\wxi_\alpha};\\
		&\wu_3^{(d)} = \wu_3^{(d)}\left(\wxi_1,\wxi_2\right),
	\end{align}
\end{subequations}
for both $d = 0\text{ and }1$. And as indicated by the equilibrium equations and constitutive laws, the order relationships of the local displacement and stress fields remain the same as those given by Eqs.~\eqref{disp_scaling} and \eqref{stress_scaling}.

Following this, as indicated in Sec.~\ref{Sec_scaling}, we will determine the appropriate orders of magnitude for both the observation time $t$ and the body force vector $\vf$ in the context of weakly curved shell. Given the proximity of the principal curvatures to zero, the influence of in-plane stresses in the equilibrium equation along the thickness direction does not significantly surpass that of the rest transverse components, so that the left end of Eq.~\eqref{equilibrium_dimensionless3} essentially starts from $\mO\left(1\right)$. If the order of the observation time $m_1=0$, we find at this point that $\frac{\partial^2\wu_3^{(d)}}{\partial\wt^2}=0$, for $d=0,1$, suggesting an absence of normal displacement during free vibration---a proposition deemed untenable. Note that, as indicated by Eq.~\eqref{disp_scaling}, the out-of-plane displacement $u_3$ predominates over the in-plane displacement components $u_\alpha$, necessitating an extended observation time to discern the vibration modes of the shell in this direction, which is the primary vibration direction during the free vibration of such shells (refer to the thin plate structure as a special case, the free vibration displacement occurs mainly in the direction perpendicular to the midplane). Accordingly, a suitable order here should be $m_1=-1$.

On the other hand, since $f_i\left(i = 1,2,3\right)$ are all components of the body force vector $\vf$, these three should be of the same order of magnitude. If the order of $\vf$ is not greater than $\mO\left(\epsilon^3\right)$, i.e., $m_2\leq3$, there will be no corresponding stress components to balance with the body force component $f_3$ in the shell thickness direction. Consequently, as implied by AA, the appropriate order for the body force in this situation is $m_2=4$.

So far, the values of two unknown parameters in Eqs.~\eqref{time_scaling} and \eqref{bodyforce_scaling} have been determined in this particular case, with $m_1=-1$ and $m_2=4$. By letting $\epsilon\to 0$, Eq.~\eqref{equilibrium_dimensionless} is then simplified to a set of equations where only the one correlated with the thickness direction contains a dimensionless inertial force term and a dimensionless body force term
\begin{subequations}\label{equilibrium_weakly}
	\begin{align}
		0 &= \frac{1}{\wa_1\wa_2}\left[\pd{}{\wxi_1}\left(\wa_2\,\wsigma_{11}^{(0)}\right)+\pd{}{\wxi_2}\left(\wa_1\,\wsigma_{12}^{(0)}\right)+\wsigma_{12}^{(0)}\pd{\wa_1}{\wxi_2}-\wsigma_{22}^{(0)}\pd{\wa_2}{\wxi_1}\right]+\pd{\wsigma_{13}^{(0)}}{\wxi_3};\label{equilibrium_weakly1}\\
		0 &= \frac{1}{\wa_1\wa_2}\left[\pd{}{\wxi_1}\left(\wa_2\,\wsigma_{12}^{(0)}\right)+\pd{}{\wxi_2}\left(\wa_1\,\wsigma_{22}^{(0)}\right)+\wsigma_{12}^{(0)}\pd{\wa_2}{\wxi_1}-\wsigma_{11}^{(0)}\pd{\wa_1}{\wxi_2}\right]+\pd{\wsigma_{23}^{(0)}}{\wxi_3};\label{equilibrium_weakly2}\\
		\wrho_{\sss\CL}\spd{\wu_3^{(0)}}{\wt}-\wf_{\sss\CL}^3 &= \frac{1}{\wa_1\wa_2}\left[\pd{}{\wxi_1}\left(\wa_2\,\wsigma_{13}^{(0)}\right)+\pd{}{\wxi_2}\left(\wa_1\,\wsigma_{23}^{(0)}\right)\right]+\pd{\wsigma_{33}^{(0)}}{\wxi_3}+\frac{\wk_1}{\epsilon}\wsigma_{11}^{(0)}+\frac{\wk_2}{\epsilon}\wsigma_{22}^{(0)},\label{equilibrium_weakly3}
	\end{align}
\end{subequations}
from the leading-order dynamic equilibrium equations above, it is obvious that the two inertial forces along the local tangential directions vanish, resulting in a constrained eigenvalue problem when considering free vibration of the shell. One can refer to the works done by \cite{GANDER1989815} and \cite{ram2010constrained} for details in obtaining the eigenvalues and corresponding eigenvectors of such a problem. As shown later in Sec.~\ref{Sec_homogenization}, the through-the-thickness integration of Eq.~\eqref{equilibrium_weakly} yields a system of equations in which only the third one incorporates the bending and twisting effects, as it involves the stress couples. Consequently, similar to thin plates, this type of shells primarily resists transverse loads through bending deformation (see the upper panel of Fig.~\ref{Fig_classification}).

\subsubsection{Normally Curved Shell}\label{Sec_normal_shell}
If the maximum principal curvature of an examined point on the shell mid-surface $\CS$ is appreciable, such as spherical ($\wk_1=\wk_2=1$) and cylindrical shells ($\wk_2=0$), etc., the order of magnitude of the non-dimensional principal curvatures should satisfy
\begin{equation}\label{curv_normal}
	\max\left(\left[\wk_\alpha\right]\right)\sim\mO\left(1\right).
\end{equation}

In the following, we attempt to derive the order relationships and possible expressions for the internal displacement and stress components under the condition set by Eq.~\eqref{curv_normal}. First, at $\mO\left(\epsilon\right)$, the constitutive relations, Eqs.~\eqref{constitutive_dimensionless1}, \eqref{constitutive_dimensionless3}, and \eqref{constitutive_dimensionless6}, give that
\begin{equation}
	\left\lbrace
	\begin{aligned}
		&-(1-\nu_{\sss\CL})\wu_3^{(0)}\wk_1-\nu_{\sss\CL}\wu_3^{(0)}\wk_2+\nu_{\sss\CL}\pd{\wu_3^{(1)}}{\wxi_3} = 0;\\
		&-\nu_{\sss\CL}\wu_3^{(0)}\wk_1-(1-\nu_{\sss\CL})\wu_3^{(0)}\wk_2+\nu_{\sss\CL}\pd{\wu_3^{(1)}}{\wxi_3} = 0;\\
		&-\nu_{\sss\CL}\wu_3^{(0)}\wk_1-\nu_{\sss\CL}\wu_3^{(0)}\wk_2+(1-\nu_{\sss\CL})\pd{\wu_3^{(1)}}{\wxi_3} = 0,
	\end{aligned}
	\right.
\end{equation}
only the trivial solution is obtained from the above equations, i.e.,
\begin{equation}\label{u3_trivial_normal}
	\wu_3^{(0)} = 0;\quad \wu_3^{(1)} = \wu_3^{(1)}\left(\wxi_1,\wxi_2\right).
\end{equation}
Furthermore, Eq.~\eqref{u3_trivial_normal} implies that, in the case of normally curved shells, the three local displacement components $u_i$ are actually of the same order of magnitude ($\mO\left(\epsilon\right)$), as the leading-order out-of-plane displacement is trivial at every point inside the shell, including umbilic points.

The first two orders of the constitutive laws related to transverse stress components (Eqs.~\eqref{constitutive_dimensionless4} and \eqref{constitutive_dimensionless5}) are also considered, that is,
\begin{subequations}
	\begin{align}
		&\mO\left(1\right): \df{1}{\wa_\alpha}\pd{\wu_3^{(0)}}{\wxi_\alpha}+\pd{\wu_\alpha^{(0)}}{\wxi_3} = 0;\label{constitutive_transverse_leading}\\
		&\mO\left(\epsilon\right): \df{\partial\wu_\alpha^{(1)}}{\partial\wxi_3} = -\df{1}{\wa_\alpha}\left(\pd{\wu_3^{(1)}}{\wxi_\alpha}+\wk_\alpha\wxi_3\pd{\wu_3^{(0)}}{\wxi_\alpha}\right)-\wu_\alpha^{(0)}\wk_\alpha,\label{constitutive_transverse_first}
	\end{align}
\end{subequations}
based on the conclusion $\wu_3^{(0)} = 0$, Eq.~\eqref{constitutive_transverse_leading} indicates that the leading-order in-plane displacements $\wu_\alpha^{(0)}$ should satisfy
\begin{equation}\label{in_disp_derivative_normal}
	\pd{\wu_\alpha^{(0)}}{\wxi_3} = 0,\quad\text{for }\alpha = 1,2,
\end{equation}
which further leads to their sole dependence on the surface parameters, i.e., $\wu_\alpha^{(0)}=\wu_\alpha^{(0)}\left(\wxi_1,\wxi_2\right)$. With the use of Eqs.~\eqref{u3_trivial_normal} and \eqref{in_disp_derivative_normal}, the left term of Eq.~\eqref{constitutive_transverse_first} is found to be independent of the normal coordinate $\wxi_3$, and another relation on the first-order in-plane displacements is then obtained by taking partial derivative w.r.t. $\wxi_3$, that is
\begin{equation}
	\spd{\wu_\alpha^{(1)}}{\wxi_3} = 0,
\end{equation}
we can accordingly assume that $\wu_\alpha^{(1)}$ take the form
\begin{equation}\label{u_alpha_first_linear}
	\wu_\alpha^{(1)} = f_\alpha^\ast\left(\wxi_1,\wxi_2\right)+f_\alpha\left(\wxi_1,\wxi_2\right)\wxi_3,
\end{equation}
where $f_\alpha^\ast\left(\wxi_1,\wxi_2\right)$ and $f_\alpha\left(\wxi_1,\wxi_2\right)$ are just functions of $\wxi_1$ and $\wxi_2$.

At this point, it should be noted that the trivial leading-order normal displacement suggests a need for a modification of the originally introduced displacement order relationship (Eq.~\eqref{disp_scaling}). Unlike the case of weakly curved shells discussed in Sec.~\ref{Sec_weaklycurved_shell}, where the out-of-plane displacement $u_3$ predominates over the other two, here all the three local displacements are of the same order of magnitude. Moreover, $\wu_3^{(1)}$ actually serves as the normal displacement component corresponding to the shell mid-surface. Thus we have
\begin{equation}\label{disp_scaling_normal}
	u_\alpha = \epsilon \rU\left(\wu_\alpha^{(0)}+\epsilon\wu_\alpha^{(1)}+\,\cdots\right),\quad u_3 = \rU\left(\wu_3^{(0)}+\epsilon\wu_3^{(1)}+\,\cdots\right) = \epsilon\rU\left(\wu_3^{(1)}+\,\cdots\right).
\end{equation}

Due to the modification of displacement scaling rules in this case (Eq.~\eqref{disp_scaling_normal}), the original order relations of the local stress components (Eq.~\eqref{stress_scaling}) should also be recalibrated. This necessitates a fresh analysis of the magnitude of each stress component. A detailed examination of Eq.~\eqref{equilibrium_concise} reveals that the transverse shear stresses $\sigma_{\alpha3}$ remain one order higher than the in-plane components $\sigma_{\alpha\beta}$, since there is no new difference here introduced by the premise of appreciable curvature (i.e., Eq.~\eqref{curv_normal}). However, this is not true for the equilibrium equation in the thickness direction (i.e., $i=3$ for Eq.~\eqref{equilibrium_concise}). To be specific, Eq.~\eqref{curv_normal} allows the terms relating to in-plane normal stresses to dominate over $\sigma_{\alpha3}$ in this equation, so that the transverse normal stress $\sigma_{33}$, like the transverse shear stresses, is only one order higher than $\sigma_{\alpha\beta}$. In general, the three local transverse stress components $\sigma_{i3}$ are of the same order of magnitude, but overall, they are one order higher than the in-plane stress components $\sigma_{\alpha\beta}$, i.e.,
\begin{equation}\label{stress_scaling_normal}
	\sigma_{\alpha\beta} = \sigma^\star\wsigma_{\alpha\beta};\quad\sigma_{i3} = \epsilon\sigma^\star\wsigma_{i3}.
\end{equation}

We then discuss further some of the intriguing conclusions that Eq.~\eqref{curv_normal} leads to: the leading-order dimensionless transverse shear stresses $\wsigma_{\alpha 3}^{(0)}$ within each layer can be theoretically demonstrated to vary (piecewise) linearly along the thickness direction and to equal to the imposed non-dimensional shear forces per area $\wload_\alpha$ at $\CS_\mathrm{t}$. The detailed proof is as follows:

With Eq.~\eqref{constitutive_transverse_first}, the formulations of the leading-order transverse shear stresses are written by
\begin{equation}\label{transverse_shear_leading}
	\begin{aligned}
		\wsigma_{\alpha3}^{(0)} &= \frac{\wE_{\sss\CL}}{2(1+\nu_{\sss\CL})}\left[\frac{1}{\wa_\alpha}\pd{\wu_3^{(2)}}{\wxi_\alpha}+\wk_\alpha\wxi_3\left(\frac{1}{\wa_\alpha}\pd{\wu_3^{(1)}}{\wxi_\alpha}+\frac{\wk_\alpha\wxi_3}{\wa_\alpha}\pd{\wu_3^{(0)}}{\wxi_\alpha}+\wu_\alpha^{(0)}\wk_\alpha\right)+\wu_\alpha^{(1)}\wk_\alpha+\pd{\wu_\alpha^{(2)}}{\wxi_3}\right]\\
		&= \frac{\wE_{\sss\CL}}{2(1+\nu_{\sss\CL})}\left(\frac{1}{\wa_\alpha}\pd{\wu_3^{(2)}}{\wxi_\alpha}+\wu_\alpha^{(1)}\wk_\alpha+\pd{\wu_\alpha^{(2)}}{\wxi_3}-\wk_\alpha\wxi_3\pd{\wu_\alpha^{(1)}}{\wxi_3}\right).
	\end{aligned}
\end{equation}

From the $\mO\left(\epsilon^2\right)$ identity of the non-dimensional constitutive law regarding $\wsigma_{33}$ (\eqref{constitutive_dimensionless6}), it can be seen that the equation $\spd{\wu_3^{(2)}}{\wxi_3} = 0$ holds, which in turn leads to the finding that the leading-order in-plane stress components $\wsigma_{\alpha\beta}^{(0)}$ of a point inside the TLS are equal to those at the point where it is projected along the normal to the midsurface. Thus, taking the derivative of both Eqs.~\eqref{equilibrium_weakly1} and \eqref{equilibrium_weakly2} w.r.t. variable $\wxi_3$ yields the identity
\begin{equation}\label{transverse_shear_linear}
	\spd{\wsigma_{\alpha3}^{(0)}}{\wxi_3} = 0,
\end{equation}
then substituting Eq.~\eqref{transverse_shear_leading} into Eq.~\eqref{transverse_shear_linear} gives
\begin{equation}
	\spd{}{\wxi_3}\left(\frac{1}{\wa_\alpha}\pd{\wu_3^{(2)}}{\wxi_\alpha}+\pd{\wu_\alpha^{(2)}}{\wxi_3}\right) = 0,
\end{equation}
and we can similarly assume that
\begin{equation}\label{transverse_shear_linear_byproduct}
	\frac{1}{\wa_\alpha}\pd{\wu_3^{(2)}}{\wxi_\alpha}+\pd{\wu_\alpha^{(2)}}{\wxi_3} = g_\alpha^\ast\left(\wxi_1,\wxi_2\right)+g_\alpha\left(\wxi_1,\wxi_2\right)\wxi_3.
\end{equation}

Incorporating Eqs.~\eqref{u_alpha_first_linear} and \eqref{transverse_shear_linear_byproduct} into Eq.~\eqref{transverse_shear_leading}, the simplified formulation for $\wsigma_{\alpha3}^{(0)}$ is derived, that is,
\begin{equation}\label{transverse_shear_piecewiselinear}
	\wsigma_{\alpha3}^{(0)} = \frac{\wE_{\sss\CL}}{2(1+\nu_{\sss\CL})}\left[G_\alpha\left(\wxi_1,\wxi_2\right)+g_\alpha\left(\wxi_1,\wxi_2\right)\wxi_3\right],
\end{equation}
where $G_\alpha\left(\wxi_1,\wxi_2\right) = \wk_\alpha f_\alpha^\ast\left(\wxi_1,\wxi_2\right)+g_\alpha^\ast\left(\wxi_1,\wxi_2\right)$. Note that, the leading-order transverse shear stress components $\wsigma_{\alpha 3}^{(0)}$ are actually piecewise linear w.r.t. the thickness coordinate $\wxi_3$, which means that $\wsigma_{\alpha 3}^{(0)}$ show straight zigzag distributions and equal to the imposed shear forces on the top and bottom surfaces of the MTS if viewed from a cross-section perpendicular to the shell mid-surface. The piecewise linear relationship of Eq.~\eqref{transverse_shear_piecewiselinear} can be partly reflected and justified by Fig.~\ref{transverse_linear_plot}, which shows the distribution of the actual transverse shear stresses in a multi-layered spherical shell. In particular, for one-layer thin shells, the leading-order transverse shear stresses can be readily determined by the surface boundary conditions given by Eq.~\eqref{bc_surface}.

Similar to Sec.~\ref{Sec_weaklycurved_shell}, here we also need to determine the specific values of $m_1$ and $m_2$. With the appreciable surface curvature, it can be inferred that the terms related to in-plane stress components $\wsigma_{11}$ and $\wsigma_{22}$ in Eq.~\eqref{equilibrium_dimensionless3} are approximately of $\mO\left(1/\epsilon\right)$, which dominate the resistance of the MTS to deformation in the thickness direction. Furthermore, as mentioned above, $\wu_3$ starts from the first order ($\mO\left(\epsilon\right)$). Therefore, the inertial force and body force terms in the dynamic equilibrium equation along the normal direction here are of the same order as those in the other two equations. The specific orders of the observation time and the body force vector are accordingly selected to be $m_1=0$ and $m_2=3$, respectively.

Without loss of generality, we assume here that both principal curvatures of $\CS$ satisfy Eq.~\eqref{curv_normal}, with the spherical shell being one of the special cases. Again, let $\epsilon\to 0$, the non-dimensional governing equations corresponding to normally curved shell are obtained
\begin{subequations}\label{equilibrium_normal}
	\begin{align}
		&\wrho_{\sss\CL}\spd{\wu_1^{(0)}}{\wt}-\wf_{\sss\CL}^1 = \frac{1}{\wa_1\wa_2}\left[\pd{}{\wxi_1}\left(\wa_2\,\wsigma_{11}^{(0)}\right)+\pd{}{\wxi_2}\left(\wa_1\,\wsigma_{12}^{(0)}\right)+\wsigma_{12}^{(0)}\pd{\wa_1}{\wxi_2}-\wsigma_{22}^{(0)}\pd{\wa_2}{\wxi_1}\right]+g_1^{\sss\CL}\left(\wxi_1,\wxi_2\right);\label{equilibrium_normal1}\\
		&\wrho_{\sss\CL}\spd{\wu_2^{(0)}}{\wt}-\wf_{\sss\CL}^2 = \frac{1}{\wa_1\wa_2}\left[\pd{}{\wxi_1}\left(\wa_2\,\wsigma_{12}^{(0)}\right)+\pd{}{\wxi_2}\left(\wa_1\,\wsigma_{22}^{(0)}\right)+\wsigma_{12}^{(0)}\pd{\wa_2}{\wxi_1}-\wsigma_{11}^{(0)}\pd{\wa_1}{\wxi_2}\right]+g_2^{\sss\CL}\left(\wxi_1,\wxi_2\right);\label{equilibrium_normal2}\\
		&\wrho_{\sss\CL}\spd{\wu_3^{(1)}}{\wt}-\wf_{\sss\CL}^3 = \wk_1\wsigma_{11}^{(0)}+\wk_2\wsigma_{22}^{(0)}+\pd{\wsigma_{33}^{(0)}}{\wxi_3},\label{equilibrium_normal3}
	\end{align}
\end{subequations}
here $g_\alpha^{\sss\CL}\left(\wxi_1,\wxi_2\right)$ are unknown functions of $\wxi_1,\wxi_2$ and remain to be solved.

Now we supplement the modifications made in this particular case with some physical elucidations. For normally curved shells, the displacement components are of the same order of magnitude. This contrasts with the situation discussed in Sec.~\ref{Sec_weaklycurved_shell} and in classical plate theories, where the normal displacement is one order larger than the in-plane components. This indicates that as the shell curvature increases, its ability to resist external loads also improves (as reflected in the order of body force, $\mO\left(\epsilon^3\right)$, which is greater than the $\mO\left(\epsilon^4\right)$ in weakly curved shells), leading to greater stiffness and a consequent reduction in normal displacement during deformation. The order of the time scale in this case, $m_1=0$, is a higher order quantity than that in Sec.~\ref{Sec_weaklycurved_shell} ($m_1=-1$). Since the three displacement components in the local coordinate system $u_i$ are of the same order, the vibration modes along all three directions can be observed simultaneously for a free vibrating shell. So in contrary to Eq.~\eqref{equilibrium_weakly}, the free vibration problem here corresponds to a complete generalized eigenvalue problem.

If we perform a through-the-thickness homogenization of the above equilibrium equations, we will arrive at a set of equations composed entirely of the in-plane stress resultants $\wT_{\alpha\beta}$. They constitute the governing equations of the membrane/momentless theory of shells (\cite{ventsel2001book}), where the shell resists external loads purely through $\wT_{\alpha\beta}$, without developing bending or twisting moments $\wM_{\alpha\beta}$. Such behavior closely resembles the deformation mode of a membrane, as illustrated in the middle panel of Fig.~\ref{Fig_classification}.

\subsubsection{Significantly Curved Shell}\label{Sec_significantlycurved_shell}
We also notice the existence of another shell deformation mode worth mentioning, in which the magnitude of the maximum principal curvature is even larger, i.e.,
\begin{equation}\label{curv_significant}
	\max\left(\left[\wk_\alpha\right]\right)\sim\mO\left(\frac{1}{\epsilon}\right).
\end{equation}

Based on the magnitude of $\max\left(\left[\wk_\alpha\right]\right)$ in this particular case, the non-dimensional constitutive relations (Eqs.~\eqref{constitutive_dimensionless1}, \eqref{constitutive_dimensionless3}, and \eqref{constitutive_dimensionless6}) imply that $\wu_3^{(d)}=0,\ \text{for }d=0,1$. Therefore, the normal displacement component here starts from $\mO\left(\epsilon^2\right)$. Then, from Eqs.~\eqref{constitutive_dimensionless4} and \eqref{constitutive_dimensionless5}, it follows that the leading and first-order terms of the in-plane displacement components satisfy ordinary differential equations parametrized by $\wxi_\alpha$, that is,
\begin{equation}\label{significant_disp_ode}
	\pd{\wu_\alpha^{(d)}}{\wxi_3}+\frac{\epsilon\wk_\alpha\wu_\alpha^{(d)}}{1-\epsilon\wk_\alpha\wxi_3}=0,\quad d=0,1.
\end{equation}
Solutions to Eq.~\eqref{significant_disp_ode} give the expressions for in-plane displacements in shells with significant curvature. Therefore, we have
\begin{equation}\label{in_plane_disp_linear_significant}
	\wu_\alpha^{(d)}=\wu_\alpha^\ast\left(\wxi_1,\wxi_2\right)\left(1-\epsilon\wk_\alpha\wxi_3\right),
\end{equation}
which also exhibit linear distributions across the shell thickness. Similar to Sec.~\ref{Sec_normal_shell}, by examining the dimensionless equilibrium equations, Eqs.~\eqref{equilibrium_dimensionless1}-\eqref{equilibrium_dimensionless3}, we can further uncover the order relationships among the stress components in this case. In general, the displacement and stress components are scaled as follows
\begin{subequations}
	\begin{align}
		&u_\alpha=\epsilon\rU\left(\wu_\alpha^{(0)}+\cdots\right),\ u_3=\epsilon^2\rU\left(\wu_3^{(2)}+\cdots\right); \label{disp_scaling_significant}\\
		&\sigma_{\alpha\beta}=\sigma^\star\wsigma_{\alpha\beta},\ \left(\sigma_{\alpha3},\sigma_{33}\right)=\epsilon\sigma^\star\left(\wsigma_{\alpha3},\wsigma_{33}\right). \label{stress_scaling_significant}
	\end{align}
\end{subequations}
The second-order normal displacement component can be further related to the leading-order in-plane components through Eq.~\eqref{constitutive_dimensionless6}. Thus, an extra equation is obtained
\begin{equation}\label{relation_normal_in_plane_disp}
	\frac{\epsilon\wk_1\wa_1\wu_3^{(2)}}{\wh_1}+\frac{\epsilon\wk_2\wa_2\wu_3^{(2)}}{\wh_2}-\frac{1-\nu_{\sss\CL}}{\nu_{\sss\CL}}\pd{\wu_3^{(2)}}{\wxi_3}=\frac{1}{\wh_1\wh_2}\left[\pd{\left(\wh_2\wu_1^{(0)}\right)}{\wxi_1}+\pd{\left(\wh_1\wu_2^{(0)}\right)}{\wxi_2}\right],
\end{equation}
which indicates how $\wu_3^{(2)}$ can be recovered from two in-plane displacement components $\wu_\alpha^{(0)}$.

To determine the order of the body force, $m_2$, we again refer to the equilibrium equations (Eq.~\eqref{equilibrium_dimensionless}). Assume $m_2=2$, the force $f_3$ contributes to shell equilibrium in the thickness direction. However, this leads to the absence of stress components in Eqs.~\eqref{equilibrium_dimensionless1} and \eqref{equilibrium_dimensionless2} that can balance it, which is physically inconsistent. Similarly, if $m_1=1$, the component $\wu_3^{(2)}$ contributes to the equilibrium in the normal direction of the structure. However, in the other two directions, there would be no stress components capable of balancing the in-plane accelerations. Therefore, for physical consistency, $m_1$ here is set to 0. In this way, the leading-order governing equations associated with significantly curved shells are obtained based on the scaling and order relationships derived above, that is,
\begin{subequations}\label{equilibrium_significant}
	\begin{align}
		\begin{split}
			\wrho_{\sss\CL}\spd{\wu_1^{(0)}}{\wt}-\wf_{\sss\CL}^1&=\frac{1}{\wh_1\wh_2}\left[\pd{}{\wxi_1}\left(\wh_2\,\wsigma_{11}^{(0)}\right)+\pd{}{\wxi_2}\left(\wh_1\,\wsigma_{12}^{(0)}\right)+\wsigma_{12}^{(0)}\pd{\wh_1}{\wxi_2}-\wsigma_{22}^{(0)}\pd{\wh_2}{\wxi_1}\right]\\
			&+\pd{\wsigma_{13}^{(0)}}{\wxi_3}-\frac{2\epsilon\wk_1\wsigma_{13}^{(0)}}{1-\epsilon\wk_1\wxi_3}-\frac{\epsilon\wk_2\wsigma_{13}^{(0)}}{1-\epsilon\wk_2\wxi_3};\label{equilibrium_significant1}
		\end{split}\\
		\begin{split}
			\wrho_{\sss\CL}\spd{\wu_2^{(0)}}{\wt}-\wf_{\sss\CL}^2&=\frac{1}{\wh_1\wh_2}\left[\pd{}{\wxi_1}\left(\wh_2\,\wsigma_{12}^{(0)}\right)+\pd{}{\wxi_2}\left(\wh_1\,\wsigma_{22}^{(0)}\right)+\wsigma_{12}^{(0)}\pd{\wh_2}{\wxi_1}-\wsigma_{11}^{(0)}\pd{\wh_1}{\wxi_2}\right]\\
			&+\pd{\wsigma_{23}^{(0)}}{\wxi_3}-\frac{\epsilon\wk_1\wsigma_{23}^{(0)}}{1-\epsilon\wk_1\wxi_3}-\frac{2\epsilon\wk_2\wsigma_{23}^{(0)}}{1-\epsilon\wk_2\wxi_3};\label{equilibrium_significant2}
		\end{split}\\
		0 &= \frac{\epsilon\wk_1}{1-\epsilon\wk_1\wxi_3}\wsigma_{11}^{(0)}+\frac{\epsilon\wk_2}{1-\epsilon\wk_2\wxi_3}\wsigma_{22}^{(0)}.\label{equilibrium_significant3}
	\end{align}
\end{subequations}

Notably, the order of the body force remains $\mO\left(\epsilon^3\right)$, indicating that as the shell curvature increases further from Eq.~\eqref{curv_normal} to Eq.~\eqref{curv_significant}, the permissible order of magnitude of the body force does not increase further. However, by combining the displacement scalings derived in Secs.~\ref{Sec_weaklycurved_shell} and \eqref{Sec_normal_shell}, it is evident that as the maximum principal curvature of the shell increases, the order of magnitude of the normal displacement component increases from $\mO\left(1\right)$ to $\mO(\epsilon^2)$, reflecting a progressively stiffened shell in the thickness direction due to enhanced geometric resistance to transverse deformation.

Moreover, it is noteworthy that in this case, the leading-order equilibrium along shell thickness is entirely determined by the in-plane stresses $\wsigma_{11}^{(0)}$ and $\wsigma_{22}^{(0)}$ (Eq.~\eqref{equilibrium_significant3}), while the transverse load $\wf_{\sss\CL}^3$ does not enter the equation until the next order. This suggests that when the curvature is significant, the in-plane stresses can achieve leading-order normal equilibrium under the influence of curvature---a purely geometric effect. The leading-order deformation mode in this case is more counterintuitive compared to the previous two types. Unlike earlier cases, the surface shear forces induce both in-plane stretching and bending/twisting effects in such shells, while equilibrium in the thickness direction is achieved through the self-balancing of two in-plane stresses. Therefore, we refer to this as a hybrid mode, as it exhibits the main characteristics of the previous two deformation modes---bending-dominated and membrane-dominated---respectively in the in-plane and thickness directions (refer to the bottom panel of Fig.~\ref{Fig_classification}).

\subsection{Unified Formulations of Leading-Order Shell Models}\label{Sec_unified_formulation}
\subsubsection{Summary of Different Deformation Modes}\label{Sec_summary_deformation_modes}
In the previous section, we identified three distinct shell deformation modes, each with its own characteristic features. These modes can be further mapped to specific regions in the $\kappa-\frac{1}{L}$ diagram, based on the order of the principal curvature, as illustrated in Fig.~\ref{Fig_deformation_region_classification}. It should be noted that Eqs.~\eqref{curv_nearflat}, \eqref{curv_normal}, and \eqref{curv_significant} primarily represent a kind of order-of-magnitude relationship, the dashed lines in the figure should not be interpreted as precise boundaries between different regions, but rather as indicative of approximate transitions. Moreover, for limiting cases where $\max\left(\left[\wk_\alpha\right]\right)\to 0$ or $\max\left(\left[\wk_\alpha\right]\right)\to\infty$, it can be shown---analogous to the analyses in Secs.~\ref{Sec_weaklycurved_shell} and \ref{Sec_significantlycurved_shell}---that the characteristic features of the corresponding deformation modes still persist.
\begin{figure}[!htbp]
\setlength{\abovecaptionskip}{0.2cm}
\setlength{\belowcaptionskip}{-0.3cm}
\centering
\tikzset{every picture/.style={line width=0.75pt}} %set default line width to 0.75pt
\begin{tikzpicture}[x=0.75pt,y=0.75pt,yscale=-1,xscale=1]
%uncomment if require: \path (0,444); %set diagram left start at 0, and has height of 444
%Shape: Polygon [id:ds37518755265648696] 
\draw  [draw opacity=0][fill={rgb, 255:red, 74; green, 144; blue, 226 }  ,fill opacity=0.06 ] (420.09,358.59) -- (200.09,358.56) -- (420.09,261) -- (420.24,293.36) -- cycle ;
%Shape: Polygon [id:ds27137394103541856] 
\draw  [draw opacity=0][fill={rgb, 255:red, 184; green, 233; blue, 134 }  ,fill opacity=0.2 ] (297.69,139.68) -- (200.09,358.56) -- (200.01,139.87) -- (269.97,139.42) -- cycle ;
%Shape: Polygon [id:ds6298443695011429] 
\draw  [draw opacity=0][fill={rgb, 255:red, 248; green, 231; blue, 28 }  ,fill opacity=0.2 ] (420.09,261) -- (200.09,358.56) -- (297.69,139.68) -- (419.88,139.68) -- cycle ;
%Straight Lines [id:da8875827824184213] 
\draw [color={rgb, 255:red, 208; green, 2; blue, 27 }  ,draw opacity=1 ] [dash pattern={on 4.5pt off 4.5pt}]  (200.09,358.56) -- (420.24,293.36) ;
%Straight Lines [id:da8061816154005408] 
\draw [color={rgb, 255:red, 208; green, 2; blue, 27 }  ,draw opacity=1 ] [dash pattern={on 4.5pt off 4.5pt}]  (200.09,358.56) -- (419.88,139.68) ;
%Straight Lines [id:da9381671983044568] 
\draw [color={rgb, 255:red, 208; green, 2; blue, 27 }  ,draw opacity=1 ] [dash pattern={on 4.5pt off 4.5pt}]  (200.09,358.56) -- (269.97,139.42) ;
%Shape: Axis 2D [id:dp5777302597060503] 
\draw  (171.44,358.56) -- (457.88,358.56)(200.09,126.72) -- (200.09,384.32) (450.88,353.56) -- (457.88,358.56) -- (450.88,363.56) (195.09,133.72) -- (200.09,126.72) -- (205.09,133.72)  ;

% Text Node
\draw (306.09,332.9) node [anchor=north west][inner sep=0.75pt]  [font=\small,rotate=-343.8]  {$\widehat{\kappa } =\epsilon $};
% Text Node
\draw (429.54,368.61) node [anchor=north west][inner sep=0.75pt]    {$1/L$};
% Text Node
\draw (179.69,128.28) node [anchor=north west][inner sep=0.75pt]  [font=\large]  {$\kappa $};
% Text Node
\draw (265.48,266.23) node [anchor=north west][inner sep=0.75pt]  [font=\small,rotate=-315.26]  {$\widehat{\kappa } =1$};
% Text Node
\draw (213.49,247.24) node [anchor=north west][inner sep=0.75pt]  [font=\small,rotate=-287.04]  {$\widehat{\kappa } =1/\epsilon $};
% Text Node
\draw (366.08,335.17) node [anchor=north west][inner sep=0.75pt]   [align=left] {{\fontfamily{ptm}\selectfont {\footnotesize Bending Mode}}};
% Text Node
\draw (209.57,122.6) node [anchor=north west][inner sep=0.75pt]   [align=left] {{\fontfamily{ptm}\selectfont {\footnotesize Hybrid Mode}}};
% Text Node
\draw (289.98,158.91) node [anchor=north west][inner sep=0.75pt]   [align=left] {{\fontfamily{ptm}\selectfont {\footnotesize Membrane Mode}}};
% Text Node
\draw (210.09,105.79) node [anchor=north west][inner sep=0.75pt]  [font=\footnotesize]  {$\widehat{\kappa } \sim \mathcal{O}( 1/\epsilon )$};
% Text Node
\draw (366.84,319.92) node [anchor=north west][inner sep=0.75pt]  [font=\footnotesize]  {$\widehat{\kappa } \sim \mathcal{O}( \epsilon )$};
% Text Node
\draw (292.16,143.39) node [anchor=north west][inner sep=0.75pt]  [font=\footnotesize]  {$\widehat{\kappa } \sim \mathcal{O}( 1)$};
% Text Node
\draw (183.52,362.19) node [anchor=north west][inner sep=0.75pt]    {$O$};
\end{tikzpicture}
\caption{Shell deformation modes mapped to the $\kappa-\frac{1}{L}$ diagram.\label{Fig_deformation_region_classification}}
\end{figure}

The core of classification is about the order-determination of displacement and transverse stress components. Depending on the shell curvature, the orders of magnitude and corresponding expressions of the displacement components vary, which we have summarized in Table~\ref{Table2_disp_summary}. It can be observed that for both the weak and normal curvature cases, if we choose the following displacement field expressions for numerical implementation:
\begin{subequations}\label{unified_disp_expression}
	\begin{align}
		\wu_\alpha^{(0)}&=\wu_\alpha^\ast-\frac{\wxi_3}{\wa_\alpha}\frac{\partial\wu_3^{(0)}}{\partial\wxi_\alpha}; \label{unified_inplane_disp_expression}\\
		\wu_3^{(0)}&\leftarrow\wu_3^{(0)}+\epsilon\wu_3^{(1)}, \label{unified_normal_disp_expression}
	\end{align}
\end{subequations}
they can automatically reduce to the appropriate forms with a tolerable error when the principal curvatures meet the condition Eq.~\eqref{curv_nearflat} or Eq.~\eqref{curv_normal}. The leftward arrow above means replacing the original $\wu_3^{(0)}$ with $\wu_3^{(0)}+\epsilon\wu_3^{(1)}$. Specifically, for weakly curved shells, $\wu_3$ starts from $\mO\left(1\right)$, making the added term $\epsilon\wu_3^{(1)}$ a higher-order correction of $\mO\left(\epsilon\right)$; while for normally curved shells, the chosen expression (Eq.~\eqref{unified_disp_expression}) simplifies exactly to the form presented in Table~\ref{Table2_disp_summary} due to $\wu_3^{(0)}=0$.
\begin{table}
	\setlength{\abovecaptionskip}{0.cm}
	\setlength{\belowcaptionskip}{0.2cm}
	\small
	\caption{Summary of the orders of magnitude and expressions of displacement components corresponding to different curvature patterns discussed in Secs.~\ref{Sec_weaklycurved_shell}, \ref{Sec_normal_shell}, and \ref{Sec_significantlycurved_shell}.\label{Table2_disp_summary}}
	\centering
	\renewcommand{\arraystretch}{1.7} % adjusting row spacing
	\begin{tabular}{c|c|c c c}
		$\wu_i$ & Curvature Pattern & $\mO\left(1\right)$ & $\mO\left(\epsilon\right)$ & $\mO\left(\epsilon^2\right)$ \\
		\hline
		 & $\max\left(\left[\wk_\alpha\right]\right)\sim\mO\left(\epsilon\right)$ & $\wu_\alpha^{(0)}=\wu_\alpha^\ast-\frac{\wxi_3}{\wa_\alpha}\frac{\partial\wu_3^\ast}{\partial\wxi_\alpha}$ & $\wu_\alpha^{(1)}=\wu_\alpha^{\ast\ast}-\frac{\wxi_3}{\wa_\alpha}\frac{\partial\wu_3^{(1)}}{\partial\wxi_\alpha}$ & $\wu_\alpha^{(2)}\left(\wxi_1,\wxi_2,\wxi_3\right)$ \\
		$\wu_\alpha$ & $\max\left(\left[\wk_\alpha\right]\right)\sim\mO\left(1\right)$ & $\wu_\alpha^{(0)}=\wu_\alpha^\ast\left(\wxi_1,\wxi_2\right)$ & linear w.r.t. $\wxi_3$ & $\wu_\alpha^{(2)}\left(\wxi_1,\wxi_2,\wxi_3\right)$ \\
		 & $\max\left(\left[\wk_\alpha\right]\right)\sim\mO\left(\frac{1}{\epsilon}\right)$ & $\wu_\alpha^{(0)}=\wu_\alpha^\ast\left(1-\epsilon\wk_\alpha\wxi_3\right)$ & $\wu_\alpha^{(1)} = \wu_\alpha^{\ast\ast}\left(1-\epsilon\wk_\alpha\wxi_3\right)$ & $\wu_\alpha^{(2)}\left(\wxi_1,\wxi_2,\wxi_3\right)$ \\
		\hline
		 & $\max\left(\left[\wk_\alpha\right]\right)\sim\mO\left(\epsilon\right)$ & $\wu_3^{(0)}=\wu_3^\ast\left(\wxi_1,\wxi_2\right)$ & $\wu_3^{(1)}=\wu_3^{(1)}\left(\wxi_1,\wxi_2\right)$ & $\wu_3^{(2)}\left(\wxi_1,\wxi_2,\wxi_3\right)$ \\
		$\wu_3$ & $\max\left(\left[\wk_\alpha\right]\right)\sim\mO\left(1\right)$ & 0 & $\wu_3^{(1)}=\wu_3^{(1)}\left(\wxi_1,\wxi_2\right)$ & linear w.r.t. $\wxi_3$ \\
		 & $\max\left(\left[\wk_\alpha\right]\right)\sim\mO\left(\frac{1}{\epsilon}\right)$ & 0 & 0 & $\wu_3^{(2)}\left(\wxi_1,\wxi_2,\wxi_3\right)$ \\
	\end{tabular}
\end{table}

However, two comments regarding the case of significant curvature are warranted here. Firstly, the magnitude of curvature, i.e., $\max\left(\left[\wk_\alpha\right]\right)\sim\mO\left(1/\epsilon\right)$, eliminates terms of $\mO\left(\epsilon\right)$ and higher in the expansion of the reciprocal of the \textit{Lam\'{e} coefficients} (refer to Eq.~\eqref{lame_coefficient_expansion}). This, in turn, prevents the leading-order stress and strain from being expressed as a (limited) polynomial function of $\wxi_3$. Secondly, if higher-order terms of $\wxi_3$ (i.e., $\wxi_3^i,\ i\geq2$) are neglected, a linear relationship can still be derived, and $\wu_3^{(2)}$ remains linear in the thickness direction. Assume that $\wu_3^{(2)}=f\left(\wxi_1,\wxi_2\right)\cdot\wxi_3+g\left(\wxi_1,\wxi_2\right)$, a unified expression for the displacement applicable to all three cases can then be obtained, that is,
\begin{subequations}
	\begin{align}
		\wu_\alpha^{(0)}&=\wu_\alpha^\ast\left(1-\epsilon\wk_\alpha\wxi_3\right)-\frac{\wxi_3}{\wa_\alpha}\frac{\partial\wu_3^{(0)}}{\partial\wxi_\alpha};\\
		\wu_3^{(0)}&\leftarrow\wu_3^{(0)}+\epsilon\wu_3^{(1)}+\epsilon^2\left(f\cdot\wxi_3+g\right).
	\end{align}
\end{subequations}
The error introduced by this linearization can be large and may result in certain deviations for the approximation of significantly curved shells. These distinguish the third, significant-curvature mode from the first two. In practical coding process, the previously non-dimensionalized variables must be rescaled according to the corresponding scaling relations (generally refer to Sec.~\ref{Sec_scaling}). Since both $\wu_3^{(0)}$ and $\wu_3^{(1)}$ in Eq.~\eqref{unified_normal_disp_expression} are independent of the thickness coordinate, their dimensional counterparts can be directly represented by $\wu_3^\ast$, which is measured on shell mid-surface. The in-plane displacement components remain linear functions of $\wxi_3$, composed of mid-surface displacement terms. As a result, the first two deformation modes can be naturally unified within a single computational framework, with only three unknowns corresponding to the midsurface displacement components. So for simplicity, the unified expressions of the first two curvature patterns are of our main concern.

By eliminating the terms containing $\pd{\wu_3}{\wxi_3}$ from the constitutive relations for the in-plane normal stress components $\wsigma_{11}$ and $\wsigma_{22}$ based on Eq.~\eqref{constitutive_dimensionless6}, we can derive the leading-order expressions for the dominant stress components $\wsigma_{\alpha\beta}^{(0)}$. Combining this with Eq.~\eqref{unified_disp_expression}, we can further relate the in-plane strain and stress components to the displacement field of the shell mid-surface. For strain components, the following expression is obtained
\begin{subequations}\label{strain_leading_midsurface}
	\begin{align}
		\begin{split}
			\varepsilon_{11}^{(0)} = &\left(\frac{1}{\wa_1}\pd{\wu_1^\ast}{\wxi_1}+\frac{\wu_2^\ast}{\wa_1\wa_2}\pd{\wa_1}{\wxi_2}-\frac{\wk_1}{\epsilon}\wu_3^\ast\right)-\wxi_3\left[\frac{1}{\wa_1}\pd{}{\wxi_1}\left(\frac{1}{\wa_1}\pd{\wu_3^\ast}{\wxi_1}\right)+\frac{1}{\wa_1\wa_2^2}\pd{\wa_1}{\wxi_2}\pd{\wu_3^\ast}{\wxi_2}\right]\\
			= &\varepsilon_{11}^\ast+\wxi_3\left(-k_{11}^\ast\right);
		\end{split}\\
		\begin{split}
			\varepsilon_{22}^{(0)} = &\left(\frac{1}{\wa_2}\pd{\wu_2^\ast}{\wxi_2}+\frac{\wu_1^\ast}{\wa_1\wa_2}\pd{\wa_2}{\wxi_1}-\frac{\wk_2}{\epsilon}\wu_3^\ast\right)-\wxi_3\left[\frac{1}{\wa_2}\pd{}{\wxi_2}\left(\frac{1}{\wa_2}\pd{\wu_3^\ast}{\wxi_2}\right)+\frac{1}{\wa_1^2\wa_2}\pd{\wa_2}{\wxi_1}\pd{\wu_3^\ast}{\wxi_1}\right]\\
			= &\varepsilon_{22}^\ast+\wxi_3\left(-k_{22}^\ast\right);
		\end{split}\\
		\begin{split}
			\gamma_{12}^{(0)} = &\left(\frac{1}{\wa_1}\pd{\wu_2^\ast}{\wxi_1}+\frac{1}{\wa_2}\pd{\wu_1^\ast}{\wxi_2}-\frac{\wu_2^\ast}{\wa_1\wa_2}\pd{\wa_2}{\wxi_1}-\frac{\wu_1^\ast}{\wa_1\wa_2}\pd{\wa_1}{\wxi_2}\right)\\
			-&2\,\wxi_3\left(\frac{1}{\wa_1\wa_2}\mpd{\wu_3^\ast}{\wxi_1}{\wxi_2}-\frac{1}{\wa_1\wa_2^2}\pd{\wa_2}{\wxi_1}\pd{\wu_3^\ast}{\wxi_2}-\frac{1}{\wa_1^2\wa_2}\pd{\wa_1}{\wxi_2}\pd{\wu_3^\ast}{\wxi_1}\right) = \gamma_{12}^\ast+\wxi_3\left(-2\,k_{12}^\ast\right),
		\end{split}
	\end{align}
\end{subequations}
here $\boldsymbol{\varepsilon}^\ast = \varepsilon_{\alpha\beta}^\ast\,\ve_{\alpha}\otimes\ve_{\beta}$ is the membrane strain tensor associated with the mid-surface $\CS$ of the shell; $\boldsymbol{k}^\ast = -k_{\alpha\beta}^\ast\,\ve_{\alpha}\otimes\ve_{\beta}$ denotes the changes of curvature of the local mid-surface infinitesimal. The in-plane stresses are able to be further related to these six general strain components of the mid-surface of the MTS, i.e.,
\begin{subequations}\label{dominant_stress_final}
	\begin{align}
		\wsigma_{11}^{(0)} = &\frac{\wE_{\sss\CL}}{1-\nu_{\sss\CL}^2}\left[\varepsilon_{11}^\ast+\nu_{\sss\CL}\,\varepsilon_{22}^\ast+\wxi_3\left(-k_{11}^\ast\right)+\nu_{\sss\CL}\,\wxi_3\left(-k_{22}^\ast\right)\right];\\
		\wsigma_{12}^{(0)} = &\frac{\wE_{\sss\CL}}{2(1+\nu_{\sss\CL})}\left[\gamma_{12}^\ast+\wxi_3\left(-2\,k_{12}^\ast\right)\right];\\
		\wsigma_{22}^{(0)} = &\frac{\wE_{\sss\CL}}{1-\nu_{\sss\CL}^2}\left[\nu_{\sss\CL}\,\varepsilon_{11}^\ast+\varepsilon_{22}^\ast+\nu_{\sss\CL}\,\wxi_3\left(-k_{11}^\ast\right)+\wxi_3\left(-k_{22}^\ast\right)\right].
	\end{align}
\end{subequations}

\subsubsection{Homogenization}\label{Sec_homogenization}
It follows that, so far, we have only obtained the corresponding governing equations based on the geometrical and load-bearing characteristics of the MTS, but what we are dealing with remains a three-dimensional equation defined in local curvilinear coordinates, which does not alleviate the complexity of the solving process. Specifically, although only the stress components $\wsigma_{\alpha\beta}^{(0)}$ mentioned above are dominant, the equilibrium of the whole structure is still maintained by all three-dimensional stress components $\wsigma_{ij}^{(0)}$ together, including the dependence of in-plane stresses on the locally complete three curvilinear coordinates $\wxi_i,\ i = 1,2,3$. An intuitive idea is to compress the information about the field functions inside the shell as much as possible into its mid-surface, thus transforming an original three-dimensional problem into one on a two-dimensional manifold.

To this end, homogenization is naturally carried out on local stress fields in the shell thickness direction to generate the stress resultant and stress couple fields, which are defined as follows
\begin{enumerate}
	\item [(1)] The non-dimensional in-plane stress resultants:
	\begin{equation}\label{stress_resultant_inplane}
		\boldsymbol{\wT} = \wT_{\alpha\beta}\,\ve_\alpha\otimes\ve_\beta;\quad\wT_{\alpha\beta} = \int_{-\frac{1}{2}}^{\frac{1}{2}}\wsigma_{\alpha\beta}\,\intd\wxi_3.
	\end{equation}
	\item [(2)] The non-dimensional transverse shear stress resultants:
	\begin{equation}
		\wN_\alpha = \int_{-\frac{1}{2}}^{\frac{1}{2}}\wsigma_{\alpha3}\,\intd\wxi_3.
	\end{equation}
	\item [(3)] The non-dimensional stress couples:
	\begin{equation}\label{stress_couple}
		\boldsymbol{\wM} = \wM_{\alpha\beta}\,\ve_\alpha\otimes\ve_\beta;\quad\wM_{\alpha\beta} = \int_{-\frac{1}{2}}^{\frac{1}{2}}\wsigma_{\alpha\beta}\,\wxi_3\,\intd\wxi_3.
	\end{equation}
\end{enumerate}

In addition the $m$-th moment of a piece-wise constant function defined in the interval $\left(-\frac{1}{2},\frac{1}{2}\right)$ is introduced to facilitate the subsequent derivation, and the expression is written by
\begin{equation}\label{mth_moment_piecewiseFunction}
	\left\langle A_{\sss\CL}\right\rangle_m = \sum_{\CL=1}^N A_{\sss\CL}\int_{\wxi_3^\CL}^{\wxi_3^{\CL+1}} (\wxi_3)^m\,\intd\wxi_3,
\end{equation}
here the interval $\left(\wxi_3^\CL,\wxi_3^{\CL+1}\right)$ denotes the dimensionless thickness range of \textit{Layer $\CL$}.

Now we set about deriving the reduced two-dimensional governing equations of the investigated MTS. First, integrating the leading order of Eq.~\eqref{equilibrium_dimensionless} with respect to $\wxi_3$ over the interval $\left(-\frac{1}{2},\frac{1}{2}\right)$ yields
\begin{subequations}\label{reduced_equilibrium_T}
	\begin{align}
		\epsilon^{-2m_1}\wCT_1^\mathrm{I} &= \frac{1}{\wa_1\wa_2}\left[\pd{}{\wxi_1}\left(\wa_2\,\wT_{11}^{(0)}\right)+\pd{}{\wxi_2}\left(\wa_1\,\wT_{12}^{(0)}\right)+\wT_{12}^{(0)}\pd{\wa_1}{\wxi_2}-\wT_{22}^{(0)}\pd{\wa_2}{\wxi_1}\right]+\wP_1;\label{reduced_equilibrium_T1}\\
		\epsilon^{-2m_1}\wCT_2^\mathrm{I} &= \frac{1}{\wa_1\wa_2}\left[\pd{}{\wxi_1}\left(\wa_2\,\wT_{12}^{(0)}\right)+\pd{}{\wxi_2}\left(\wa_1\,\wT_{22}^{(0)}\right)+\wT_{12}^{(0)}\pd{\wa_2}{\wxi_1}-\wT_{11}^{(0)}\pd{\wa_1}{\wxi_2}\right]+\wP_2;\label{reduced_equilibrium_T2}\\
		\epsilon^{-2m_1-2}\,\wCT_3^\mathrm{I} &= \frac{1}{\wa_1\wa_2}\left[\pd{}{\wxi_1}\left(\wa_2\,\wN_1^{(0)}\right)+\pd{}{\wxi_2}\left(\wa_1\,\wN_2^{(0)}\right)\right]+\frac{\wk_1}{\epsilon}\wT_{11}^{(0)}+\frac{\wk_2}{\epsilon}\wT_{22}^{(0)}+\wP_3,\label{reduced_equilibrium_T3}
	\end{align}
\end{subequations}
here, $\wP_\alpha=\wload_\alpha+\epsilon^{m_2-3}\left\langle\wf_{\sss\CL}^\alpha\right\rangle_0$ and $\wP_3=\wload_3+\epsilon^{m_2-4}\left\langle\wf_{\sss\CL}^3\right\rangle_0$ represent the homogenized aggregate effects of the in-plane and transverse forces, respectively. The superscript ``$\mathrm{I}$'' here implies that the corresponding variable relates to the inertial force, and the new introduced notations $\wCT_i^\mathrm{I}$ represent the homogenized inertial force components, whose specific definitions are given below
\begin{equation}\label{homo_inertial_force}
	\wCT_i^\mathrm{I} = \sum_{\CL=1}^N\int_{\wxi_3^\CL}^{\wxi_3^{\CL+1}}\wrho_{\sss\CL}\,\ddot{\wu}_i\,\intd\wxi_3 = \left\langle\wrho_{\sss\mathcal{L}}\right\rangle_0\spd{\wu_i^\ast}{\wt}+\left\langle\wrho_{\sss\mathcal{L}}\right\rangle_1\spd{\wphi_i^\ast}{\wt},
\end{equation}
where $\left\langle\bigcdot\right\rangle_m,\,m=0,1$, take the definition of Eq.~\eqref{mth_moment_piecewiseFunction}. The term $\frac{\partial^2\boldsymbol{\wu}^\ast}{\partial\wt^2} = \left[\frac{\partial^2\wu_1^\ast}{\partial\wt^2},\frac{\partial^2\wu_2^\ast}{\partial\wt^2},\frac{\partial^2\wu_3^\ast}{\partial\wt^2}\right]^\rT$ represents the dimensionless acceleration vector, and $\frac{\partial^2\boldsymbol{\wphi}^\ast}{\partial\wt^2} = \left[-\frac{1}{\wa_1}\frac{\partial\ddot{\wu}_3^\ast}{\partial\wxi_1},-\frac{1}{\wa_2}\frac{\partial\ddot{\wu}_3^\ast}{\partial\wxi_2},0\right]^\rT$ represents the dimensionless angular acceleration vector.

Note that Eq.~\eqref{reduced_equilibrium_T3} contains the transverse shear stress resultants $\wN_\alpha^{(0)}$, which actually account for the out-of-plane effects, so in order to obtain a set of fully two-dimensional governing equations, another equation related to the stress couples $\wM_{\alpha\beta}^{(0)}$ should be determined. Here the leading order of Eqs.~\eqref{equilibrium_dimensionless1} and \eqref{equilibrium_dimensionless2} are multiplied by the normal coordinate $\wxi_3$, and their corresponding first-order moments are then given by
\begin{subequations}
	\begin{align}
		\epsilon^{-2m_1}\wCM_1^\mathrm{I} &= \frac{1}{\wa_1\wa_2}\left[\pd{}{\wxi_1}\left(\wa_2\,\wM_{11}^{(0)}\right)+\pd{}{\wxi_2}\left(\wa_1\,\wM_{12}^{(0)}\right)+\wM_{12}^{(0)}\pd{\wa_1}{\wxi_2}-\wM_{22}^{(0)}\pd{\wa_2}{\wxi_1}\right]-\wN_1^{(0)}+\wQ_1;\\
		\epsilon^{-2m_1}\wCM_2^\mathrm{I} &= \frac{1}{\wa_1\wa_2}\left[\pd{}{\wxi_1}\left(\wa_2\,\wM_{12}^{(0)}\right)+\pd{}{\wxi_2}\left(\wa_1\,\wM_{22}^{(0)}\right)+\wM_{12}^{(0)}\pd{\wa_2}{\wxi_1}-\wM_{11}^{(0)}\pd{\wa_1}{\wxi_2}\right]-\wN_2^{(0)}+\wQ_2,
	\end{align}
\end{subequations}
where $\wQ_\alpha=\frac{\wload_\alpha}{2}+\epsilon^{m_2-3}\left\langle\wf_{\sss\CL}^\alpha\right\rangle_1$. Likewise, the notations $\wCM_i^\mathrm{I}$ here represent the homogenized inertial force moments, which are defined by
\begin{equation}\label{homo_inertial_moment}
	\wCM_i^\mathrm{I} = \sum_{\CL=1}^N\int_{\wxi_3^\CL}^{\wxi_3^{\CL+1}}\wrho_{\sss\CL}\,\ddot{\wu}_i\,\wxi_3\,\intd\wxi_3 = \left\langle\wrho_{\sss\mathcal{L}}\right\rangle_1\spd{\wu_i^\ast}{\wt}+\left\langle\wrho_{\sss\mathcal{L}}\right\rangle_2\spd{\wphi_i^\ast}{\wt}.
\end{equation}

Combined with Eq.~\eqref{reduced_equilibrium_T3}, the terms regarding $\wN_\alpha^{(0)}$ can be eliminated, thus resulting in a relation between the leading-order stress couples $\wM_{\alpha\beta}^{(0)}$, i.e.,
\begin{equation}\label{reduced_equilibrium_M}
	\begin{split}
		&\frac{1}{\wa_1\wa_2}\left\{\pd{}{\wxi_1}\left(\frac{\wa_2}{\wa_1}\pd{\wM_{11}^{(0)}}{\wxi_1}\right)+\pd{}{\wxi_2}\left(\frac{\wa_1}{\wa_2}\pd{\wM_{22}^{(0)}}{\wxi_2}\right)+\pd{}{\wxi_1}\left[\frac{\left(\wM_{11}^{(0)}-\wM_{22}^{(0)}\right)}{\wa_1}\pd{\wa_2}{\wxi_1}\right]+2\mpd{\wM_{12}^{(0)}}{\wxi_1}{\wxi_2}+\right.\\
		&\left.\pd{}{\wxi_2}\left[\frac{\left(\wM_{22}^{(0)}-\wM_{11}^{(0)}\right)}{\wa_2}\pd{\wa_1}{\wxi_2}\right]+\pd{}{\wxi_1}\left(\frac{2\wM_{12}^{(0)}}{\wa_1}\pd{\wa_1}{\wxi_2}\right)+\pd{}{\wxi_2}\left(\frac{2\wM_{12}^{(0)}}{\wa_2}\pd{\wa_2}{\wxi_1}\right)+\pd{}{\wxi_1}\left(\wa_2\wQ_1\right)+\right.\\
		&\left.\pd{}{\wxi_2}\left(\wa_1\wQ_2\right)-\epsilon^{-2m_1}\left[\pd{\left(\wa_2\wCM_1^\mathrm{I}\right)}{\wxi_1}+\pd{\left(\wa_1\wCM_2^\mathrm{I}\right)}{\wxi_2}\right]\right\}+\frac{\wk_1}{\epsilon}\wT_{11}^{(0)}+\frac{\wk_2}{\epsilon}\wT_{22}^{(0)}+\wP_3 = \epsilon^{-2m_1-2}\,\wCT_3^\mathrm{I}.
	\end{split}
\end{equation}

So far, Eqs.~\eqref{reduced_equilibrium_T1}, \eqref{reduced_equilibrium_T2}, and \eqref{reduced_equilibrium_M} together constitute the homogenized dynamic equilibrium equations of the MTS, which are a set of equations containing the in-plane stress resultants and couples (both just functions of the in-plane dominant stress components $\wsigma_{\alpha\beta}^{(0)}$). Based on the specific expressions of the leading-order in-plane stress components Eq.~\eqref{dominant_stress_final}, the involved homogenized quantities defined by Eqs.~\eqref{stress_resultant_inplane} and \eqref{stress_couple} can be expressed by
\begin{subequations}\label{stress_resultant_couple}
	\begin{align}
		\wT_{\alpha\beta}^{(0)} &= \CA_{\alpha\beta\gamma\lambda}\,\varepsilon_{\gamma\lambda}^\ast+\CB_{\alpha\beta\gamma\lambda}\left(-k_{\gamma\lambda}^\ast\right);\\
		\wM_{\alpha\beta}^{(0)} &= \CB_{\alpha\beta\gamma\lambda}\,\varepsilon_{\gamma\lambda}^\ast+\CD_{\alpha\beta\gamma\lambda}\left(-k_{\gamma\lambda}^\ast\right),
	\end{align}
\end{subequations}
where the introduced material matrices are actually the different order moments of the original (non-dimensional) elasticity matrix, that is,
\begin{equation}\label{stiffness_matrices_ABD}
	\CA_{\alpha\beta\gamma\lambda} = \left\langle\bmC_{\alpha\beta\gamma\lambda}\right\rangle_0;\quad\CB_{\alpha\beta\gamma\lambda} = \left\langle\bmC_{\alpha\beta\gamma\lambda}\right\rangle_1;\quad\CD_{\alpha\beta\gamma\lambda} = \left\langle\bmC_{\alpha\beta\gamma\lambda}\right\rangle_2,
\end{equation}
it can be found that, in a physical sense, the matrices $\vCA$, $\vCD$ and $\vCB$ capture the local extensional, bending and coupling effects of the shell mid-surface infinitesimal, respectively. For isotropic cases, we have $\CA_{\alpha\beta\gamma\lambda} = \CA_{\beta\alpha\gamma\lambda} = \CA_{\gamma\lambda\alpha\beta},\ \CB_{\alpha\beta\gamma\lambda} = \CB_{\beta\alpha\gamma\lambda} = \CB_{\gamma\lambda\alpha\beta},\ \CD_{\alpha\beta\gamma\lambda} = \CD_{\beta\alpha\gamma\lambda} = \CD_{\gamma\lambda\alpha\beta}$. Their specific matrix forms are given as follows
\begin{equation}
	\vCA = \begin{bmatrix}
		\CA_1 & \CA_2 & 0\\ & \CA_1 & 0\\ sym. & & \frac{\CA_1-\CA_2}{2}\\
	\end{bmatrix};\quad\vCB = \begin{bmatrix}
		\CB_1 & \CB_2 & 0\\ & \CB_1 & 0\\ sym. & & \frac{\CB_1-\CB_2}{2}\\
	\end{bmatrix};\quad\vCD = \begin{bmatrix}
		\CD_1 & \CD_2 & 0\\ & \CD_1 & 0\\ sym. & & \frac{\CD_1-\CD_2}{2}\\
	\end{bmatrix},
\end{equation}
and with Eq.~\eqref{mth_moment_piecewiseFunction}, the corresponding entries can be expressed by
\begin{subequations}\label{stiffness_ABD_entries}
	\begin{align}
		&\CA_1 = \left\langle\frac{E_{\sss\CL}}{1-\nu_{\sss\CL}^2}\right\rangle_0,\ \CA_2 = \left\langle\frac{E_{\sss\CL}\nu_{\sss\CL}}{1-\nu_{\sss\CL}^2}\right\rangle_0;\\
		&\CB_1 = \left\langle\frac{E_{\sss\CL}}{1-\nu_{\sss\CL}^2}\right\rangle_1,\ \CB_2 = \left\langle\frac{E_{\sss\CL}\nu_{\sss\CL}}{1-\nu_{\sss\CL}^2}\right\rangle_1;\\
		&\CD_1 = \left\langle\frac{E_{\sss\CL}}{1-\nu_{\sss\CL}^2}\right\rangle_2,\ \CD_2 = \left\langle\frac{E_{\sss\CL}\nu_{\sss\CL}}{1-\nu_{\sss\CL}^2}\right\rangle_2.
	\end{align}
\end{subequations}

\subsubsection{Model Implementation in an Isogeometric Manner}\label{Sec_IGA_implementation}
The isogeometric discretization (\cite{hughes2005isogeometric}) is considered for the simulation process due to the fact that the non-uniform rational B-spline (NURBS) basis function is able to provide smoothness of any order ($C^\infty-$continuity) across elements while accurately representing complex surfaces. And these properties prove particularly advantageous in the rotation-free Kirchhoff-Love shell models (\cite{kiendl2009isogeometric}, \cite{nguyen2011rotation}, \cite{kiendl2015isogeometric}), where at least $C^1-$continuity is required for the approximate basis functions.

As indicated, the original geometric representation of the shell mid-surface Eq.~\eqref{representation_midsurface} can be alternatively achieved by means of NURBS surfaces in the form of
\begin{equation}\label{NURBS_representation}
	\vr\left(\xi_1,\xi_2\right) = \sum_{i=1}^n\sum_{j=1}^mR_{i,j}^{p,q}\left(\xi_1,\xi_2\right)\vP_{i,j},
\end{equation}
where $\xi_1,\,\xi_2$ are the NURBS parameters, as illustrated in Fig.~\ref{Fig_midsurface_nurbs}. $\vP_{i,j}$ represents a vector comprising the coordinates of $n\times m$ control points in the $\xi_1$-parameter and $\xi_2$-parameter directions, while $R_{i,j}^{p,q}$ stand for the NURBS basis functions (piecewise rational polynomials), with $p,q$ denoting the degrees of the corresponding B-spline basis functions. For further insights into the mathematics behind the NURBS surfaces, one may refer to the work of \cite{Piegl1997}.

\begin{figure}[!htbp]
	\setlength{\abovecaptionskip}{0.cm}
	\setlength{\belowcaptionskip}{-0.cm}
	\centering
\tikzset{every picture/.style={line width=0.75pt}} %set default line width to 0.75pt
\begin{tikzpicture}[x=0.75pt,y=0.75pt,yscale=-1,xscale=1]
%uncomment if require: \path (0,444); %set diagram left start at 0, and has height of 444
%Curve Lines [id:da7560499061400372] 
\draw [fill={rgb, 255:red, 74; green, 144; blue, 226 }  ,fill opacity=0.7 ][line width=0.75]  [dash pattern={on 4.5pt off 4.5pt}]  (405.72,94.06) .. controls (428.99,81.91) and (548.03,104.06) .. (534.08,106) .. controls (520.14,107.94) and (481.7,213.84) .. (473.88,205.13) .. controls (466.06,196.42) and (365.61,179.99) .. (355.18,189.04) .. controls (344.75,198.09) and (381.03,116.66) .. (405.72,94.06) -- cycle ;
%Straight Lines [id:da9470539388774353] 
\draw    (428.85,155.77) -- (465,162.19) ;
\draw [shift={(466.97,162.54)}, rotate = 190.06] [color={rgb, 255:red, 0; green, 0; blue, 0 }  ][line width=0.75]    (10.93,-3.29) .. controls (6.95,-1.4) and (3.31,-0.3) .. (0,0) .. controls (3.31,0.3) and (6.95,1.4) .. (10.93,3.29)   ;
%Straight Lines [id:da9685837943313702] 
\draw    (428.85,155.77) -- (445.56,124.09) ;
\draw [shift={(446.5,122.32)}, rotate = 117.82] [color={rgb, 255:red, 0; green, 0; blue, 0 }  ][line width=0.75]    (10.93,-3.29) .. controls (6.95,-1.4) and (3.31,-0.3) .. (0,0) .. controls (3.31,0.3) and (6.95,1.4) .. (10.93,3.29)   ;
%Straight Lines [id:da05119896336027163] 
\draw    (518.49,195.56) -- (509.32,205.72) ;
\draw [shift={(507.97,207.2)}, rotate = 312.11] [fill={rgb, 255:red, 0; green, 0; blue, 0 }  ][line width=0.08]  [draw opacity=0] (12,-3) -- (0,0) -- (12,3) -- cycle    ;
%Straight Lines [id:da9875196247013553] 
\draw    (518.49,195.56) -- (535.52,195.3) ;
\draw [shift={(537.52,195.27)}, rotate = 179.1] [fill={rgb, 255:red, 0; green, 0; blue, 0 }  ][line width=0.08]  [draw opacity=0] (12,-3) -- (0,0) -- (12,3) -- cycle    ;
%Straight Lines [id:da06360195652309408] 
\draw    (518.49,195.56) -- (518.44,179.58) ;
\draw [shift={(518.44,177.58)}, rotate = 89.82] [fill={rgb, 255:red, 0; green, 0; blue, 0 }  ][line width=0.08]  [draw opacity=0] (12,-3) -- (0,0) -- (12,3) -- cycle    ;
%Straight Lines [id:da6622559153762282] 
\draw [color={rgb, 255:red, 208; green, 2; blue, 27 }  ,draw opacity=0.8 ]   (428.85,155.77) -- (445.94,158.81) ;
\draw [shift={(447.91,159.15)}, rotate = 190.06] [color={rgb, 255:red, 208; green, 2; blue, 27 }  ,draw opacity=0.8 ][line width=0.75]    (10.93,-3.29) .. controls (6.95,-1.4) and (3.31,-0.3) .. (0,0) .. controls (3.31,0.3) and (6.95,1.4) .. (10.93,3.29)   ;
%Straight Lines [id:da6387227895950764] 
\draw [color={rgb, 255:red, 208; green, 2; blue, 27 }  ,draw opacity=0.8 ]   (428.85,155.77) -- (436.74,140.82) ;
\draw [shift={(437.67,139.05)}, rotate = 117.82] [color={rgb, 255:red, 208; green, 2; blue, 27 }  ,draw opacity=0.8 ][line width=0.75]    (10.93,-3.29) .. controls (6.95,-1.4) and (3.31,-0.3) .. (0,0) .. controls (3.31,0.3) and (6.95,1.4) .. (10.93,3.29)   ;
%Shape: Rectangle [id:dp6002186032367596] 
\draw  [fill={rgb, 255:red, 74; green, 144; blue, 226 }  ,fill opacity=0.7 ] (145.41,104.34) -- (226.61,104.34) -- (226.61,184.35) -- (145.41,184.35) -- cycle ;
%Curve Lines [id:da5258990211986496] 
\draw    (250.85,120.61) .. controls (272.51,91.05) and (311.46,92.07) .. (353.43,109.34) ;
\draw [shift={(356,110.41)}, rotate = 203.07] [fill={rgb, 255:red, 0; green, 0; blue, 0 }  ][line width=0.08]  [draw opacity=0] (10.72,-5.15) -- (0,0) -- (10.72,5.15) -- (7.12,0) -- cycle    ;
%Straight Lines [id:da5920861403360995] 
\draw    (428.85,155.77) -- (462.89,139.74) ;
\draw [shift={(464.7,138.89)}, rotate = 154.78] [color={rgb, 255:red, 0; green, 0; blue, 0 }  ][line width=0.75]    (10.93,-3.29) .. controls (6.95,-1.4) and (3.31,-0.3) .. (0,0) .. controls (3.31,0.3) and (6.95,1.4) .. (10.93,3.29)   ;
%Straight Lines [id:da44403402536336545] 
\draw [color={rgb, 255:red, 5; green, 37; blue, 255 }  ,draw opacity=1 ]   (428.85,155.77) -- (444.96,148.18) ;
\draw [shift={(446.77,147.33)}, rotate = 154.78] [color={rgb, 255:red, 5; green, 37; blue, 255 }  ,draw opacity=1 ][line width=0.75]    (10.93,-3.29) .. controls (6.95,-1.4) and (3.31,-0.3) .. (0,0) .. controls (3.31,0.3) and (6.95,1.4) .. (10.93,3.29)   ;
%Straight Lines [id:da9822939770696089] 
\draw    (428.92,156.22) -- (416.18,123.75) ;
\draw [shift={(415.45,121.89)}, rotate = 68.58] [color={rgb, 255:red, 0; green, 0; blue, 0 }  ][line width=0.75]    (10.93,-3.29) .. controls (6.95,-1.4) and (3.31,-0.3) .. (0,0) .. controls (3.31,0.3) and (6.95,1.4) .. (10.93,3.29)   ;
%Straight Lines [id:da8270035206468058] 
\draw [color={rgb, 255:red, 5; green, 37; blue, 255 }  ,draw opacity=1 ]   (428.92,156.22) -- (422.91,140.92) ;
\draw [shift={(422.18,139.05)}, rotate = 68.58] [color={rgb, 255:red, 5; green, 37; blue, 255 }  ,draw opacity=1 ][line width=0.75]    (10.93,-3.29) .. controls (6.95,-1.4) and (3.31,-0.3) .. (0,0) .. controls (3.31,0.3) and (6.95,1.4) .. (10.93,3.29)   ;
%Straight Lines [id:da1875103827362985] 
\draw    (122.25,206.18) -- (145.06,206.22) ;
\draw [shift={(147.06,206.23)}, rotate = 180.12] [fill={rgb, 255:red, 0; green, 0; blue, 0 }  ][line width=0.08]  [draw opacity=0] (12,-3) -- (0,0) -- (12,3) -- cycle    ;
%Straight Lines [id:da8736700099517034] 
\draw    (122.25,206.18) -- (122.65,184.77) ;
\draw [shift={(122.69,182.77)}, rotate = 91.07] [fill={rgb, 255:red, 0; green, 0; blue, 0 }  ][line width=0.08]  [draw opacity=0] (12,-3) -- (0,0) -- (12,3) -- cycle    ;
%Curve Lines [id:da9668170326210486] 
\draw [fill={rgb, 255:red, 74; green, 144; blue, 226 }  ,fill opacity=0.7 ]   (269.13,304.38) .. controls (288.86,274.56) and (257.56,241.71) .. (277.97,243.39) .. controls (298.38,245.06) and (330.97,233.4) .. (344.98,244.73) .. controls (358.98,256.05) and (354.5,306.73) .. (346,310.41) .. controls (337.49,314.1) and (255.52,320.47) .. (269.13,304.38) -- cycle ;
%Curve Lines [id:da28253118716093173] 
\draw    (255.33,278.31) .. controls (194.65,288.69) and (184.88,258.92) .. (184.55,239.87) ;
\draw [shift={(259.11,277.63)}, rotate = 169.14] [fill={rgb, 255:red, 0; green, 0; blue, 0 }  ][line width=0.08]  [draw opacity=0] (10.72,-5.15) -- (0,0) -- (10.72,5.15) -- (7.12,0) -- cycle    ;
%Curve Lines [id:da6953580443806051] 
\draw    (451.71,245.67) .. controls (436.91,288.29) and (392.89,286.49) .. (377.97,278.33) ;
\draw [shift={(452.79,242.3)}, rotate = 106.45] [fill={rgb, 255:red, 0; green, 0; blue, 0 }  ][line width=0.08]  [draw opacity=0] (10.72,-5.15) -- (0,0) -- (10.72,5.15) -- (7.12,0) -- cycle    ;

% Text Node
\draw (445.48,106.16) node [anchor=north west][inner sep=0.75pt]  [font=\footnotesize]  {$\widehat{\xi _{2}}$};
% Text Node
\draw (537.62,185.85) node [anchor=north west][inner sep=0.75pt]  [font=\footnotesize]  {$\widetilde{\boldsymbol{e_{2}}}$};
% Text Node
\draw (521.87,166.88) node [anchor=north west][inner sep=0.75pt]  [font=\footnotesize]  {$\widetilde{\boldsymbol{e_{3}}}$};
% Text Node
\draw (493.82,204.94) node [anchor=north west][inner sep=0.75pt]  [font=\footnotesize]  {$\widetilde{\boldsymbol{e_{1}}}$};
% Text Node
\draw (424.21,131.4) node [anchor=north west][inner sep=0.75pt]  [font=\scriptsize]  {$\boldsymbol{e_{2}}$};
% Text Node
\draw (438.97,161.82) node [anchor=north west][inner sep=0.75pt]  [font=\scriptsize]  {$\boldsymbol{e_{1}}$};
% Text Node
\draw (468.1,155.13) node [anchor=north west][inner sep=0.75pt]  [font=\footnotesize]  {$\widehat{\xi _{1}}$};
% Text Node
\draw (507.86,105.04) node [anchor=north west][inner sep=0.75pt]  [font=\small]  {$S$};
% Text Node
\draw (418.48,159.45) node [anchor=north west][inner sep=0.75pt]  [font=\small]  {$P$};
% Text Node
\draw (149.39,201.35) node [anchor=north west][inner sep=0.75pt]  [font=\footnotesize]  {$\xi _{1}$};
% Text Node
\draw (105.03,174.2) node [anchor=north west][inner sep=0.75pt]  [font=\footnotesize]  {$\xi _{2}$};
% Text Node
\draw (269.97,77.46) node [anchor=north west][inner sep=0.75pt]  [font=\small]  {$\mathbf{r}( \xi _{1} ,\xi _{2})$};
% Text Node
\draw (133.52,217.76) node [anchor=north west][inner sep=0.75pt]  [font=\normalsize] [align=left] {{\fontfamily{ptm}\selectfont {\small Parametric Space}}};
% Text Node
\draw (399.35,217.79) node [anchor=north west][inner sep=0.75pt]  [font=\normalsize] [align=left] {{\fontfamily{ptm}\selectfont {\small Physical Space}}};
% Text Node
\draw (133.74,184.73) node [anchor=north west][inner sep=0.75pt]  [font=\small]  {$0$};
% Text Node
\draw (227.42,186.13) node [anchor=north west][inner sep=0.75pt]  [font=\small]  {$1$};
% Text Node
\draw (132.61,99) node [anchor=north west][inner sep=0.75pt]  [font=\small]  {$1$};
% Text Node
\draw (468.2,129.44) node [anchor=north west][inner sep=0.75pt]  [font=\footnotesize]  {$\xi _{1}$};
% Text Node
\draw (404.34,107.6) node [anchor=north west][inner sep=0.75pt]  [font=\footnotesize]  {$\xi _{2}$};
% Text Node
\draw (406.51,138.85) node [anchor=north west][inner sep=0.75pt]  [font=\scriptsize]  {$\boldsymbol{e_{2}^{\mathrm{p}}}$};
% Text Node
\draw (443.8,145.56) node [anchor=north west][inner sep=0.75pt]  [font=\scriptsize]  {$\boldsymbol{e_{1}^{\mathrm{p}}}$};
% Text Node
\draw (348,313.81) node [anchor=north west][inner sep=0.75pt]  [font=\footnotesize]  {$\widehat{\xi _{1}}$};
% Text Node
\draw (252.99,241.47) node [anchor=north west][inner sep=0.75pt]  [font=\footnotesize]  {$\widehat{\xi _{2}}$};
% Text Node
\draw (230.49,332.94) node [anchor=north west][inner sep=0.75pt]  [font=\normalsize] [align=left] {{\fontfamily{ptm}\selectfont {\small Principal Parametric Space}}};
% Text Node
\draw (177.59,139.42) node [anchor=north west][inner sep=0.75pt]  [font=\small]  {$\mathcal{P}$};
% Text Node
\draw (304.17,269.66) node [anchor=north west][inner sep=0.75pt]  [font=\small]  {$\widehat{\mathcal{P}}$};
% Text Node
\draw (130.17,275.74) node [anchor=north west][inner sep=0.75pt]  [font=\small]  {$\mathscr{L} :\mathcal{P}\rightarrow \widehat{\mathcal{P}}$};
% Text Node
\draw (412.81,279.16) node [anchor=north west][inner sep=0.75pt]  [font=\small]  {$\mathbf{r}\left(\widehat{\xi _{1}} ,\widehat{\xi _{2}}\right)$};
\end{tikzpicture}
	\caption{Illustration of the mapping relationship between the NURBS parametric space $\CP$ and shell mid-surface in the physical space $\CS$. The one-to-one mapping $\mathscr{L}: \CP\to\wCP$ defines a reparameterization for the representation of shell mid-surface, which bridges the gap between the principal-parameter-based analysis and the NURBS-based computation.\label{Fig_midsurface_nurbs}}
\end{figure}

Note that a new set of surface parameters $\left(\xi_1,\xi_2\right)$ has been introduced with the adoption of the NURBS representation Eq.~\eqref{NURBS_representation}, which generally differ from the coordinates $\left(\wxi_1,\wxi_2\right)$ employed in formulating the equations of the MTS outlined in Secs.~\ref{Sec_problem_settings} and \ref{Sec_asymptotic_analysis}. This is because, mathematically, a set of surface parameters constitutes a mesh coinciding with the principal curvature network of the surface if and only if they satisfy the aforementioned diagonal condition, i.e., Eq.~\eqref{condition_curvature_lines}. Hence, the NURBS parameters $\left(\xi_1,\xi_2\right)$ used for generating the geometry do not necessarily parameterize the ``lines of curvature'' of the surface, as indicated in red and blue in Fig.~\ref{Fig_midsurface_nurbs}, the tangent vectors $\left(\ve_1^\mathrm{p},\ve_2^\mathrm{p}\right)$ associated with the NURBS parameters may not align with those $\left(\ve_1,\ve_2\right)$ associated with the principal parameters.

By choosing the ``lines of curvature'' as parametric curves, the characteristics of various shell types are more straightforward to interpret, and the derivation of stress-strain relations appears to be simpler and more direct, which has already been demonstrated by \cite{reissner1941new}.

Given all these advantages, why bother with reparameterization? There are two main reasons. Firstly, for the mid-surface geometry described by Eq.~\eqref{NURBS_representation}, the calculation of derivatives of the surface with respect to NURBS parameters is straightforward, whereas that based on the principal parameters presents some unwanted complexities. Additionally, when working with basis vectors $\left(\ve_1,\ve_2\right)$, special boundary conditions (e.g., symmetry boundaries, etc.) are not easy to handle, which is not an issue if it is considered in the parameter directions $\left(\ve_1^\mathrm{p},\ve_2^\mathrm{p}\right)$ or more directly, in the global Cartesian coordinate $\left(\tve_1,\tve_2,\tve_3\right)$. As a consequence, the partial derivatives originally taken with respect to the principal parameters $\wxi_\alpha$ should be transformed to those in terms of the NURBS parameters $\xi_\alpha$. Moreover, the displacement components in the local orthogonal curvilinear coordinate system have to be transformed into the global Cartesian coordinate system to facilitate the numerical implementation, such as the imposition of boundary conditions and the avoidance of stress singularities caused by isolated umbilical points.

Let $\mathscr{L}:\ \left(\xi_1,\xi_2\right)\to\left(\wxi_1,\wxi_2\right)$ be a one-to-one regular map, whose expression is given by
\begin{equation}\label{reparameterization}
	\wxi_\alpha = \wxi_\alpha\left(\xi_1,\xi_2\right),
\end{equation}
thus defining a reparameterization for the shell mid-surface: $\vr=\vr\left(\wxi_1\left(\xi_1,\xi_2\right),\wxi_2\left(\xi_1,\xi_2\right)\right)$. The mapping relationship between two sets of tangent vectors of the surface at a particular point $P$, e.g., the principal directions $\vr_{\wxi_\alpha}$ and the parameter directions $\vr_{\xi_\alpha}$, can be obtained by the chain rule, that is,
\begin{equation}\label{mapping_principal_parameter}
	\vr_{\wxi_\beta} = \pd{\xi_\alpha}{\wxi_\beta}\cdot\vr_{\xi_\alpha},
\end{equation}
the Einstein summation convention is adopted here. From differential geometry, the principal direction corresponds to the eigenvector of the \textit{Weingarten map}, and $\pd{\xi_\alpha}{\wxi_\beta}$ actually serve as the combination coefficients for two parameter vectors $\vr_{\xi_\alpha}$.

Once the mapping relation is obtained, the partial derivatives in the original shell governing equations (Eqs.~\eqref{reduced_equilibrium_T1}, \eqref{reduced_equilibrium_T2}, and \eqref{reduced_equilibrium_M}) and the strain-displacement relations (Eq.~\eqref{strain_leading_midsurface}) can be transformed into derivatives with respect to the NURBS parameter coordinates $\left(\xi_1,\xi_2\right)$ through the same coefficients as in Eq.~\eqref{mapping_principal_parameter}, i.e.,
\begin{equation}\label{derivative_chain_rule}
	\left(\bigcdot\right)_{,\,\wxi_\beta} = J_{\alpha\beta}\cdot\left(\bigcdot\right)_{,\,\xi_\alpha},
\end{equation}
for simplicity, the mapping Jacobian $\left[\pd{\xi_\alpha}{\wxi_\beta}\right]$ is denoted by $J$ hereafter.

For the numerical implementation of shell equations, we start this section from the principle of virtual work, which states the equilibrium of internal and external virtual work under any admissible variation, i.e., $\delta W = \delta W_\text{int}-\delta W_\text{ext} = 0$. Mathematically, this is given by
\begin{equation}\label{variational_derivative}
	\delta W\left(\vu,\delta\vu\right) = \left.\frac{\mathrm{d}}{\mathrm{d}t}\right|_{t=0}W\left(\vu+t\,\delta\vu\right) := D_{\delta\vu}W\left(\vu\right) = 0,\quad\delta\vu\in\mathcal{V},
\end{equation}
here the operator $D_{\delta\vu}$ is known as the G$\hat{\text{a}}$teaux derivative. The test function space $\mathcal{V}=\{v\,|\,v\in H^2\left(\Omega\right);\,v=0,\text{ on }\partial_\mathrm{u}\Omega\}$ and $H^2\left(\Omega\right)$ denotes the Sobolev space, where the function elements should be square-integrable up to the second derivatives, i.e., $H^2\left(\Omega\right) = \{v\,|\,D^\alpha v\in L^2\left(\Omega\right),\,|\alpha|\leq2\}$, with the notation $D^\alpha = \frac{\partial^{|\alpha|}}{\partial x_1^{\alpha_1}\cdots\partial x_n^{\alpha_n}},\ |\alpha| = \sum_{i=1}^{n}\alpha_i$.

Since two parametric coordinates are involved here, as shown in Fig.~\ref{Fig_midsurface_nurbs}, the global-to-local transformation should be taken into account when investigating the specific variational formulation. For an MTS, the leading-order internal virtual work is defined by
\begin{equation}\label{internal_virtual_work}
	\begin{split}
		\delta W_\text{int}^0 &= \int_{\Omega}\left[\boldsymbol{\sigma}:\delta\veps+\rho_{\sss\CL}\ddot{\vu}\cdot\delta\vu\right]_{\widetilde{u}_i}\,\intd\Omega\\
		&= \int_{\wCP}\left[\vT:\delta\veps^\ast+\vM:\delta\boldsymbol{k}^\ast+\boldsymbol{\CT}^\mathrm{I}\cdot\delta\vu^\ast+\boldsymbol{\CM}^\mathrm{I}\cdot\delta\vphi^\ast\right]_{u_i^\ast}\left|J_{\Omega\to\left(\wCP\times\wxi_3\right)}\right|\,\intd\wCP\\
		&= \int_\CP\left[\vT:\delta\veps^\ast+\vM:\delta\boldsymbol{k}^\ast+\boldsymbol{\CT}^\mathrm{I}\cdot\delta\vu^\ast+\boldsymbol{\CM}^\mathrm{I}\cdot\delta\vphi^\ast\right]_{\widetilde{u}_i^\ast}\left|J_{\Omega\to\left(\CP\times\xi_3\right)}\right|\,\intd\CP,
	\end{split}
\end{equation}
and the leading-order external virtual work is defined by
\begin{equation}\label{external_virtual_work}
	\begin{split}
		\delta W_\text{ext}^0 &= \int_{\Omega}\left[\vf\cdot\delta\vu\right]_{\widetilde{u}_i}\,\intd\Omega+\int_{\partial_\mathrm{t}\Omega}\left[\vp\cdot\left.\delta\vu\right|_{\partial_\mathrm{t}\Omega}\right]_{\widetilde{u}_i}\,\intd \CS_\mathrm{t}\\
		&= \int_{\wCP}\left[\left\langle\vf\right\rangle_0\cdot\delta\vu^\ast\right]_{u_i^\ast}\left|J_{\Omega\to\left(\wCP\times\wxi_3\right)}\right|\,\intd{\wCP}+\int_{\wCP}\left[\vp\cdot\left(\delta\vu^\ast+\frac{1}{2}\delta\vphi^\ast\right)\right]_{u_i^\ast}\left|J_{\partial_\mathrm{t}\Omega\to\wCP}\right|\,\intd\wCP\\
		&= \int_\CP\left[\left\langle\vf\right\rangle_0\cdot\delta\vu^\ast\right]_{\widetilde{u}_i^\ast}\left|J_{\Omega\to\left(\CP\times\xi_3\right)}\right|\,\intd{\CP}+\int_{\CP}\left[\vp\cdot\left(\delta\vu^\ast+\frac{1}{2}\delta\vphi^\ast\right)\right]_{\widetilde{u}_i^\ast}\left|J_{\partial_\mathrm{t}\Omega\to\CP}\right|\,\intd\CP,
	\end{split}
\end{equation}
where the subscript outside the square bracket indicates the specific variables of the functions contained within the brackets. $\widetilde{u}_i,\,u_i^\ast,\,\text{and }\widetilde{u}_i^\ast$ are identified as the displacement components in the Cartesian coordinate system, along the local principal directions within the mid-surface, and along the global coordinate axes within the mid-surface, respectively. Therefore, we have the following relationship
\begin{equation}\label{displacement_components}
	\vu = \widetilde{u}_i\tve_i;\quad\vu^\ast = u_i^\ast\ve_i = \widetilde{u}_i^\ast\tve_i.
\end{equation}
where $\vu$ and $\vu^\ast$ represent the displacement field and that of the shell mid-surface $\CS$. The third identities of Eqs.~\eqref{internal_virtual_work} and \eqref{external_virtual_work} denote the local-to-global displacement transformation with an intention of avoiding the difficulties and singularities introduced by the isolated umbilical point, which is inherent in the surface geometry such as the ellipsoidal surface.

Some remarks regarding the main variables involved in the above equations are summarized as follows. The in-plane stress resultant tensor $\vT$ and the stress couple tensor $\vM$ can be obtained simply by re-dimensionalizing Eq.~\eqref{stress_resultant_couple}, and the expressions for the corresponding virtual membrane strain $\delta\boldsymbol{\varepsilon}^\ast$ and the changes of curvature $\delta\boldsymbol{k}^\ast$ are obtained by first multiplying Eq.~\eqref{strain_leading_midsurface} by the coefficient $\frac{h\rU}{L^2}$ and then taking the variation. Furthermore, $\boldsymbol{\CT}^\mathrm{I}$ and $\boldsymbol{\CM}^\mathrm{I}$ stand for the thickness-averaged inertial force and moment vectors, respectively, whose non-dimensional expressions are given in Eqs.~\eqref{homo_inertial_force} and \eqref{homo_inertial_moment}. $\delta\vu^\ast$ and $\delta\vphi^\ast$ here denote the (mid-surface) virtual displacement and the virtual rotation angle of the normal, their expressions are written by
\begin{subequations}\label{displacement_rotation_variation}
	\begin{align}
		\delta\vu^\ast &= \left[\delta u_1^\ast,\delta u_2^\ast,\delta u_3^\ast\right]^\rT;\\
		\delta\vphi^\ast &= \left[-\frac{h}{a_1}\pd{\left(\delta u_3^\ast\right)}{\wxi_1},-\frac{h}{a_2}\pd{\left(\delta u_3^\ast\right)}{\wxi_2},0\right]^\rT.
	\end{align}
\end{subequations}
The Jacobian determinant involved in Eq.~\eqref{internal_virtual_work} satisfies: $\left|J_{\Omega\to\left(\CP\times\xi_3\right)}\right| = \left|J_{\Omega\to\left(\wCP\times\wxi_3\right)}\right|\left|J_{\wCP\to\CP}\right|$, which measures the volume change of a volume infinitesimal from the physical space $\Omega$ to the NURBS parametric space $\left(\CP\times\xi_3\right)$, and $\left|J_{\partial_\mathrm{t}\Omega\to\CP}\right| = \left|J_{\partial_\mathrm{t}\Omega\to\wCP}\right|\left|J_{\wCP\to\CP}\right|$ in Eq.~\eqref{external_virtual_work} measures the area ratio between the actual surface $\partial_\mathrm{t}\Omega\subseteq\CS_\mathrm{t}$, on which the surface force $\vt$ is imposed, and its counterpart in the NURBS parametric space $\CP$. The specific expressions for several Jacobian determinants mentioned above will be derived in \ref{Appendix1}.

Up to this point, the weak forms have been presented pertaining to the so-called strong forms of the boundary value problem defined by Eqs.~\eqref{bc_surface}-\eqref{bc_interface_sigma}, \eqref{reduced_equilibrium_T1}-\eqref{reduced_equilibrium_T2}, and \eqref{reduced_equilibrium_M}. However, for the sake of numerical implementation, an appropriate discretization approach should be adopted to further convert the weak form Eq.~\eqref{variational_derivative}, or equivalently Eqs.~\eqref{internal_virtual_work}-\eqref{external_virtual_work}, into a system of linear algebraic equations. In the context of shell problems, the isogeometric discretization is deemed ``appropriate'' due to its capacity to facilitate straightforward calculation of the second derivatives contained in the changes of curvature $\boldsymbol{k}^\ast$, while circumventing the need for additional rotational degrees of freedom.

In this way, the displacement field $\vu^\ast$ of the mid-surface, as well as the virtual displacement field $\delta\vu^\ast$, is discretized by directly applying the NURBS basis functions $R_{i,j}^{p,q}$, which characterize the geometry (Eq.~\eqref{NURBS_representation}), to the shape functions that approximate the solution fields, i.e.,
\begin{subequations}\label{discretized_displacement}
	\begin{align}
		\vu^\ast\left(\xi_1,\xi_2,t\right) &= \sum_{l=1}^{n_\mathrm{cp}^\re}R_l\left(\xi_1,\xi_2\right)\vu_l^\re\left(t\right)\Rightarrow\vu^\ast = \vR\cdot\vu^\re;\\
		\delta\vu^\ast\left(\xi_1,\xi_2,t\right) &= \sum_{l=1}^{n_\mathrm{cp}^\re}R_l\left(\xi_1,\xi_2\right)\delta\vu_l^\re\left(t\right)\Rightarrow\delta\vu^\ast = \vR\cdot\delta\vu^\re,
	\end{align}
\end{subequations}
where $n_\mathrm{cp}^\re$ denotes the total number of control points in an element $\CS^\re$, and $\vu_l^\re$ is the element nodal displacement vector at the $l$-th control point, with its components corresponding to the global Cartesian coordinates.

Based on the discretized nodal displacement vector provided by Eq.~\eqref{discretized_displacement}, the integrands of the internal and external virtual work can be re-expressed in matrix forms, and notably, both strain and acceleration terms within an element, along with their variations, can be related to displacements of the shell mid-surface $\vu^\ast$, which are further expressed by the corresponding nodal displacements $\vu_l^\re\,\left(l=1,\,\cdots,n_\mathrm{cp}^\re\right)$ of the element control points. First, by restoring the dimensions of Eqs.~\eqref{homo_inertial_force} and \eqref{homo_inertial_moment}, we can rewrite the inertial force and moment vectors into a combined matrix form, that is,
\begin{equation}\label{inertial_force_moment_matrix}
	\begin{bmatrix}
		\vCT^\mathrm{I}\\\vCM^\mathrm{I}\\
	\end{bmatrix} = \begin{bmatrix}
		\vP_1 & \vP_2\\
		\vP_2 & \vP_3\\
	\end{bmatrix}\cdot\begin{bmatrix}
		\ddot{\vu}^\ast\\\ddot{\vphi}^\ast\\
	\end{bmatrix} = \vP^\re\cdot\left[\partial_1\right]\vQ\vR\ddot{\vu}^\re = \vP^\re\cdot\vN\ddot{\vu}^\re,
\end{equation}
here $\vP^\re$ represents the element density matrix consisting of three introduced diagonal matrices $\vP_1$, $\vP_2$ and $\vP_3$, which are specifically given by $\vP_i=diag\left(\left\langle\rho_{\sss\mathcal{L}}\right\rangle_{i-1},\left\langle\rho_{\sss\mathcal{L}}\right\rangle_{i-1},\left\langle\rho_{\sss\mathcal{L}}\right\rangle_{i-1}\right),\ i=1,2,3$. Based on Eq.~\eqref{displacement_rotation_variation}, the operator matrix $\left[\partial_1\right]$ takes the form:
\begin{equation}\label{opertor_matrix2}
	\setlength{\arraycolsep}{2.2pt}
	\renewcommand{\arraystretch}{1}
	\left[\partial_1\right] = \begin{bmatrix}
		1 & 0 & 0 & 0 & 0 & 0\\ 0 & 1 & 0 & 0 & 0 & 0\\ 0 & 0 & 1 & -\frac{h}{a_1}\frac{\partial}{\partial\wxi_1} & -\frac{h}{a_2}\frac{\partial}{\partial\wxi_2} & 0\\
	\end{bmatrix}^\rT.
\end{equation}

From Eq.~\eqref{inertial_force_moment_matrix}, we defined an augmented shape function matrix $\vN = \left[\partial_1\right]\vQ\vR$, which links the element nodal displacement vector $\wvu^\re$ to the displacement-rotation vector $\left[\left(\vu^\ast\right)^\rT,\left(\vphi^\ast\right)^\rT\right]^\rT$, or more precisely, their second derivatives with respect to time $t$. Second, incorporating the stress resultants and couples, we obtain:
\begin{equation}
	\begin{bmatrix}
		\vT\\\vM\\
	\end{bmatrix} = \begin{bmatrix}
		\vCA & \vCB\\
		\vCB & \vCD\\
	\end{bmatrix}\cdot\begin{bmatrix}
		\veps^\ast\\\boldsymbol{k}^\ast\\
	\end{bmatrix} = \vD^\re\cdot\left[\partial_2\right]\vQ\vR\vu^\re = \vD^\re\cdot\vB\vu^\re,
\end{equation}
where $\vD^\re$ is the in-plane elasticity matrix composed of the extensional, coupling, and bending stiffness $\vCA$, $\vCB$ and $\vCD$, whose components are provided in Eq.~\eqref{stiffness_ABD_entries}. Matrix $\vB = \left[\partial_2\right]\vQ\vR$ links the mid-surface strain to the global nodal displacement vector $\wvu^\rn$ and is referred to as the strain-displacement matrix. Using Eq.~\eqref{strain_leading_midsurface}, the operator matrix $\left[\partial_2\right]$ is expressed by
\begin{equation}\label{opertor_matrix1}
	\setlength{\arraycolsep}{2.5pt}
	\renewcommand{\arraystretch}{1.35}
	\left[\partial_2\right] = \begin{bmatrix}
		\frac{1}{a_1}\frac{\partial}{\partial\wxi_1} & \frac{1}{a_1a_2}\frac{\partial a_1}{\partial\wxi_2} & -k_1\\ \frac{1}{a_1a_2}\frac{\partial a_2}{\partial\wxi_1} & \frac{1}{a_2}\frac{\partial}{\partial\wxi_2} & -k_2\\ \frac{1}{a_2}\frac{\partial}{\partial\wxi_2}-\frac{1}{a_1a_2}\frac{\partial a_1}{\partial\wxi_2} & \frac{1}{a_1}\frac{\partial}{\partial\wxi_1}-\frac{1}{a_1a_2}\frac{\partial a_2}{\partial\wxi_1} & 0\\ 0 & 0 & -\frac{h}{a_1}\frac{\partial}{\partial\wxi_1}\left(\frac{1}{a_1}\frac{\partial}{\partial\wxi_1}\right)-\frac{h}{a_1a_2^2}\frac{\partial a_1}{\partial\wxi_2}\frac{\partial}{\partial\wxi_2}\\ 0 & 0 & -\frac{h}{a_2}\frac{\partial}{\partial\wxi_2}\left(\frac{1}{a_2}\frac{\partial}{\partial\wxi_2}\right)-\frac{h}{a_1^2a_2}\frac{\partial a_2}{\partial\wxi_1}\frac{\partial}{\partial\wxi_1}\\ 0 & 0 & -2h\left(\frac{1}{a_1a_2}\frac{\partial^2}{\partial\wxi_1\wxi_2}-\frac{1}{\wa_1\wa_2^2}\frac{\partial a_2}{\partial\wxi_1}\frac{\partial}{\partial\wxi_2}-\frac{1}{\wa_1^2\wa_2}\frac{\partial a_1}{\partial\wxi_2}\frac{\partial}{\partial\wxi_1}\right)\\
	\end{bmatrix},
\end{equation}
it is worth mentioning that the partial derivatives taken w.r.t. the principal coordinates $\wxi_\alpha$ must be reconsidered according to Eq.~\eqref{derivative_chain_rule}. The corresponding local-to-global displacement transformation matrix $\vQ$ is derived from the relationship $u_i^\ast = \widetilde{u}_j^\ast \tve_j\cdot\ve_i$. Thus, the orthogonal matrix takes the form
\begin{equation}\label{local2global_matrix}
	\vQ = \begin{bmatrix}
		\ve_1 & \ve_2 & \ve_3
	\end{bmatrix}^\rT.
\end{equation}

Finally, we derive the element body and surface force vectors $\vf^\re$ and $\vp^\re$. From Eq.~\eqref{external_virtual_work}, the external work done by the body force is simplified to a surface integral defined on the shell mid-surface $\CS$ after homogenization along the thickness direction. In this way, the element body force vector reads
\begin{equation}\label{element_body_force_vector}
	\vf^\re = \left[\left\langle f_1\right\rangle_0,\left\langle f_2\right\rangle_0,\left\langle f_3\right\rangle_0\right]^\rT.
\end{equation}
Combining Eqs.~\eqref{external_virtual_work} and \eqref{displacement_rotation_variation}, the element surface force vector $\vp^\re$ takes the following form
\begin{equation}\label{element_surface_force_vector}
	\vp^\re = \left[p_1,p_2,p_3-\frac{h}{2}\left(\frac{p_1}{a_1}\pd{}{\wxi_1}-\frac{p_2}{a_2}\pd{}{\wxi_2}\right)\right]^\rT.
\end{equation}

With all revelent variables now discretized and presented in matrices, we can finally show the linear algebraic equations through the discretized form of the principle of virtual work. Generally, such equations can be expressed either in the well-known global form as $\textbf{M}\ddot{\vu}^\rn + \vK\vu^\rn = \vF$ or in the discretized form of elements as:
\begin{equation}\label{linear_algebraic_assembly}
	\sum_{e=1}^\text{nel}\left(\textbf{M}^\re\ddot{\vu}^\re+\vK^\re\vu^\re = \vF^\re\right),
\end{equation}
where ``$\text{nel}$'' denotes the total number of elements $\CS^\re$, and $\vu^\re$ is the nodal displacement vector of all control points in an element. The summation operation refers to the assembly process of elements. Therefore, the corresponding global matrices and vectors are constructed as $\textbf{M} = \sum_{e=1}^\text{nel}\left(\textbf{M}^\re\right)$, $\vK = \sum_{e=1}^\text{nel}\left(\vK^\re\right)$, $\vF = \sum_{e=1}^\text{nel}\left(\vF^\re\right)$ and $\vu^\rn = \sum_{e=1}^\text{nel}\left(\vu^\re\right)$. The specific expressions for the element mass and stiffness matrix, as well as the nodal force vector in Eq.~\eqref{linear_algebraic_assembly}, are derived from the matrices defined above. They are
\begin{subequations}\label{mass_stiffness_force}
	\begin{align}
		\textbf{M}^\re &= \int_{\CP^\re}\vN^\rT\vP^\re\vN\left|J_{\Omega\to\left(\CP\times\xi_3\right)}\right|\,\intd\CP;\label{mass_matrix_def}\\
		\vK^\re &= \int_{\CP^\re}\vB^\rT\vD^\re\vB\left|J_{\Omega\to\left(\CP\times\xi_3\right)}\right|\,\intd\CP;\label{stiffness_matrix_def}\\
		\vF^\re &= \int_{\CP^\re}\left[\vR^\rT\vQ^\rT\vf^\re\left|J_{\Omega\to\left(\CP\times\xi_3\right)}\right|+\vR^\rT\vQ^\rT\vp^\re\left|J_{\partial_\mathrm{t}\Omega\to\CP}\right|\right]\,\intd\CP.\label{force_vector_def}
	\end{align}
\end{subequations}

\subsection{Remarks on Shear Loads and Transverse Shear Effects}\label{Sec_remarks_on_shear}
We begin this subsection by discussing the role of shear loads. In cases of moderate curvature (i.e., when $\max\left(\left[\wk_\alpha\right]\right)\sim\mO\left(\epsilon^d\right)$, with $d\geq0$), Eqs.~\eqref{reduced_equilibrium_T1}, \eqref{reduced_equilibrium_T2}, and \eqref{reduced_equilibrium_M} indicate that shear loads applied to the shell surface directly influence the in-plane dominant stress components, inducing their redistribution to balance the shear forces. In such scenarios, it is essential to accurately compute the in-plane stress components, while the higher-order transverse components (relative to in-plane components) can be obtained by integrating the equilibrium equations. However, for cases with larger principal curvatures (as discussed in Sec.~\ref{Sec_significantlycurved_shell}), we find that even after homogenizing Eq.~\eqref{equilibrium_significant}, non-eliminable transverse shear stresses remain, playing a role in balancing the applied shear forces. In this context, the deformation pattern becomes more complex, and the surface shear forces may induce bending moments. The maximum admissible magnitude of external surface shear loads that a shell can sustain is governed by the parameter $m_2$, as determined in Secs.~\ref{Sec_weaklycurved_shell}-\ref{Sec_significantlycurved_shell}.

Next, we examine the transverse shear effects within the shell. From the non-dimensional constitutive relations (Eq.~\eqref{constitutive_dimensionless4}-\eqref{constitutive_dimensionless5}), it is evident that the transverse shear strains $\gamma_{\alpha3}^{(0)}$ vanish asymptotically in the leading-order model, and thus transverse shear effects need not be considered at this level. However, as we progress to higher orders, the transverse shear strain components become nonzero and coupled with the corresponding (higher-order) in-plane displacements. In this way, shear effects naturally enter the solution of higher-order equations. Therefore, unlike CST that enforce $\gamma_{\alpha3}=0$ a priori, the asymptotic framework permits a hierarchical progression in which $\gamma_{\alpha3}$ is zero only in an asymptotic sense. It should be emphasized that the conclusions drawn earlier are based on an implicit premise: the elastic moduli of the individual layers comprising the shell are of comparable magnitude ($\wE_{\sss\CL}\sim\mO\left(1\right)$). Under this condition, transverse shear effects only appear in the equations at higher orders. However, when considering soft-hard laminated structures, where the elastic modulus of the intermediate layer is significantly lower than that of the adjacent layers, the conclusions differ.

Consider a weakly curved, three-layered thin shell in which the elastic modulus of the middle layer is two orders of magnitude higher than that of the adjacent layers, i.e., $\wE_2\sim\mO\left(\epsilon^2\cdot\wE_{1\,\text{or}\,3}\right)$. To ensure an $\mO\left(1\right)$ elastic modulus in the middle layer, $\wE_2$ should be rescaled, that is,
\begin{equation}\label{soft_modulus_rescaling}
	\wE_2^\prime=\frac{1}{\epsilon^2}\cdot\wE_2\sim\mO\left(1\right).
\end{equation}
According to Eq.~\eqref{soft_modulus_rescaling}, the stress scaling relations within the middle layer are expected to be $\sigma_{\alpha\beta}=\epsilon^2\sigma^\star\wsigma_{\alpha\beta},\ \sigma_{\alpha3}=\epsilon^3\sigma^\star\wsigma_{\alpha3},\ \sigma_{33}=\epsilon^4\sigma^\star\wsigma_{33}$. However, the transverse shear stresses in the top and bottom layers start at $\mO\left(\epsilon^3\right)$ (as shown in Eq.~\eqref{stress_scaling}), which is significantly larger than the $\mO\left(\epsilon^5\right)$ in the middle layer. Therefore, to maintain equilibrium across the material interfaces, $\sigma_{\alpha3}$ in the softer middle layer must be of the same order of magnitude as those in the adjacent stiffer layers. The (significant) moduli difference inherently reshapes the scaling relations of the stress components within the soft layer, which now become: $\sigma_{\alpha\beta}\sim\mO\left(\epsilon^4\right),\ \sigma_{\alpha3}\sim\mO\left(\epsilon^3\right),\ \sigma_{33}\sim\mO\left(\epsilon^4\right)$. This revised scaling rule supports the commonly held understanding that soft materials are more prone to bear shear deformation in response to external loads, thus resulting in much larger transverse shear effects. Morewover, by combining Eqs.~\eqref{constitutive_dimensionless} and \eqref{soft_modulus_rescaling}, we find that the transverse shear stresses appear in the expressions for the in-plane displacements at the leading order, i.e.,
\begin{equation}\label{soft_in_plane_disp}
	\wu_\alpha^{(0)}=\wu_\alpha^{\mathrm{mid}\ast}\left(\wxi_1,\wxi_2\right)+\left(-\frac{1}{\wa_\alpha}\pd{\wu_3^{(0)}}{\wxi_\alpha}+\frac{2\left(1+\nu_2\right)}{\wE_2^\prime}\wsigma_{\alpha3}^{(0)}\right)\cdot\wxi_3,
\end{equation}
where the superscript ``$\mathrm{mid}$'' denotes quantities associated with the middle layer. Eq.~\eqref{soft_in_plane_disp} clearly captures the significant transverse shear effects in soft materials, which is in stark contrast to the model discussed earlier in this paper.

It is also worth noting that the influence of transverse normal strain emerges at higher orders. CST often assume inextensibility of the normal; however, according to the 3D constitutive law, $\pd{\wu_3}{\wxi_3}\sim\nu\left(\varepsilon_{11}+\varepsilon_{22}\right)$, leading to an inherent contradiction known as Poisson locking. As shown in Eq.~\eqref{constitutive_dimensionless6}, when the transverse normal strain is accounted for at $\mO\left(\epsilon^2\right)$, we have
\begin{equation}
	\pd{\wu_3^{(2)}}{\wxi_3}=\frac{-\nu_{\sss\CL}}{\wa_1(1-\nu_{\sss\CL})}\left(\pd{\wu_1^{(0)}}{\wxi_1}+\frac{\wu_2^{(0)}}{\wa_2}\pd{\wa_1}{\wxi_2}-\frac{\wk_1}{\epsilon}\wu_3^{(0)}\wa_1\right)+\frac{-\nu_{\sss\CL}}{\wa_2(1-\nu_{\sss\CL})}\left(\pd{\wu_2^{(0)}}{\wxi_2}+\frac{\wu_1^{(0)}}{\wa_1}\pd{\wa_2}{\wxi_1}-\frac{\wk_2}{\epsilon}\wu_3^{(0)}\wa_2\right).
\end{equation}
The above equation naturally satisfies the required consistency, and in subsequent analyses, $\pd{\wu_3^{(2)}}{\wxi_3}$ can be eliminated using this expression, rather than being artificially set to 0. This ensures that Poisson locking is inherently avoided at each order of the asymptotic expansion.

\section{Numerical Validations}\label{Sec_numerical_example}
Based on the previous theoretical derivation, several numerical examples will be covered in this section to demonstrate the reliability of the proposed isogeometric-based asymptotic framework. The numerical implementation of the asymptotic expressions for the MTS, derived in Sec.~\ref{Sec_asymptotic_analysis}, is performed on the MATLAB\textsuperscript{\textregistered} R2022b platform with several related IGA functions provided by \cite{du2020nliga}. The results are further validated by checking their deviations from those computed by direct numerical simulation based on the \cite{COMSOL}. In this particular section, we first introduce a modified shell obstacle course consisting of (1) Pinched cylinder with end diaphragms (Fig.~\ref{shell_obstacle}(a)), (2) Scordelis-Lo roof (Fig.~\ref{shell_obstacle}(b)), and (3) Pinched sphere with a $18^\circ$ hole (Fig.~\ref{shell_obstacle}(c)) to evaluate the performance of the asymptotic formulations for multi-layered shells. We then focus on the transverse shear stresses in normally curved shells and the role of applied shear forces, providing displacement and stress fields under shear deformations, without considering the SCF or kinematic assumptions. Moreover, free vibration problems are also included here to assess the current framework in the context of dynamic behavior.

\subsection{Shell Obstacle Course}
\begin{figure}[!htbp]
	\setlength{\abovecaptionskip}{0.cm}
	\setlength{\belowcaptionskip}{-0.cm}
	\centering
	\tikzset{every picture/.style={line width=0.75pt}} %set default line width to 0.75pt
	\begin{tikzpicture}[x=0.75pt,y=0.75pt,yscale=-1,xscale=1]
		%uncomment if require: \path (0,310); %set diagram left start at 0, and has height of 310
		%Image [id:dp8709915181364853] 
		\draw (134.97,158.91) node  {\includegraphics[width=159.98pt,height=120pt]{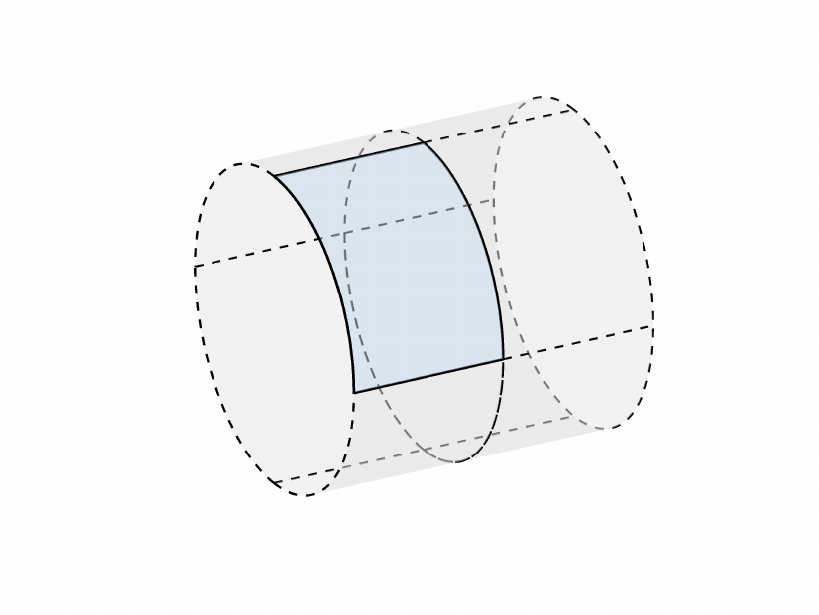}};
		%Straight Lines [id:da23271697343454378] 
		\draw    (99.75,165.91) -- (129.88,189.17) ;
		\draw [shift={(131.46,190.39)}, rotate = 217.66] [fill={rgb, 255:red, 0; green, 0; blue, 0 }  ][line width=0.08]  [draw opacity=0] (8.4,-2.1) -- (0,0) -- (8.4,2.1) -- cycle    ;
		%Straight Lines [id:da9686359253590711] 
		\draw    (99.75,165.91) -- (98.92,99.53) ;
		\draw [shift={(98.89,97.53)}, rotate = 89.28] [fill={rgb, 255:red, 0; green, 0; blue, 0 }  ][line width=0.08]  [draw opacity=0] (8.4,-2.1) -- (0,0) -- (8.4,2.1) -- cycle    ;
		%Straight Lines [id:da3620066124854153] 
		\draw [color={rgb, 255:red, 155; green, 155; blue, 155 }  ,draw opacity=1 ]   (99.75,165.91) -- (213.51,141.1) ;
		\draw [shift={(215.46,140.68)}, rotate = 167.7] [fill={rgb, 255:red, 155; green, 155; blue, 155 }  ,fill opacity=1 ][line width=0.08]  [draw opacity=0] (8.4,-2.1) -- (0,0) -- (8.4,2.1) -- cycle    ;
		%Straight Lines [id:da11043807448197507] 
		\draw    (99.75,165.91) -- (84.79,129.38) ;
		\draw [shift={(84.03,127.53)}, rotate = 67.73] [fill={rgb, 255:red, 0; green, 0; blue, 0 }  ][line width=0.08]  [draw opacity=0] (7.2,-1.8) -- (0,0) -- (7.2,1.8) -- cycle    ;
		%Straight Lines [id:da4798665317675008] 
		\draw [color={rgb, 255:red, 74; green, 74; blue, 74 }  ,draw opacity=1 ]   (108.03,213.25) -- (190.32,194.68) ;
		\draw [shift={(190.32,194.68)}, rotate = 167.28] [color={rgb, 255:red, 74; green, 74; blue, 74 }  ,draw opacity=1 ][line width=0.75]    (0,3.35) -- (0,-3.35)(6.56,-1.97) .. controls (4.17,-0.84) and (1.99,-0.18) .. (0,0) .. controls (1.99,0.18) and (4.17,0.84) .. (6.56,1.97)   ;
		\draw [shift={(108.03,213.25)}, rotate = 347.28] [color={rgb, 255:red, 74; green, 74; blue, 74 }  ,draw opacity=1 ][line width=0.75]    (0,3.35) -- (0,-3.35)(6.56,-1.97) .. controls (4.17,-0.84) and (1.99,-0.18) .. (0,0) .. controls (1.99,0.18) and (4.17,0.84) .. (6.56,1.97)   ;
		%Straight Lines [id:da2688431899586592] 
		\draw [color={rgb, 255:red, 74; green, 74; blue, 74 }  ,draw opacity=1 ]   (123.41,175.12) -- (156.94,167.41) ;
		\draw [shift={(158.89,166.96)}, rotate = 167.05] [fill={rgb, 255:red, 74; green, 74; blue, 74 }  ,fill opacity=1 ][line width=0.08]  [draw opacity=0] (8.4,-2.1) -- (0,0) -- (8.4,2.1) -- cycle    ;
		\draw [shift={(121.46,175.57)}, rotate = 347.05] [fill={rgb, 255:red, 74; green, 74; blue, 74 }  ,fill opacity=1 ][line width=0.08]  [draw opacity=0] (8.4,-2.1) -- (0,0) -- (8.4,2.1) -- cycle    ;
		%Shape: Circle [id:dp5844444835462954] 
		\draw  [fill={rgb, 255:red, 0; green, 0; blue, 0 }  ,fill opacity=1 ] (138.2,116.3) .. controls (138.2,115.51) and (138.85,114.86) .. (139.64,114.87) .. controls (140.43,114.87) and (141.08,115.51) .. (141.07,116.31) .. controls (141.07,117.1) and (140.43,117.74) .. (139.63,117.74) .. controls (138.84,117.74) and (138.2,117.09) .. (138.2,116.3) -- cycle ;
		%Straight Lines [id:da2597717734984193] 
		\draw [color={rgb, 255:red, 208; green, 2; blue, 27 }  ,draw opacity=1 ]   (140.07,96.23) -- (139.7,113.3) ;
		\draw [shift={(139.64,116.3)}, rotate = 271.23] [fill={rgb, 255:red, 208; green, 2; blue, 27 }  ,fill opacity=1 ][line width=0.08]  [draw opacity=0] (5.36,-2.57) -- (0,0) -- (5.36,2.57) -- cycle    ;
		%Straight Lines [id:da8129274877759363] 
		\draw [color={rgb, 255:red, 208; green, 2; blue, 27 }  ,draw opacity=0.6 ]   (138.57,215.38) -- (138.57,198.98) ;
		\draw [shift={(138.57,195.98)}, rotate = 89.99] [fill={rgb, 255:red, 208; green, 2; blue, 27 }  ,fill opacity=0.6 ][line width=0.08]  [draw opacity=0] (5.36,-2.57) -- (0,0) -- (5.36,2.57) -- cycle    ;
		%Straight Lines [id:da1300097349239535] 
		\draw    (75.58,187.97) -- (81.1,171.34) ;
		\draw [shift={(81.73,169.44)}, rotate = 108.37] [fill={rgb, 255:red, 0; green, 0; blue, 0 }  ][line width=0.08]  [draw opacity=0] (4.8,-1.2) -- (0,0) -- (4.8,1.2) -- cycle    ;
		%Straight Lines [id:da08304225098694928] 
		\draw [color={rgb, 255:red, 155; green, 155; blue, 155 }  ,draw opacity=1 ]   (176.11,121.62) -- (179.66,111.09) ;
		\draw [shift={(180.3,109.2)}, rotate = 108.66] [fill={rgb, 255:red, 155; green, 155; blue, 155 }  ,fill opacity=1 ][line width=0.08]  [draw opacity=0] (4.8,-1.2) -- (0,0) -- (4.8,1.2) -- cycle    ;
		%Image [id:dp11982717115998986] 
		\draw (355.75,135.73) node  {\includegraphics[width=159.98pt,height=120pt]{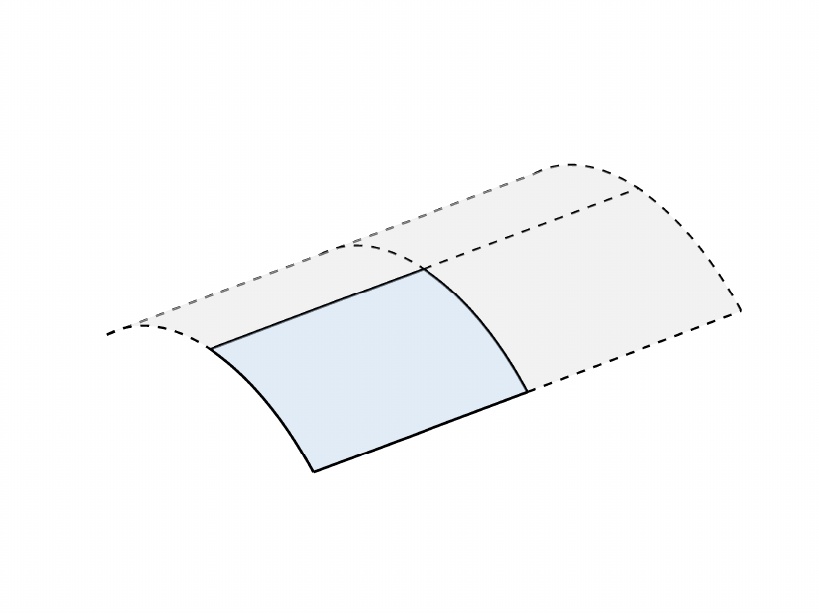}};
		%Straight Lines [id:da08683241802666197] 
		\draw    (303.77,161.34) -- (363.1,200.68) ;
		\draw [shift={(364.77,201.79)}, rotate = 213.55] [fill={rgb, 255:red, 0; green, 0; blue, 0 }  ][line width=0.08]  [draw opacity=0] (7.2,-1.8) -- (0,0) -- (7.2,1.8) -- cycle    ;
		%Straight Lines [id:da14823772636865207] 
		\draw [color={rgb, 255:red, 155; green, 155; blue, 155 }  ,draw opacity=1 ]   (303.77,161.34) -- (439.32,108.91) ;
		\draw [shift={(441.18,108.19)}, rotate = 158.85] [fill={rgb, 255:red, 155; green, 155; blue, 155 }  ,fill opacity=1 ][line width=0.08]  [draw opacity=0] (7.2,-1.8) -- (0,0) -- (7.2,1.8) -- cycle    ;
		%Straight Lines [id:da8784752875427995] 
		\draw    (304.27,168.84) -- (303.95,117.84) ;
		\draw [shift={(303.93,115.84)}, rotate = 89.64] [fill={rgb, 255:red, 0; green, 0; blue, 0 }  ][line width=0.08]  [draw opacity=0] (7.2,-1.8) -- (0,0) -- (7.2,1.8) -- cycle    ;
		%Straight Lines [id:da8178902959626368] 
		\draw  [dash pattern={on 4.5pt off 4.5pt}]  (304.22,162.74) -- (303.88,222.07) ;
		%Straight Lines [id:da4913131353753917] 
		\draw    (303.88,222.07) -- (324.44,170.88) ;
		\draw [shift={(325.18,169.02)}, rotate = 111.88] [fill={rgb, 255:red, 0; green, 0; blue, 0 }  ][line width=0.08]  [draw opacity=0] (4.8,-1.2) -- (0,0) -- (4.8,1.2) -- cycle    ;
		%Straight Lines [id:da132765716573501] 
		\draw [color={rgb, 255:red, 0; green, 0; blue, 0 }  ,draw opacity=1 ]   (333.27,184.21) -- (388.68,162.94) ;
		\draw [shift={(388.68,162.94)}, rotate = 159.01] [color={rgb, 255:red, 0; green, 0; blue, 0 }  ,draw opacity=1 ][line width=0.75]    (0,3.35) -- (0,-3.35)(6.56,-1.97) .. controls (4.17,-0.84) and (1.99,-0.18) .. (0,0) .. controls (1.99,0.18) and (4.17,0.84) .. (6.56,1.97)   ;
		\draw [shift={(333.27,184.21)}, rotate = 339.01] [color={rgb, 255:red, 0; green, 0; blue, 0 }  ,draw opacity=1 ][line width=0.75]    (0,3.35) -- (0,-3.35)(6.56,-1.97) .. controls (4.17,-0.84) and (1.99,-0.18) .. (0,0) .. controls (1.99,0.18) and (4.17,0.84) .. (6.56,1.97)   ;
		%Straight Lines [id:da5449282935480271] 
		\draw [color={rgb, 255:red, 0; green, 0; blue, 0 }  ,draw opacity=1 ]   (388.68,162.94) -- (444.43,141.44) ;
		\draw [shift={(444.43,141.44)}, rotate = 158.91] [color={rgb, 255:red, 0; green, 0; blue, 0 }  ,draw opacity=1 ][line width=0.75]    (0,3.35) -- (0,-3.35)(6.56,-1.97) .. controls (4.17,-0.84) and (1.99,-0.18) .. (0,0) .. controls (1.99,0.18) and (4.17,0.84) .. (6.56,1.97)   ;
		\draw [shift={(388.68,162.94)}, rotate = 338.91] [color={rgb, 255:red, 0; green, 0; blue, 0 }  ,draw opacity=1 ][line width=0.75]    (0,3.35) -- (0,-3.35)(6.56,-1.97) .. controls (4.17,-0.84) and (1.99,-0.18) .. (0,0) .. controls (1.99,0.18) and (4.17,0.84) .. (6.56,1.97)   ;
		%Shape: Circle [id:dp009203393084980638] 
		\draw  [fill={rgb, 255:red, 0; green, 0; blue, 0 }  ,fill opacity=1 ] (385.35,157.74) .. controls (385.35,156.95) and (386,156.31) .. (386.79,156.31) .. controls (387.59,156.31) and (388.23,156.95) .. (388.23,157.75) .. controls (388.23,158.54) and (387.58,159.18) .. (386.79,159.18) .. controls (385.99,159.18) and (385.35,158.54) .. (385.35,157.74) -- cycle ;
		%Straight Lines [id:da8141483486234522] 
		\draw  [dash pattern={on 4.5pt off 4.5pt}]  (277.43,143.52) -- (303.88,222.07) ;
		%Curve Lines [id:da3778879281781802] 
		\draw [color={rgb, 255:red, 155; green, 155; blue, 155 }  ,draw opacity=1 ]   (292.87,180.7) .. controls (295.84,178.11) and (297.67,177.78) .. (301.22,179.11) ;
		\draw [shift={(303.93,180.27)}, rotate = 204.78] [fill={rgb, 255:red, 155; green, 155; blue, 155 }  ,fill opacity=1 ][line width=0.08]  [draw opacity=0] (3.57,-1.72) -- (0,0) -- (3.57,1.72) -- cycle    ;
		\draw [shift={(290.66,182.8)}, rotate = 315.04] [fill={rgb, 255:red, 155; green, 155; blue, 155 }  ,fill opacity=1 ][line width=0.08]  [draw opacity=0] (3.57,-1.72) -- (0,0) -- (3.57,1.72) -- cycle    ;
		%Straight Lines [id:da04288252916607016] 
		\draw [color={rgb, 255:red, 208; green, 2; blue, 27 }  ,draw opacity=1 ]   (293.7,115.8) -- (293.7,132.8) ;
		\draw [shift={(293.7,135.8)}, rotate = 270] [fill={rgb, 255:red, 208; green, 2; blue, 27 }  ,fill opacity=1 ][line width=0.08]  [draw opacity=0] (5.36,-2.57) -- (0,0) -- (5.36,2.57) -- cycle    ;
		%Straight Lines [id:da8350586590846112] 
		\draw [color={rgb, 255:red, 208; green, 2; blue, 27 }  ,draw opacity=1 ]   (359.79,105.83) -- (359.79,122.83) ;
		\draw [shift={(359.79,125.83)}, rotate = 270] [fill={rgb, 255:red, 208; green, 2; blue, 27 }  ,fill opacity=1 ][line width=0.08]  [draw opacity=0] (5.36,-2.57) -- (0,0) -- (5.36,2.57) -- cycle    ;
		%Straight Lines [id:da21708642350614893] 
		\draw [color={rgb, 255:red, 208; green, 2; blue, 27 }  ,draw opacity=1 ]   (336.78,100.14) -- (336.78,117.14) ;
		\draw [shift={(336.78,120.14)}, rotate = 270] [fill={rgb, 255:red, 208; green, 2; blue, 27 }  ,fill opacity=1 ][line width=0.08]  [draw opacity=0] (5.36,-2.57) -- (0,0) -- (5.36,2.57) -- cycle    ;
		%Straight Lines [id:da6139633057034461] 
		\draw [color={rgb, 255:red, 208; green, 2; blue, 27 }  ,draw opacity=1 ]   (386.79,137.75) -- (386.79,154.75) ;
		\draw [shift={(386.79,157.75)}, rotate = 270] [fill={rgb, 255:red, 208; green, 2; blue, 27 }  ,fill opacity=1 ][line width=0.08]  [draw opacity=0] (5.36,-2.57) -- (0,0) -- (5.36,2.57) -- cycle    ;
		%Straight Lines [id:da8151612850205554] 
		\draw [color={rgb, 255:red, 208; green, 2; blue, 27 }  ,draw opacity=1 ]   (317.53,121.97) -- (317.53,138.97) ;
		\draw [shift={(317.53,141.97)}, rotate = 270] [fill={rgb, 255:red, 208; green, 2; blue, 27 }  ,fill opacity=1 ][line width=0.08]  [draw opacity=0] (5.36,-2.57) -- (0,0) -- (5.36,2.57) -- cycle    ;
		%Straight Lines [id:da655500759509047] 
		\draw [color={rgb, 255:red, 208; green, 2; blue, 27 }  ,draw opacity=1 ]   (345.03,153.3) -- (345.03,170.3) ;
		\draw [shift={(345.03,173.3)}, rotate = 270] [fill={rgb, 255:red, 208; green, 2; blue, 27 }  ,fill opacity=1 ][line width=0.08]  [draw opacity=0] (5.36,-2.57) -- (0,0) -- (5.36,2.57) -- cycle    ;
		%Straight Lines [id:da5656154042455728] 
		\draw [color={rgb, 255:red, 208; green, 2; blue, 27 }  ,draw opacity=1 ]   (376.86,84.72) -- (376.86,101.72) ;
		\draw [shift={(376.86,104.72)}, rotate = 270] [fill={rgb, 255:red, 208; green, 2; blue, 27 }  ,fill opacity=1 ][line width=0.08]  [draw opacity=0] (5.36,-2.57) -- (0,0) -- (5.36,2.57) -- cycle    ;
		%Straight Lines [id:da6603238126411846] 
		\draw [color={rgb, 255:red, 208; green, 2; blue, 27 }  ,draw opacity=1 ]   (399.28,90.39) -- (399.28,107.39) ;
		\draw [shift={(399.28,110.39)}, rotate = 270] [fill={rgb, 255:red, 208; green, 2; blue, 27 }  ,fill opacity=1 ][line width=0.08]  [draw opacity=0] (5.36,-2.57) -- (0,0) -- (5.36,2.57) -- cycle    ;
		%Straight Lines [id:da12856186609066378] 
		\draw [color={rgb, 255:red, 208; green, 2; blue, 27 }  ,draw opacity=1 ]   (426.45,122.3) -- (426.45,139.3) ;
		\draw [shift={(426.45,142.3)}, rotate = 270] [fill={rgb, 255:red, 208; green, 2; blue, 27 }  ,fill opacity=1 ][line width=0.08]  [draw opacity=0] (5.36,-2.57) -- (0,0) -- (5.36,2.57) -- cycle    ;
		%Straight Lines [id:da15580447670274578] 
		\draw    (276.27,168.26) -- (297.23,145.83) ;
		\draw [shift={(298.6,144.37)}, rotate = 133.07] [fill={rgb, 255:red, 0; green, 0; blue, 0 }  ][line width=0.08]  [draw opacity=0] (4.8,-1.2) -- (0,0) -- (4.8,1.2) -- cycle    ;
		%Straight Lines [id:da7365760484747139] 
		\draw    (419.67,95.62) -- (422.82,109.34) ;
		\draw [shift={(423.27,111.29)}, rotate = 257.06] [fill={rgb, 255:red, 0; green, 0; blue, 0 }  ][line width=0.08]  [draw opacity=0] (4.8,-1.2) -- (0,0) -- (4.8,1.2) -- cycle    ;
		%Image [id:dp5159123089987041] 
		\draw (572.35,142.66) node  {\includegraphics[width=159.97pt,height=120pt]{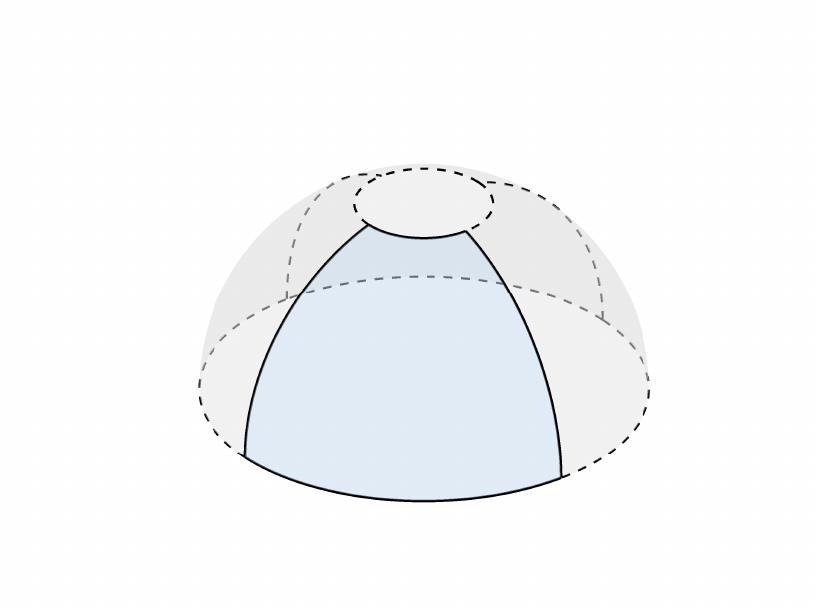}};
		%Straight Lines [id:da835293999565955] 
		\draw [color={rgb, 255:red, 0; green, 0; blue, 0 }  ,draw opacity=0.8 ]   (491.88,195.93) -- (576.08,163.84) ;
		\draw [shift={(490.01,196.64)}, rotate = 339.14] [fill={rgb, 255:red, 0; green, 0; blue, 0 }  ,fill opacity=0.8 ][line width=0.08]  [draw opacity=0] (9.6,-2.4) -- (0,0) -- (9.6,2.4) -- cycle    ;
		%Straight Lines [id:da1130651595139629] 
		\draw [color={rgb, 255:red, 0; green, 0; blue, 0 }  ,draw opacity=0.8 ]   (576.08,163.84) -- (643.36,207.9) ;
		\draw [shift={(645.03,209)}, rotate = 213.22] [fill={rgb, 255:red, 0; green, 0; blue, 0 }  ,fill opacity=0.8 ][line width=0.08]  [draw opacity=0] (9.6,-2.4) -- (0,0) -- (9.6,2.4) -- cycle    ;
		%Straight Lines [id:da36868591200313694] 
		\draw [color={rgb, 255:red, 0; green, 0; blue, 0 }  ,draw opacity=0.8 ]   (576.08,163.84) -- (575.51,90.79) ;
		\draw [shift={(575.49,88.79)}, rotate = 89.55] [fill={rgb, 255:red, 0; green, 0; blue, 0 }  ,fill opacity=0.8 ][line width=0.08]  [draw opacity=0] (9.6,-2.4) -- (0,0) -- (9.6,2.4) -- cycle    ;
		%Straight Lines [id:da9983226941000649] 
		\draw [color={rgb, 255:red, 208; green, 2; blue, 27 }  ,draw opacity=1 ][line width=0.75]    (627.79,197.84) -- (614.02,188.85) ;
		\draw [shift={(611.5,187.21)}, rotate = 33.15] [fill={rgb, 255:red, 208; green, 2; blue, 27 }  ,fill opacity=1 ][line width=0.08]  [draw opacity=0] (5.36,-2.57) -- (0,0) -- (5.36,2.57) -- cycle    ;
		%Straight Lines [id:da0016567267537355157] 
		\draw [color={rgb, 255:red, 208; green, 2; blue, 27 }  ,draw opacity=1 ][line width=0.75]    (529.53,181.59) -- (512.36,188.23) ;
		\draw [shift={(509.57,189.31)}, rotate = 338.86] [fill={rgb, 255:red, 208; green, 2; blue, 27 }  ,fill opacity=1 ][line width=0.08]  [draw opacity=0] (5.36,-2.57) -- (0,0) -- (5.36,2.57) -- cycle    ;
		%Straight Lines [id:da7629227920844508] 
		\draw [color={rgb, 255:red, 208; green, 2; blue, 27 }  ,draw opacity=0.75 ][line width=0.75]    (622.59,146.07) -- (635.79,140.93) ;
		\draw [shift={(638.59,139.84)}, rotate = 158.75] [fill={rgb, 255:red, 208; green, 2; blue, 27 }  ,fill opacity=0.75 ][line width=0.08]  [draw opacity=0] (5.36,-2.57) -- (0,0) -- (5.36,2.57) -- cycle    ;
		%Straight Lines [id:da30563480272405386] 
		\draw [color={rgb, 255:red, 208; green, 2; blue, 27 }  ,draw opacity=0.65 ][line width=0.75]    (526.7,131.09) -- (537.77,138.53) ;
		\draw [shift={(540.26,140.2)}, rotate = 213.91] [fill={rgb, 255:red, 208; green, 2; blue, 27 }  ,fill opacity=0.65 ][line width=0.08]  [draw opacity=0] (5.36,-2.57) -- (0,0) -- (5.36,2.57) -- cycle    ;
		%Straight Lines [id:da21333047557783913] 
		\draw [color={rgb, 255:red, 155; green, 155; blue, 155 }  ,draw opacity=1 ]   (576.08,163.84) -- (521.77,155.97) ;
		\draw [shift={(519.79,155.68)}, rotate = 8.25] [fill={rgb, 255:red, 155; green, 155; blue, 155 }  ,fill opacity=1 ][line width=0.08]  [draw opacity=0] (7.2,-1.8) -- (0,0) -- (7.2,1.8) -- cycle    ;
		%Straight Lines [id:da9351889601865728] 
		\draw [color={rgb, 255:red, 155; green, 155; blue, 155 }  ,draw opacity=1 ] [dash pattern={on 4.5pt off 4.5pt}]  (558.9,114.13) -- (576.08,163.84) ;
		%Curve Lines [id:da7763458636085163] 
		\draw [color={rgb, 255:red, 155; green, 155; blue, 155 }  ,draw opacity=1 ]   (565.01,122.81) .. controls (567.48,120.69) and (569.59,120.1) .. (573.27,121.65) ;
		\draw [shift={(575.94,122.99)}, rotate = 209.54] [fill={rgb, 255:red, 155; green, 155; blue, 155 }  ,fill opacity=1 ][line width=0.08]  [draw opacity=0] (3.57,-1.72) -- (0,0) -- (3.57,1.72) -- cycle    ;
		\draw [shift={(562.84,124.9)}, rotate = 314.01] [fill={rgb, 255:red, 155; green, 155; blue, 155 }  ,fill opacity=1 ][line width=0.08]  [draw opacity=0] (3.57,-1.72) -- (0,0) -- (3.57,1.72) -- cycle    ;
		%Shape: Circle [id:dp7760808390944338] 
		\draw  [fill={rgb, 255:red, 0; green, 0; blue, 0 }  ,fill opacity=1 ] (528.1,181.59) .. controls (528.1,180.79) and (528.74,180.15) .. (529.54,180.15) .. controls (530.33,180.15) and (530.97,180.8) .. (530.97,181.59) .. controls (530.97,182.39) and (530.33,183.03) .. (529.53,183.03) .. controls (528.74,183.03) and (528.1,182.38) .. (528.1,181.59) -- cycle ;
		
		% Text Node
		\draw (122.32,188.55) node [anchor=north west][inner sep=0.75pt]  [font=\fontsize{0.47em}{0.56em}\selectfont]  {$x$};
		% Text Node
		\draw (209.46,145.7) node [anchor=north west][inner sep=0.75pt]  [font=\fontsize{0.47em}{0.56em}\selectfont]  {$y$};
		% Text Node
		\draw (100.89,100.93) node [anchor=north west][inner sep=0.75pt]  [font=\fontsize{0.47em}{0.56em}\selectfont]  {$z$};
		% Text Node
		\draw (84.89,153.13) node [anchor=north west][inner sep=0.75pt]  [font=\fontsize{0.47em}{0.56em}\selectfont]  {$R$};
		% Text Node
		\draw (151.18,207.36) node [anchor=north west][inner sep=0.75pt]  [font=\fontsize{0.47em}{0.56em}\selectfont]  {$L$};
		% Text Node
		\draw (129.97,163.4) node [anchor=north west][inner sep=0.75pt]  [font=\fontsize{0.47em}{0.56em}\selectfont,rotate=-347.4]  {$L/2$};
		% Text Node
		\draw (142.07,99.63) node [anchor=north west][inner sep=0.75pt]  [font=\fontsize{0.47em}{0.56em}\selectfont]  {$P$};
		% Text Node
		\draw (132.27,120.95) node [anchor=north west][inner sep=0.75pt]  [font=\fontsize{0.47em}{0.56em}\selectfont]  {$A$};
		% Text Node
		\draw (128.21,212.62) node [anchor=north west][inner sep=0.75pt]  [font=\fontsize{0.47em}{0.56em}\selectfont]  {$P$};
		% Text Node
		\draw (10.63,58.4) node [anchor=north west][inner sep=0.75pt]  [font=\fontsize{0.5em}{0.6em}\selectfont]  {$ \begin{array}{l}
				R=300\\
				L=600\\
				h=3.0\\
				E=[ 3,1.5,3] \times 10^{6}\\
				\nu =[ 0.3,0.3,0.3]\\
				P=1.0
			\end{array}$};
		% Text Node
		\draw (172.88,124.84) node   [align=left] {\begin{minipage}[lt]{14.22pt}\setlength\topsep{0pt}
				\begin{center}
					{\fontfamily{ptm}\selectfont {\fontsize{0.47em}{0.56em}\selectfont \textit{\textcolor[rgb]{0.5,0.5,0.5}{Rigid }}}}
				\end{center}
				
		\end{minipage}};
		% Text Node
		\draw (172.91,131.96) node  [font=\normalsize,rotate=-0.03] [align=left] {\begin{minipage}[lt]{23.62pt}\setlength\topsep{0pt}
				\begin{center}
					{\fontsize{0.47em}{0.56em}\selectfont \textit{{\fontfamily{ptm}\selectfont \textcolor[rgb]{0.5,0.5,0.5}{diaphragm}}}}
				\end{center}
				
		\end{minipage}};
		% Text Node
		\draw (69.57,191.72) node   [align=left] {\begin{minipage}[lt]{14.22pt}\setlength\topsep{0pt}
				\begin{center}
					{\fontfamily{ptm}\selectfont {\fontsize{0.47em}{0.56em}\selectfont \textit{Rigid }}}
				\end{center}
				
		\end{minipage}};
		% Text Node
		\draw (68.94,199.15) node  [rotate=-0.03] [align=left] {\begin{minipage}[lt]{23.62pt}\setlength\topsep{0pt}
				\begin{center}
					{\fontsize{0.47em}{0.56em}\selectfont \textit{{\fontfamily{ptm}\selectfont diaphragm}}}
				\end{center}
				
		\end{minipage}};
		% Text Node
		\draw (352.66,200.13) node [anchor=north west][inner sep=0.75pt]  [font=\fontsize{0.47em}{0.56em}\selectfont]  {$x$};
		% Text Node
		\draw (440.18,111.59) node [anchor=north west][inner sep=0.75pt]  [font=\fontsize{0.47em}{0.56em}\selectfont]  {$y$};
		% Text Node
		\draw (305.93,119.24) node [anchor=north west][inner sep=0.75pt]  [font=\fontsize{0.47em}{0.56em}\selectfont]  {$z$};
		% Text Node
		\draw (314.82,198.37) node [anchor=north west][inner sep=0.75pt]  [font=\fontsize{0.47em}{0.56em}\selectfont]  {$R$};
		% Text Node
		\draw (354.43,178.69) node [anchor=north west][inner sep=0.75pt]  [font=\fontsize{0.47em}{0.56em}\selectfont,rotate=-338.3]  {$L/2$};
		% Text Node
		\draw (409.59,157.69) node [anchor=north west][inner sep=0.75pt]  [font=\fontsize{0.47em}{0.56em}\selectfont,rotate=-338.3]  {$L/2$};
		% Text Node
		\draw (372.82,151.98) node [anchor=north west][inner sep=0.75pt]  [font=\fontsize{0.47em}{0.56em}\selectfont]  {$B$};
		% Text Node
		\draw (292.44,168.7) node [anchor=north west][inner sep=0.75pt]  [font=\fontsize{0.47em}{0.56em}\selectfont]  {$\phi $};
		% Text Node
		\draw (378.86,88.12) node [anchor=north west][inner sep=0.75pt]  [font=\fontsize{0.47em}{0.56em}\selectfont]  {$g$};
		% Text Node
		\draw (266.63,172.9) node   [align=left] {\begin{minipage}[lt]{14.22pt}\setlength\topsep{0pt}
				\begin{center}
					{\fontfamily{ptm}\selectfont {\fontsize{0.47em}{0.56em}\selectfont \textit{Rigid }}}
				\end{center}
				
		\end{minipage}};
		% Text Node
		\draw (209.77,58.4) node [anchor=north west][inner sep=0.75pt]  [font=\fontsize{0.5em}{0.6em}\selectfont]  {$ \begin{array}{l}
				R=25\\
				L=50\\
				\phi =40^{\circ }\\
				h=0.25\\
				E=[ 4.32,2,4.32] \times 10^{8}\\
				\nu =[ 0.0,0.0,0.0]\\
				g=90
			\end{array}$};
		% Text Node
		\draw (338.65,109.36) node [anchor=north west][inner sep=0.75pt]  [font=\fontsize{0.57em}{0.68em}\selectfont,rotate=-338.76] [align=left] {{\fontfamily{ptm}\selectfont \textit{free}}};
		% Text Node
		\draw (350.27,160.28) node [anchor=north west][inner sep=0.75pt]  [font=\fontsize{0.57em}{0.68em}\selectfont,rotate=-338.98] [align=left] {{\fontfamily{ptm}\selectfont \textit{free}}};
		% Text Node
		\draw (264.75,180.35) node   [align=left] {\begin{minipage}[lt]{23.62pt}\setlength\topsep{0pt}
				\begin{center}
					{\fontsize{0.47em}{0.56em}\selectfont \textit{{\fontfamily{ptm}\selectfont diaphragm}}}
				\end{center}
				
		\end{minipage}};
		% Text Node
		\draw (422.34,88.8) node   [align=left] {\begin{minipage}[lt]{23.62pt}\setlength\topsep{0pt}
				\begin{center}
					{\fontsize{0.47em}{0.56em}\selectfont \textit{{\fontfamily{ptm}\selectfont diaphragm}}}
				\end{center}
				
		\end{minipage}};
		% Text Node
		\draw (421.48,82) node   [align=left] {\begin{minipage}[lt]{14.22pt}\setlength\topsep{0pt}
				\begin{center}
					{\fontfamily{ptm}\selectfont {\fontsize{0.47em}{0.56em}\selectfont \textit{Rigid }}}
				\end{center}
				
		\end{minipage}};
		% Text Node
		\draw (492.01,200.04) node [anchor=north west][inner sep=0.75pt]  [font=\fontsize{0.47em}{0.56em}\selectfont]  {$x$};
		% Text Node
		\draw (511.57,192.71) node [anchor=north west][inner sep=0.75pt]  [font=\fontsize{0.47em}{0.56em}\selectfont]  {$P$};
		% Text Node
		\draw (616.68,199.04) node [anchor=north west][inner sep=0.75pt]  [font=\fontsize{0.47em}{0.56em}\selectfont]  {$P$};
		% Text Node
		\draw (635.52,209.43) node [anchor=north west][inner sep=0.75pt]  [font=\fontsize{0.47em}{0.56em}\selectfont]  {$y$};
		% Text Node
		\draw (577.49,92.19) node [anchor=north west][inner sep=0.75pt]  [font=\fontsize{0.47em}{0.56em}\selectfont]  {$z$};
		% Text Node
		\draw (547.02,151.7) node [anchor=north west][inner sep=0.75pt]  [font=\fontsize{0.47em}{0.56em}\selectfont]  {$R$};
		% Text Node
		\draw (564.22,111.73) node [anchor=north west][inner sep=0.75pt]  [font=\fontsize{0.47em}{0.56em}\selectfont]  {$\phi $};
		% Text Node
		\draw (526.23,187.77) node [anchor=north west][inner sep=0.75pt]  [font=\fontsize{0.47em}{0.56em}\selectfont]  {$C$};
		% Text Node
		\draw (573.88,116.93) node [anchor=north west][inner sep=0.75pt]  [font=\fontsize{0.53em}{0.64em}\selectfont,rotate=-334.43] [align=left] {{\fontfamily{ptm}\selectfont \textit{free}}};
		% Text Node
		\draw (565.31,182.46) node [anchor=north west][inner sep=0.75pt]  [font=\fontsize{0.6em}{0.7em}\selectfont,rotate=-4.37] [align=left] {{\fontfamily{ptm}\selectfont \textit{sym.}}};
		% Text Node
		\draw (447.34,58.4) node [anchor=north west][inner sep=0.75pt]  [font=\fontsize{0.5em}{0.6em}\selectfont]  {$ \begin{array}{l}
				R=10\\
				\phi =18^{\circ }\\
				h=0.04\\
				E=[ 6.825,4,6.825] \times 10^{7}\\
				\nu =[ 0.3,0.3,0.3]\\
				P=2.0
			\end{array}$};
		% Text Node
		\draw (82.32,234) node [anchor=north west][inner sep=0.75pt]   [align=left] {{\fontsize{0.8em}{0.96em}\selectfont {\fontfamily{ptm}\selectfont (a) Pinched cylinder}}};
		% Text Node
		\draw (301.32,234) node [anchor=north west][inner sep=0.75pt]   [align=left] {{\fontsize{0.8em}{0.96em}\selectfont {\fontfamily{ptm}\selectfont (b) Scordelis-Lo roof}}};
		% Text Node
		\draw (542.82,234) node [anchor=north west][inner sep=0.75pt]   [align=left] {{\fontsize{0.8em}{0.96em}\selectfont {\fontfamily{ptm}\selectfont (c) Sphere}}};
	\end{tikzpicture}
	\caption{Illustration of the mid-surface geometries and boundary conditions of the modified shell obstacle course for three-layered thin shells. (a) pinched cylinder with two ends attached to rigid diaphragms; (b) scordelis-lo roof subject to a dead weight of $90$ per unit area; (c) a whole sphere with a $18^\circ$ hole. Only the blue parts are considered here due to the intrinsic symmetry of the shell geometries and the particular boundary conditions.\label{shell_obstacle}}
\end{figure}

Three benchmark problems, as shown in Fig.~\ref{shell_obstacle}, are considered in this section to assess the performance of the proposed shell element under complex strain states. Note that, the original problems devised by \cite{BELYTSCHKO1985221} and \cite{macneal1985proposed} are modified to three-layered shells with the thickness proportions of $\lambda_1 = \lambda_3 = 0.35,\ \lambda_2 = 0.3$. The basic information is presented on the top-left of each panel of Fig.~\ref{shell_obstacle}. We here study the vertical displacement of three prescribed points A, B and C, their specific values are compared with the results obtained from fine-mesh FEA and related studies are also carried out to check the convergence performance with the refinement of the discretized mesh.

\begin{figure}[!htbp]
	\setlength{\abovecaptionskip}{0.cm}
	\setlength{\belowcaptionskip}{-0.cm}
	\centering
	\subfigure[Cylindrical shell: Convergence study of $u_3^\mathrm{A}$]{
		\includegraphics[height=6.5cm]{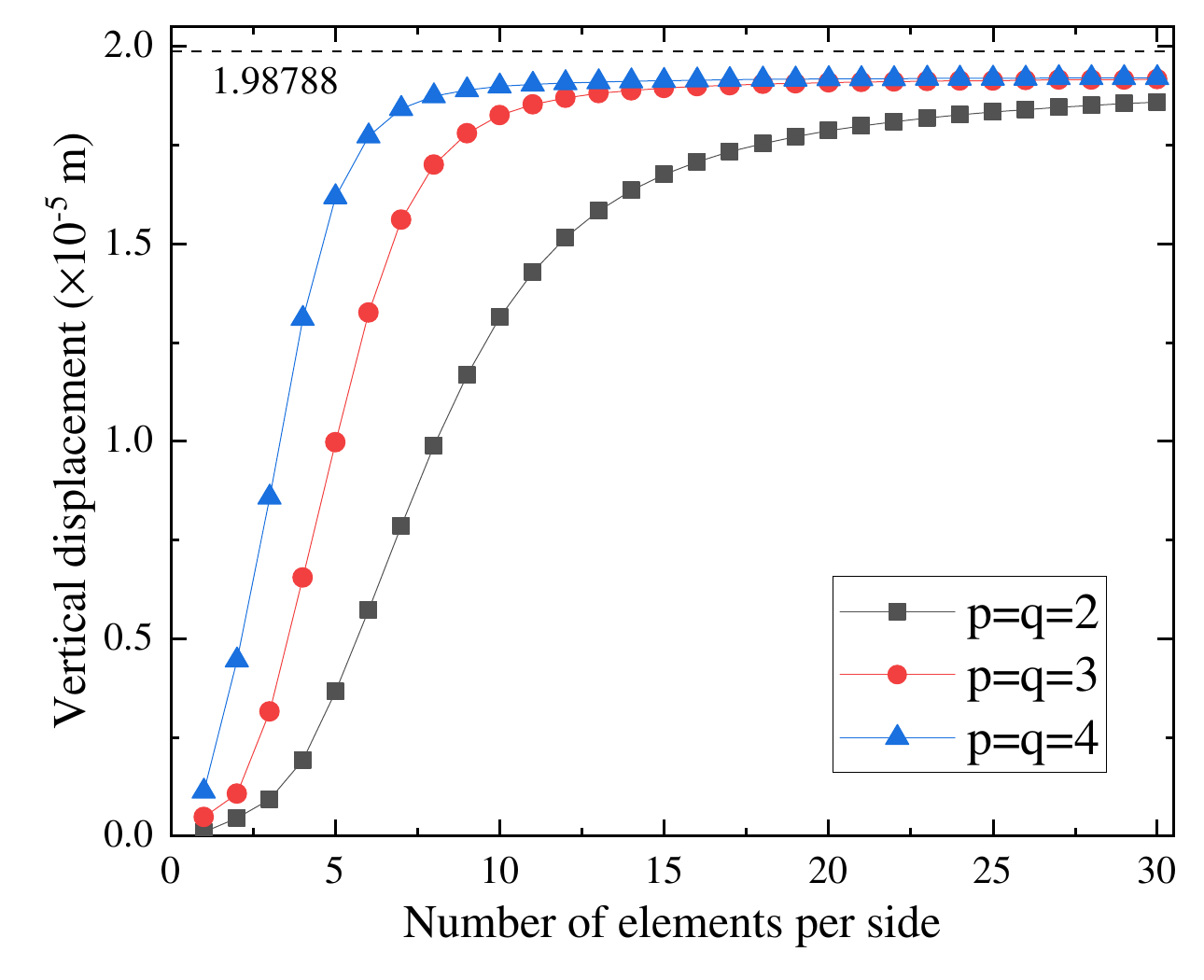}\hspace{0.2cm}
	}
	\subfigure[Scordelis-Lo roof: Convergence study of $u_3^\mathrm{B}$]{
		\includegraphics[height=6.6cm]{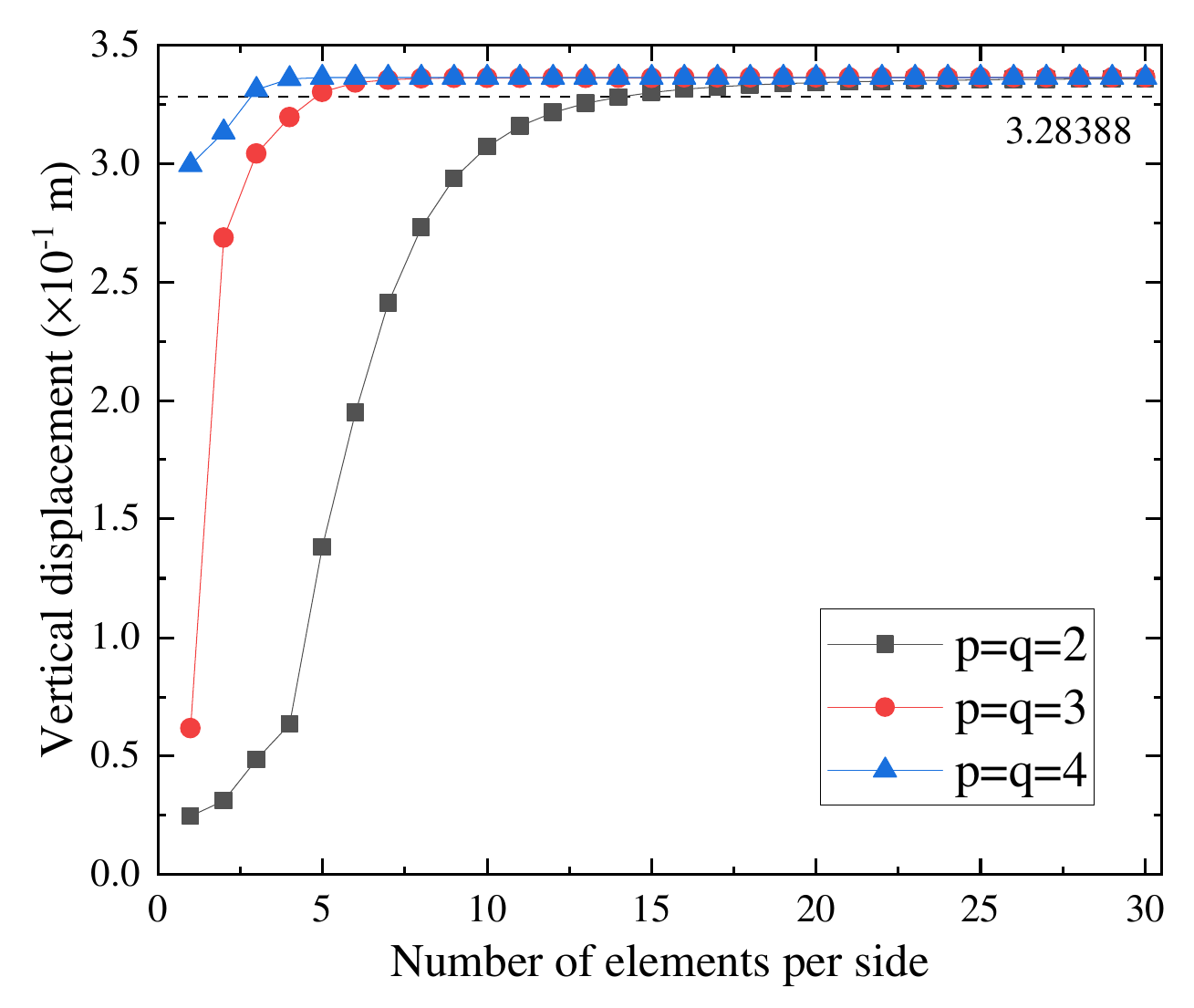}
	}
	\subfigure[Spherical shell with holes: Convergence study of $u_1^\mathrm{C}$]{
		\includegraphics[height=6.5cm]{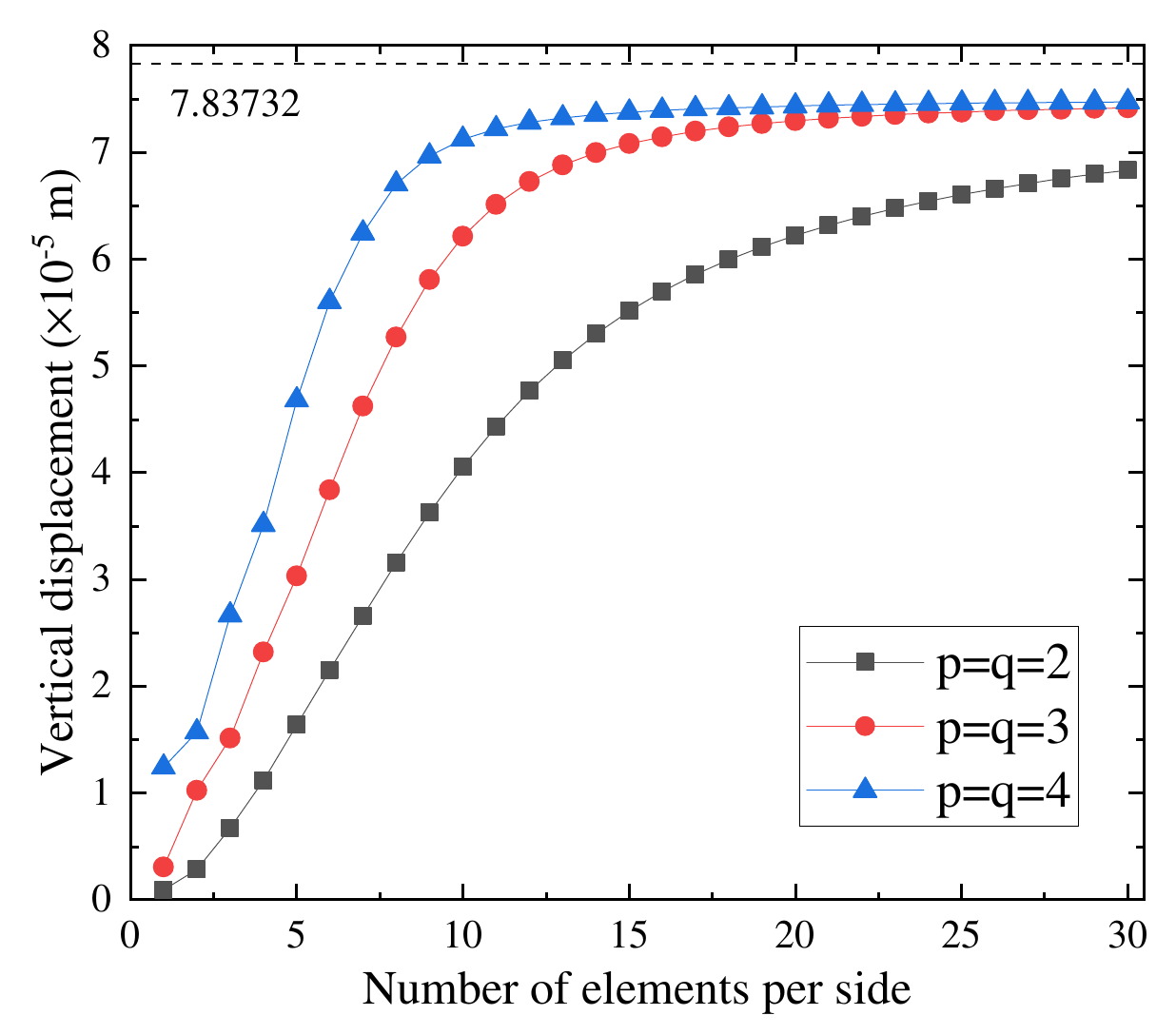}
	}
	\caption{Convergence study for the modified shell obstacle course shown in Fig.~\ref{shell_obstacle}. The p-refinement is carried out by choosing the NURBS basis functions with degrees $p=q=2,3,4$. The h-refinement is achieved by increasing the number of elements per side. And the horizontal dashed line represents the reference value obtained from fine-mesh FEA with 3D solid elements.\label{convergence_plot}}
\end{figure}

For simplicity, only one eighth of the cylindrical and spherical shell, and a quarter of the Scordelis-Lo roof, refer to the blue parts in Fig.~\ref{shell_obstacle}, will be modeled owing to the symmetric nature of the shell geometry and their imposed boundary conditions. Fig.~\ref{convergence_plot} shows the convergence behavior of the proposed method in three obstacle problems. We can find that, all three cases present good convergence behavior corresponding to the reference line obtained from the fine-mesh FEA, with relative errors roughly of $\mO\left(h/L\right)$.

\subsection{Undamped Free Vibration}
We then investigate the mode and angular frequency of the MTS under the case of free vibration. It has been proved in Sec.~\ref{Sec_weaklycurved_shell} that, free vibration of the weakly curved shell actually corresponds to a constrained eigenvalue problem, which resembles the behavior of thin plates. For normally curved shells, as shown by Eq.~\eqref{equilibrium_normal}, all three inertial forces participate in the governing equations. To obtain the general eigenvalue problem of this case, the field solution is firstly written as follows
\begin{equation}\label{vibration_separation}
	\vu^\re\left(\xi_1,\xi_2,t\right) = \vU^\re\left(\xi_1,\xi_2\right)\sin\left(\omega t+\phi\right),
\end{equation}
where $\vU^\re$ represents the element mode shape vector and $\omega$ is called the angular (circular) frequency. Substituting Eq.~\eqref{vibration_separation} into the discretized governing equations (Eq.~\eqref{linear_algebraic_assembly}) and setting $\vF^\re=\boldsymbol{0}$ due to undamped vibration, the eigenvalue problem is finally expressed by
\begin{equation}\label{eigenvalue_problem}
	\vK\vU = \omega^2\textbf{M}\vU,
\end{equation}
from which the angular frequency for undamped free vibration of shells can be obtained. And $\vU = \sum_{e=1}^\text{nel}\left(\vU^\re\right)$ is the corresponding global mode shape vector.

In this section, we consider the free vibration behavior of two types of shells, three-layered cylindrical shells with end diaphragms and spherical shells with a $18^\circ$ hole. The material density of each layer is selected to be $\rho_{\sss\CL} = \left[7800, 5000, 7800\right] \mathrm{kg/m^3}$ from bottom to top. The small parameter $\epsilon$ takes two values $0.1$ and $0.01$ for both cases, while the remaining geometry and material parameters are the same as those listed in Fig.~\ref{shell_obstacle}.

Table.~\ref{Table3_frequency} shows the first eight angular frequency $\omega$ for two types of shells and aspect ratios. The values obtained based on the derived asymptotic expressions are compared to those given by the fine-mesh FEA, and we can find that, the proposed method based on AA exhibits a good accuracy in capturing the angular frequency of the free vibration of the MTS, and the accuracy increases as MTS becomes thinner (from $\epsilon=0.1$ to $\epsilon=0.01$).

\begin{table}
	\setlength{\abovecaptionskip}{0.cm}
	\setlength{\belowcaptionskip}{0.2cm}
	\caption{Comparison of the first eight angular frequency $\omega\left(\mathrm{rad/s}\right)$ for the free vibration of spherical and cylindrical three-layered shells.\label{Table3_frequency}}
	\footnotesize
	\centering
	\renewcommand{\arraystretch}{1.2} % adjusting row spacing
	\begin{tabular}{c c c c c c c c c c}
		\hline
		$\epsilon=h/L$ & Method & \multicolumn{8}{l}{Mode Sequence}\\
		\cline{3-10}
		& & 1 & 2 & 3 & 4 & 5 & 6 & 7 & 8\\
		\hline
		\multicolumn{10}{l}{(a) One-eighth cylindrical shell shown in Fig.~\ref{shell_obstacle}(a)}\\
		0.1 & Present Method & 0.0249 & 0.0378 & 0.0622 & 0.0748 & 0.0777 & 0.0778 & 0.0847 & 0.1071\\
		& Fine-mesh & 0.0240 & 0.0345 & 0.0622 & 0.0706 & 0.0723 & 0.0764 & 0.0785 & 0.1063\\
		& Relative Error ($\%$) & 4.06 & 9.56 & 0.00 & 5.95 & 7.53 & 1.82 & 7.92 & 0.77\\
		0.01 & Present Method & 0.0089 & 0.0090 & 0.0139 & 0.0212 & 0.0219 & 0.0241 & 0.0268 & 0.0278\\
		& Fine-mesh & 0.0087 & 0.0090 & 0.0137 & 0.0210 & 0.0219 & 0.0239 & 0.0267 & 0.0275\\
		& Relative Error ($\%$) & 2.28 & 1.00 & 1.67 & 1.21 & 0.05 & 0.82 & 0.41 & 0.97\\
		\hline
		\multicolumn{10}{l}{(b) One-eighth spherical shell shown in Fig.~\ref{shell_obstacle}(c)}\\
		0.1 & Present Method & 12.4083 & 17.9600 & 22.3795 & 25.3521 & 26.4880 & 34.0364 & 36.5192 & 36.8108\\
		& Fine-mesh & 11.8187 & 17.8066 & 21.6047 & 24.5352 & 25.7001 & 31.8979 & 34.4589 & 34.7435\\
		& Relative Error ($\%$) & 4.99 & 0.86 & 3.59 & 3.33 & 3.07 & 6.70 & 5.98 & 5.95\\
		0.01 & Present Method & 2.8879 & 12.9907 & 17.6279 & 18.5418 & 21.4444 & 21.8301 & 21.9999 & 22.5321\\
		& Fine-mesh & 2.8322 & 12.8322 & 17.6262 & 18.5360 & 21.4363 & 21.8221 & 21.9874 & 22.5152\\
		& Relative Error ($\%$) & 1.97 & 1.24 & 0.00 & 0.03 & 0.04 & 0.04 & 0.06 & 0.08\\
		\hline
	\end{tabular}
\end{table}

\subsection{Shear Effect Studies}\label{Sec_shear_studies}
\subsubsection{Transverse Shear Stress Distribution in Normally Curved Shells}
\begin{figure}[!htbp]
	\setlength{\abovecaptionskip}{0.cm}
	\setlength{\belowcaptionskip}{-0.cm}
	\centering
	\tikzset{every picture/.style={line width=0.55pt}} %set default line width to 0.55pt
	\begin{tikzpicture}[x=0.75pt,y=0.75pt,yscale=-1,xscale=1]
		%uncomment if require: \path (0,327); %set diagram left start at 0, and has height of 327
		%Image [id:dp1404151556835216] 
		\draw (192.91,180.25) node  {\includegraphics[width=240pt,height=180pt]{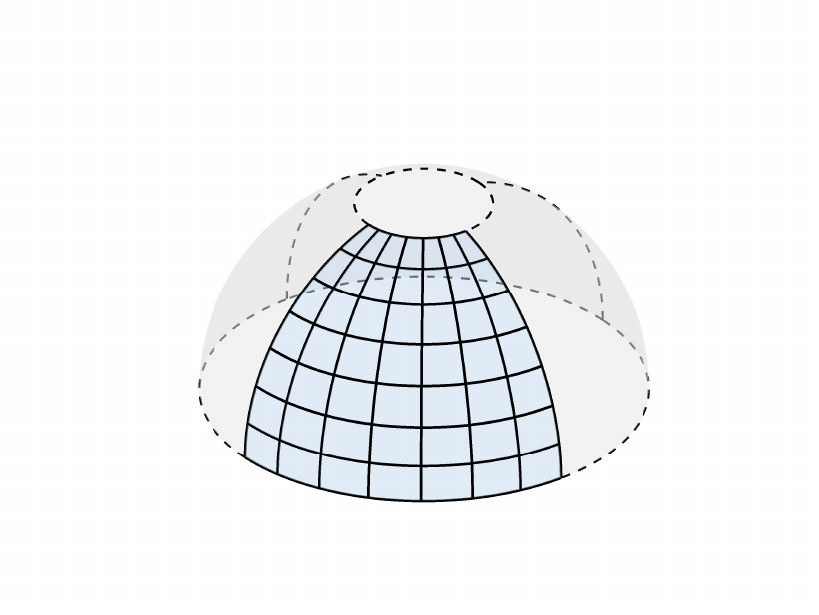}};
		%Straight Lines [id:da5347645740556974] 
		\draw [color={rgb, 255:red, 0; green, 0; blue, 0 }  ,draw opacity=0.5 ]   (197.91,212.12) -- (282.61,267.1) ;
		\draw [shift={(284.29,268.19)}, rotate = 212.99] [fill={rgb, 255:red, 0; green, 0; blue, 0 }  ,fill opacity=0.5 ][line width=0.08]  [draw opacity=0] (9.6,-2.4) -- (0,0) -- (9.6,2.4) -- cycle    ;
		%Straight Lines [id:da5552407676304671] 
		\draw [color={rgb, 255:red, 0; green, 0; blue, 0 }  ,draw opacity=0.5 ][fill={rgb, 255:red, 0; green, 0; blue, 0 }  ,fill opacity=0.2 ]   (197.91,212.12) -- (90.15,254.6) ;
		\draw [shift={(88.29,255.33)}, rotate = 338.48] [fill={rgb, 255:red, 0; green, 0; blue, 0 }  ,fill opacity=0.5 ][line width=0.08]  [draw opacity=0] (9.6,-2.4) -- (0,0) -- (9.6,2.4) -- cycle    ;
		%Straight Lines [id:da7221169164962666] 
		\draw [color={rgb, 255:red, 0; green, 0; blue, 0 }  ,draw opacity=0.5 ]   (197.91,212.12) -- (198.81,96.72) ;
		\draw [shift={(198.82,94.72)}, rotate = 90.44] [fill={rgb, 255:red, 0; green, 0; blue, 0 }  ,fill opacity=0.5 ][line width=0.08]  [draw opacity=0] (9.6,-2.4) -- (0,0) -- (9.6,2.4) -- cycle    ;
		%Straight Lines [id:da5314672882500637] 
		\draw [color={rgb, 255:red, 208; green, 2; blue, 27 }  ,draw opacity=1 ]   (138.73,196.96) -- (149.68,202.95) ;
		\draw [shift={(152.31,204.39)}, rotate = 208.68] [fill={rgb, 255:red, 208; green, 2; blue, 27 }  ,fill opacity=1 ][line width=0.08]  [draw opacity=0] (5.36,-2.57) -- (0,0) -- (5.36,2.57) -- cycle    ;
		%Straight Lines [id:da2687858883830687] 
		\draw [color={rgb, 255:red, 208; green, 2; blue, 27 }  ,draw opacity=1 ]   (176.62,148.14) -- (185.77,152.73) ;
		\draw [shift={(188.46,154.07)}, rotate = 206.62] [fill={rgb, 255:red, 208; green, 2; blue, 27 }  ,fill opacity=1 ][line width=0.08]  [draw opacity=0] (5.36,-2.57) -- (0,0) -- (5.36,2.57) -- cycle    ;
		%Straight Lines [id:da026260827887159888] 
		\draw [color={rgb, 255:red, 5; green, 37; blue, 246 }  ,draw opacity=1 ]   (138.73,196.96) -- (143.77,186.35) ;
		\draw [shift={(145.06,183.64)}, rotate = 115.43] [fill={rgb, 255:red, 5; green, 37; blue, 246 }  ,fill opacity=1 ][line width=0.08]  [draw opacity=0] (5.36,-2.57) -- (0,0) -- (5.36,2.57) -- cycle    ;
		%Straight Lines [id:da7409849165836684] 
		\draw [color={rgb, 255:red, 5; green, 37; blue, 246 }  ,draw opacity=1 ]   (176.62,148.14) -- (184.65,141.49) ;
		\draw [shift={(186.96,139.57)}, rotate = 140.33] [fill={rgb, 255:red, 5; green, 37; blue, 246 }  ,fill opacity=1 ][line width=0.08]  [draw opacity=0] (5.36,-2.57) -- (0,0) -- (5.36,2.57) -- cycle    ;
		%Straight Lines [id:da2705835096066964] 
		\draw [color={rgb, 255:red, 208; green, 2; blue, 27 }  ,draw opacity=1 ]   (155.18,168.9) -- (164.4,174.1) ;
		\draw [shift={(167.01,175.58)}, rotate = 209.45] [fill={rgb, 255:red, 208; green, 2; blue, 27 }  ,fill opacity=1 ][line width=0.08]  [draw opacity=0] (5.36,-2.57) -- (0,0) -- (5.36,2.57) -- cycle    ;
		%Straight Lines [id:da2634437957397828] 
		\draw [color={rgb, 255:red, 5; green, 37; blue, 246 }  ,draw opacity=1 ]   (155.18,168.9) -- (161.49,160.25) ;
		\draw [shift={(163.26,157.83)}, rotate = 126.14] [fill={rgb, 255:red, 5; green, 37; blue, 246 }  ,fill opacity=1 ][line width=0.08]  [draw opacity=0] (5.36,-2.57) -- (0,0) -- (5.36,2.57) -- cycle    ;
		%Straight Lines [id:da7116938456973321] 
		\draw [color={rgb, 255:red, 208; green, 2; blue, 27 }  ,draw opacity=1 ]   (128.73,239.35) -- (138.94,245.06) ;
		\draw [shift={(141.56,246.53)}, rotate = 209.23] [fill={rgb, 255:red, 208; green, 2; blue, 27 }  ,fill opacity=1 ][line width=0.08]  [draw opacity=0] (5.36,-2.57) -- (0,0) -- (5.36,2.57) -- cycle    ;
		%Straight Lines [id:da05848622122736491] 
		\draw [color={rgb, 255:red, 5; green, 37; blue, 246 }  ,draw opacity=1 ]   (128.73,239.35) -- (129.6,226.52) ;
		\draw [shift={(129.81,223.53)}, rotate = 93.92] [fill={rgb, 255:red, 5; green, 37; blue, 246 }  ,fill opacity=1 ][line width=0.08]  [draw opacity=0] (5.36,-2.57) -- (0,0) -- (5.36,2.57) -- cycle    ;
		%Straight Lines [id:da1982175756554474] 
		\draw [color={rgb, 255:red, 208; green, 2; blue, 27 }  ,draw opacity=1 ]   (215.03,150.6) -- (223.43,146.38) ;
		\draw [shift={(226.11,145.04)}, rotate = 153.32] [fill={rgb, 255:red, 208; green, 2; blue, 27 }  ,fill opacity=1 ][line width=0.08]  [draw opacity=0] (5.36,-2.57) -- (0,0) -- (5.36,2.57) -- cycle    ;
		%Straight Lines [id:da6304088370615655] 
		\draw [color={rgb, 255:red, 5; green, 37; blue, 246 }  ,draw opacity=1 ]   (215.03,150.6) -- (209.3,143.83) ;
		\draw [shift={(207.36,141.54)}, rotate = 49.79] [fill={rgb, 255:red, 5; green, 37; blue, 246 }  ,fill opacity=1 ][line width=0.08]  [draw opacity=0] (5.36,-2.57) -- (0,0) -- (5.36,2.57) -- cycle    ;
		%Straight Lines [id:da9045455997331757] 
		\draw [color={rgb, 255:red, 208; green, 2; blue, 27 }  ,draw opacity=1 ]   (231.47,174.09) -- (241.13,169.13) ;
		\draw [shift={(243.8,167.77)}, rotate = 152.87] [fill={rgb, 255:red, 208; green, 2; blue, 27 }  ,fill opacity=1 ][line width=0.08]  [draw opacity=0] (5.36,-2.57) -- (0,0) -- (5.36,2.57) -- cycle    ;
		%Straight Lines [id:da990437714148853] 
		\draw [color={rgb, 255:red, 5; green, 37; blue, 246 }  ,draw opacity=1 ]   (231.47,174.09) -- (226.37,163.71) ;
		\draw [shift={(225.05,161.02)}, rotate = 63.85] [fill={rgb, 255:red, 5; green, 37; blue, 246 }  ,fill opacity=1 ][line width=0.08]  [draw opacity=0] (5.36,-2.57) -- (0,0) -- (5.36,2.57) -- cycle    ;
		%Straight Lines [id:da9742017851537768] 
		\draw [color={rgb, 255:red, 208; green, 2; blue, 27 }  ,draw opacity=1 ]   (244.43,203.55) -- (255.53,198.47) ;
		\draw [shift={(258.26,197.23)}, rotate = 155.45] [fill={rgb, 255:red, 208; green, 2; blue, 27 }  ,fill opacity=1 ][line width=0.08]  [draw opacity=0] (5.36,-2.57) -- (0,0) -- (5.36,2.57) -- cycle    ;
		%Straight Lines [id:da9065982014850045] 
		\draw [color={rgb, 255:red, 5; green, 37; blue, 246 }  ,draw opacity=1 ]   (244.43,203.55) -- (241.28,191.63) ;
		\draw [shift={(240.51,188.73)}, rotate = 75.2] [fill={rgb, 255:red, 5; green, 37; blue, 246 }  ,fill opacity=1 ][line width=0.08]  [draw opacity=0] (5.36,-2.57) -- (0,0) -- (5.36,2.57) -- cycle    ;
		%Straight Lines [id:da999919883225056] 
		\draw [color={rgb, 255:red, 208; green, 2; blue, 27 }  ,draw opacity=1 ]   (251.93,247.34) -- (264.76,241.72) ;
		\draw [shift={(267.51,240.52)}, rotate = 156.37] [fill={rgb, 255:red, 208; green, 2; blue, 27 }  ,fill opacity=1 ][line width=0.08]  [draw opacity=0] (5.36,-2.57) -- (0,0) -- (5.36,2.57) -- cycle    ;
		%Straight Lines [id:da6011681026034439] 
		\draw [color={rgb, 255:red, 5; green, 37; blue, 246 }  ,draw opacity=1 ]   (251.93,247.34) -- (252.21,233.02) ;
		\draw [shift={(252.26,230.02)}, rotate = 91.1] [fill={rgb, 255:red, 5; green, 37; blue, 246 }  ,fill opacity=1 ][line width=0.08]  [draw opacity=0] (5.36,-2.57) -- (0,0) -- (5.36,2.57) -- cycle    ;
		%Shape: Ellipse [id:dp0563769826877889] 
		\draw  [dash pattern={on 4.5pt off 4.5pt}][line width=0.75]  (300.75,127.81) .. controls (300.06,116.45) and (314.05,106.37) .. (331.98,105.28) .. controls (349.92,104.2) and (365.02,112.52) .. (365.7,123.88) .. controls (366.39,135.23) and (352.41,145.32) .. (334.47,146.4) .. controls (316.53,147.49) and (301.44,139.16) .. (300.75,127.81) -- cycle ;
		%Curve Lines [id:da5443697109931533] 
		\draw    (305.06,116.66) .. controls (328.33,107.03) and (330.13,124.8) .. (360.47,113.8) ;
		%Curve Lines [id:da04548758946516274] 
		\draw    (305.51,137.74) .. controls (328.78,128.12) and (331.06,146.04) .. (361.4,135.04) ;
		%Straight Lines [id:da07295162427899404] 
		\draw [color={rgb, 255:red, 208; green, 2; blue, 27 }  ,draw opacity=1 ]   (197.38,256.24) -- (210.9,256.34) ;
		\draw [shift={(213.9,256.36)}, rotate = 180.42] [fill={rgb, 255:red, 208; green, 2; blue, 27 }  ,fill opacity=1 ][line width=0.08]  [draw opacity=0] (5.36,-2.57) -- (0,0) -- (5.36,2.57) -- cycle    ;
		%Straight Lines [id:da3163566247296514] 
		\draw [color={rgb, 255:red, 5; green, 37; blue, 246 }  ,draw opacity=1 ]   (197.38,256.24) -- (197.55,241.97) ;
		\draw [shift={(197.59,238.97)}, rotate = 90.69] [fill={rgb, 255:red, 5; green, 37; blue, 246 }  ,fill opacity=1 ][line width=0.08]  [draw opacity=0] (5.36,-2.57) -- (0,0) -- (5.36,2.57) -- cycle    ;
		%Straight Lines [id:da6489990821367448] 
		\draw [color={rgb, 255:red, 208; green, 2; blue, 27 }  ,draw opacity=1 ]   (197.45,211.19) -- (211.02,211.19) ;
		\draw [shift={(214.02,211.19)}, rotate = 180] [fill={rgb, 255:red, 208; green, 2; blue, 27 }  ,fill opacity=1 ][line width=0.08]  [draw opacity=0] (5.36,-2.57) -- (0,0) -- (5.36,2.57) -- cycle    ;
		%Straight Lines [id:da5886966439081442] 
		\draw [color={rgb, 255:red, 5; green, 37; blue, 246 }  ,draw opacity=1 ]   (197.45,211.19) -- (197.69,196.9) ;
		\draw [shift={(197.74,193.9)}, rotate = 90.96] [fill={rgb, 255:red, 5; green, 37; blue, 246 }  ,fill opacity=1 ][line width=0.08]  [draw opacity=0] (5.36,-2.57) -- (0,0) -- (5.36,2.57) -- cycle    ;
		%Shape: Ellipse [id:dp832179653272296] 
		\draw  [line width=0.75]  (271.09,191.94) .. controls (270.94,189.42) and (272.85,187.26) .. (275.35,187.11) .. controls (277.86,186.96) and (280.01,188.87) .. (280.16,191.39) .. controls (280.32,193.9) and (278.41,196.07) .. (275.9,196.22) .. controls (273.4,196.37) and (271.24,194.45) .. (271.09,191.94) -- cycle ;
		%Straight Lines [id:da4128589825930553] 
		\draw  [dash pattern={on 4.5pt off 4.5pt}]  (275.35,187.11) -- (291.56,147.06) ;
		%Straight Lines [id:da48716363313364996] 
		\draw  [dash pattern={on 4.5pt off 4.5pt}]  (291.56,217.39) -- (275.9,196.22) ;
		%Shape: Circle [id:dp02499891302357926] 
		\draw  [color={rgb, 255:red, 0; green, 0; blue, 0 }  ,draw opacity=1 ][fill={rgb, 255:red, 0; green, 0; blue, 0 }  ,fill opacity=1 ] (187.19,203.72) .. controls (187.19,202.6) and (188.1,201.69) .. (189.23,201.69) .. controls (190.35,201.69) and (191.26,202.61) .. (191.26,203.73) .. controls (191.26,204.86) and (190.34,205.77) .. (189.22,205.77) .. controls (188.09,205.76) and (187.18,204.85) .. (187.19,203.72) -- cycle ;
		%Shape: Ellipse [id:dp4945971474762443] 
		\draw  [dash pattern={on 4.5pt off 4.5pt}][line width=0.75]  (463.01,125.82) .. controls (462.32,114.46) and (476.31,104.38) .. (494.24,103.29) .. controls (512.18,102.21) and (527.28,110.54) .. (527.96,121.89) .. controls (528.65,133.25) and (514.67,143.33) .. (496.73,144.41) .. controls (478.79,145.5) and (463.7,137.17) .. (463.01,125.82) -- cycle ;
		%Curve Lines [id:da9937462439856672] 
		\draw    (467.14,114.49) .. controls (490.41,104.86) and (492.21,122.63) .. (522.55,111.63) ;
		%Curve Lines [id:da5131995694728178] 
		\draw    (468.91,136.9) .. controls (491.14,130.9) and (490.39,146.99) .. (520.73,135.99) ;
		%Curve Lines [id:da895750918466806] 
		\draw    (463.01,125.82) .. controls (486.28,116.19) and (503.88,137.02) .. (527.96,121.89) ;
		%Shape: Ellipse [id:dp7956296995154255] 
		\draw  [dash pattern={on 4.5pt off 4.5pt}][line width=0.75]  (301.72,223.21) .. controls (301.04,211.85) and (315.02,201.77) .. (332.96,200.69) .. controls (350.89,199.6) and (365.99,207.93) .. (366.68,219.28) .. controls (367.36,230.64) and (353.38,240.72) .. (335.44,241.8) .. controls (317.51,242.89) and (302.41,234.56) .. (301.72,223.21) -- cycle ;
		%Curve Lines [id:da5148676570608566] 
		\draw    (311.33,207.15) .. controls (334.6,197.52) and (340.68,212.46) .. (357.52,206.1) ;
		%Curve Lines [id:da1454115423142246] 
		\draw    (302.35,226.29) .. controls (346.11,215.02) and (330.47,235.02) .. (366.29,222.29) ;
		%Curve Lines [id:da5233826800464814] 
		\draw    (302.63,217.57) .. controls (348.29,204.47) and (330.83,225.02) .. (364.59,212.84) ;
		%Curve Lines [id:da4152470747062864] 
		\draw    (309.08,235.2) .. controls (329.74,225.56) and (339.93,243.56) .. (359.93,233.2) ;
		%Shape: Ellipse [id:dp4123004115749007] 
		\draw  [dash pattern={on 4.5pt off 4.5pt}][line width=0.75]  (463.1,221.68) .. controls (462.42,210.32) and (476.4,200.24) .. (494.33,199.15) .. controls (512.27,198.07) and (527.37,206.39) .. (528.05,217.75) .. controls (528.74,229.1) and (514.76,239.19) .. (496.82,240.27) .. controls (478.88,241.36) and (463.79,233.03) .. (463.1,221.68) -- cycle ;
		%Curve Lines [id:da25480928120754265] 
		\draw    (477.81,202.67) .. controls (500.38,198.67) and (505.81,205.81) .. (515.24,202.95) ;
		%Curve Lines [id:da22543061571637124] 
		\draw    (463,218.58) .. controls (499.24,210.64) and (494.69,226.64) .. (527.42,216.27) ;
		%Curve Lines [id:da5354292325776258] 
		\draw    (467.1,210.4) .. controls (502.87,203.36) and (500.69,217.18) .. (522.51,207.36) ;
		%Curve Lines [id:da03246637849200873] 
		\draw    (465,228.39) .. controls (488.87,221.73) and (503.78,238.09) .. (526.33,225.36) ;
		%Curve Lines [id:da10035823567996949] 
		\draw    (472.09,235.12) .. controls (483.05,231.18) and (497.6,242.09) .. (513.42,236.27) ;
		%Straight Lines [id:da12967639496134042] 
		\draw [color={rgb, 255:red, 155; green, 155; blue, 155 }  ,draw opacity=0.8 ]   (197.91,212.12) -- (112.89,211.19) ;
		\draw [shift={(110.89,211.17)}, rotate = 0.62] [fill={rgb, 255:red, 155; green, 155; blue, 155 }  ,fill opacity=0.8 ][line width=0.08]  [draw opacity=0] (7.2,-1.8) -- (0,0) -- (7.2,1.8) -- cycle    ;
		
		% Text Node
		\draw (269.12,266.96) node [anchor=north west][inner sep=0.75pt]  [font=\fontsize{0.71em}{0.85em}\selectfont]  {$y$};
		% Text Node
		\draw (96.18,257.07) node [anchor=north west][inner sep=0.75pt]  [font=\fontsize{0.71em}{0.85em}\selectfont]  {$x$};
		% Text Node
		\draw (200.82,98.12) node [anchor=north west][inner sep=0.75pt]  [font=\fontsize{0.71em}{0.85em}\selectfont]  {$z$};
		% Text Node
		\draw (265.85,243.85) node [anchor=north west][inner sep=0.75pt]  [font=\fontsize{0.59em}{0.71em}\selectfont,color={rgb, 255:red, 208; green, 2; blue, 27 }  ,opacity=1 ]  {$p_{1}$};
		% Text Node
		\draw (255.75,226.64) node [anchor=north west][inner sep=0.75pt]  [font=\fontsize{0.59em}{0.71em}\selectfont,color={rgb, 255:red, 5; green, 37; blue, 246 }  ,opacity=1 ]  {$p_{2}$};
		% Text Node
		\draw (360.68,130.95) node [anchor=north west][inner sep=0.75pt]  [font=\fontsize{0.71em}{0.85em}\selectfont]  {$ \begin{array}{l}
				h=0.2\\
				E=6.825\times 10^{7}\\
				\nu =0.3
			\end{array}$};
		% Text Node
		\draw (360.7,228.95) node [anchor=north west][inner sep=0.75pt]  [font=\fontsize{0.71em}{0.85em}\selectfont]  {$ \begin{array}{l}
				h=0.2\\
				\lambda =[ 0.35,0.3,0.35]\\
				E=[ 6.825,4,6.825] \times 10^{7}\\
				\nu =[ 0.3,0.3,0.3]
			\end{array}$};
		% Text Node
		\draw (523.41,227.9) node [anchor=north west][inner sep=0.75pt]  [font=\fontsize{0.71em}{0.85em}\selectfont]  {$ \begin{array}{l}
				h=0.2\\
				\lambda =[ 0.25,0.2,0.35,0.2]\\
				E=[ 6.825,3.6,6.825,4.2] \times 10^{7}\\
				\nu =[ 0.3,0.3,0.3]
			\end{array}$};
		% Text Node
		\draw (185.78,261.67) node [anchor=north west][inner sep=0.75pt]  [font=\fontsize{0.93em}{1.12em}\selectfont] [align=left] {{\fontfamily{ptm}\selectfont {\footnotesize sym.}}};
		% Text Node
		\draw (523.48,131) node [anchor=north west][inner sep=0.75pt]  [font=\fontsize{0.71em}{0.85em}\selectfont]  {$ \begin{array}{l}
				h=0.2\\
				\lambda =[ 0.65,0.35]\\
				E=[ 3,6.825] \times 10^{7}\\
				\nu =[ 0.3,0.3]
			\end{array}$};
		% Text Node
		\draw (168.37,198.44) node [anchor=north west][inner sep=0.75pt]  [font=\fontsize{0.53em}{0.64em}\selectfont]  {$D$};
		% Text Node
		\draw (367.74,110.52) node [anchor=north west][inner sep=0.75pt]  [font=\fontsize{0.83em}{1em}\selectfont] [align=left] {{\fontfamily{ptm}\selectfont One layer}};
		% Text Node
		\draw (530.8,108.5) node [anchor=north west][inner sep=0.75pt]  [font=\fontsize{0.83em}{1em}\selectfont] [align=left] {{\fontfamily{ptm}\selectfont Two layers}};
		% Text Node
		\draw (367.7,208.17) node [anchor=north west][inner sep=0.75pt]  [font=\fontsize{0.83em}{1em}\selectfont] [align=left] {{\fontfamily{ptm}\selectfont Three layers}};
		% Text Node
		\draw (532.66,207.2) node [anchor=north west][inner sep=0.75pt]  [font=\fontsize{0.83em}{1em}\selectfont] [align=left] {{\fontfamily{ptm}\selectfont Four layers}};
		% Text Node
		\draw (122.37,202.44) node [anchor=north west][inner sep=0.75pt]  [font=\fontsize{0.53em}{0.64em}\selectfont]  {$R$};
	\end{tikzpicture}
	\caption{Illustration of the mid-surface of a multi-layered spherical shell with a $18^\circ$ hole at each pole. Its radius is selected to be $R=4$. The applied shear forces point in a set of orthogonal principal directions with magnitudes $p_1=5000\,\mathrm{N/m^2}$ and $p_2=8000\,\mathrm{N/m^2}$, respectively. The right four panels present the parameters needed for modeling shells with different layers $\CL=1,2,3,4$. Due to the symmetry of shell geometry and loading conditions, only one eighth of the structure (gridded part) is modeled here.\label{spherical_linear_study}}
\end{figure}

For normally curved shells whose maximum non-dimensional principal curvature satisfies Eq.~\eqref{curv_normal}, the leading-order transverse stress components in each layer have been proved to be linear functions of the normal variable $\wxi_3$ (Eq.~\eqref{transverse_shear_piecewiselinear}). In this section, we consider an eighth of a multi-layered spherical shell, as shown in the gridded part of Fig.~\ref{spherical_linear_study}. Two external shear forces $p_1=5000\,\mathrm{N/m^2}$ and $p_2=8000\,\mathrm{N/m^2}$ are applied on the top surface of the shell. In order to justify Eq.~\eqref{transverse_shear_piecewiselinear} from a fully computational aspect, we devise to extract the local transverse shear stresses given by the fine-mesh FEA along the surface normal at point D ($\left(\xi_1,\xi_2\right)=\left(0.5,0.5\right)$) and then check the actual variation of $\sigma_{\alpha 3}$ in the shell thickness direction.
%\vspace{-0.5cm}

\begin{figure}[!htbp]
	\setlength{\abovecaptionskip}{0.cm}
	\setlength{\belowcaptionskip}{-0.cm}
	\centering
	\subfigure[Variation of $\sigma_{\alpha 3}$ in one-layered spherical shell]{
		\includegraphics[width=8cm]{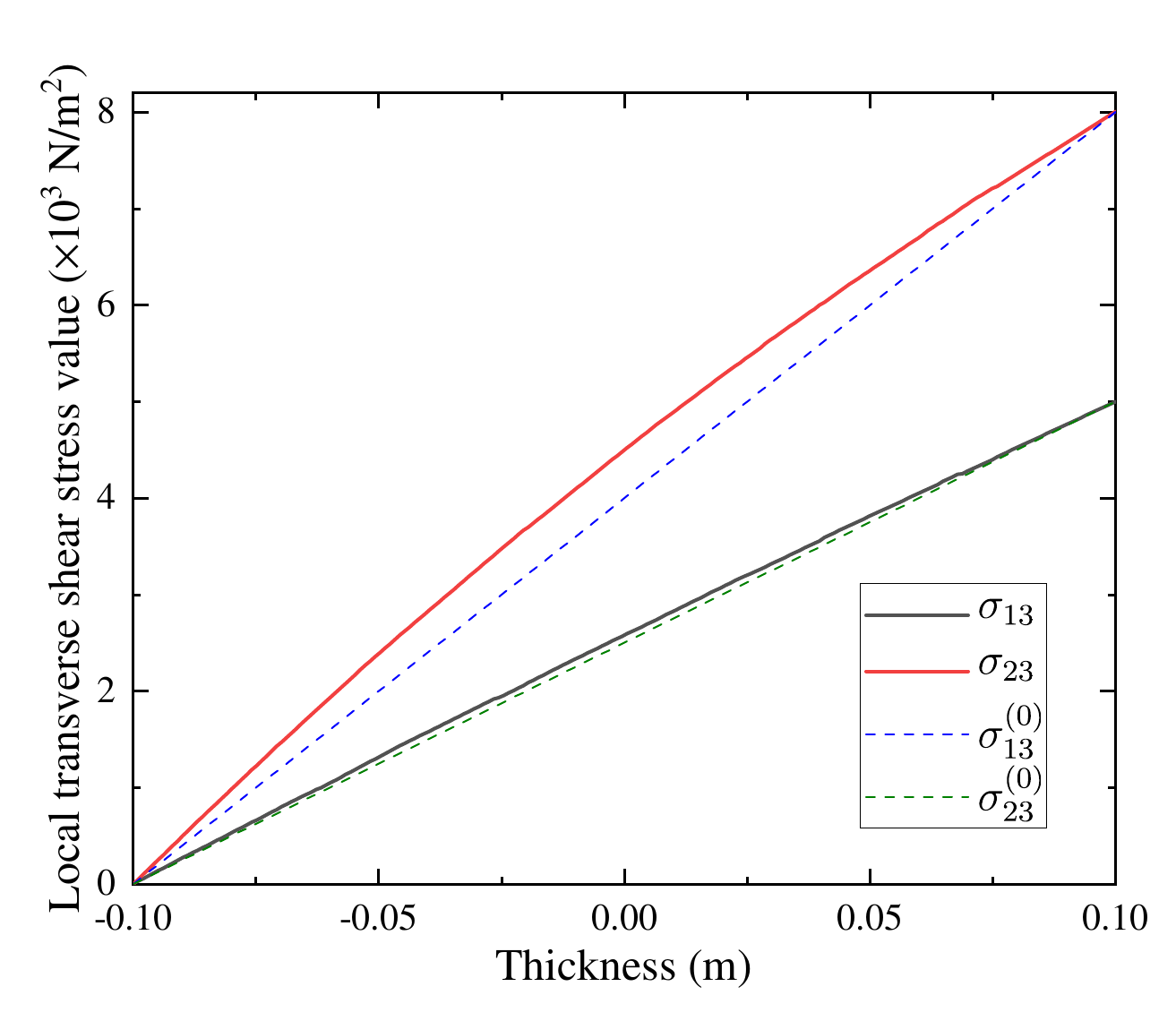}\hspace{0.3cm}
		\label{transverse_linear_plot1L}
	}
	\subfigure[Variation of $\sigma_{\alpha 3}$ in two-layered spherical shell]{
		\includegraphics[width=8cm]{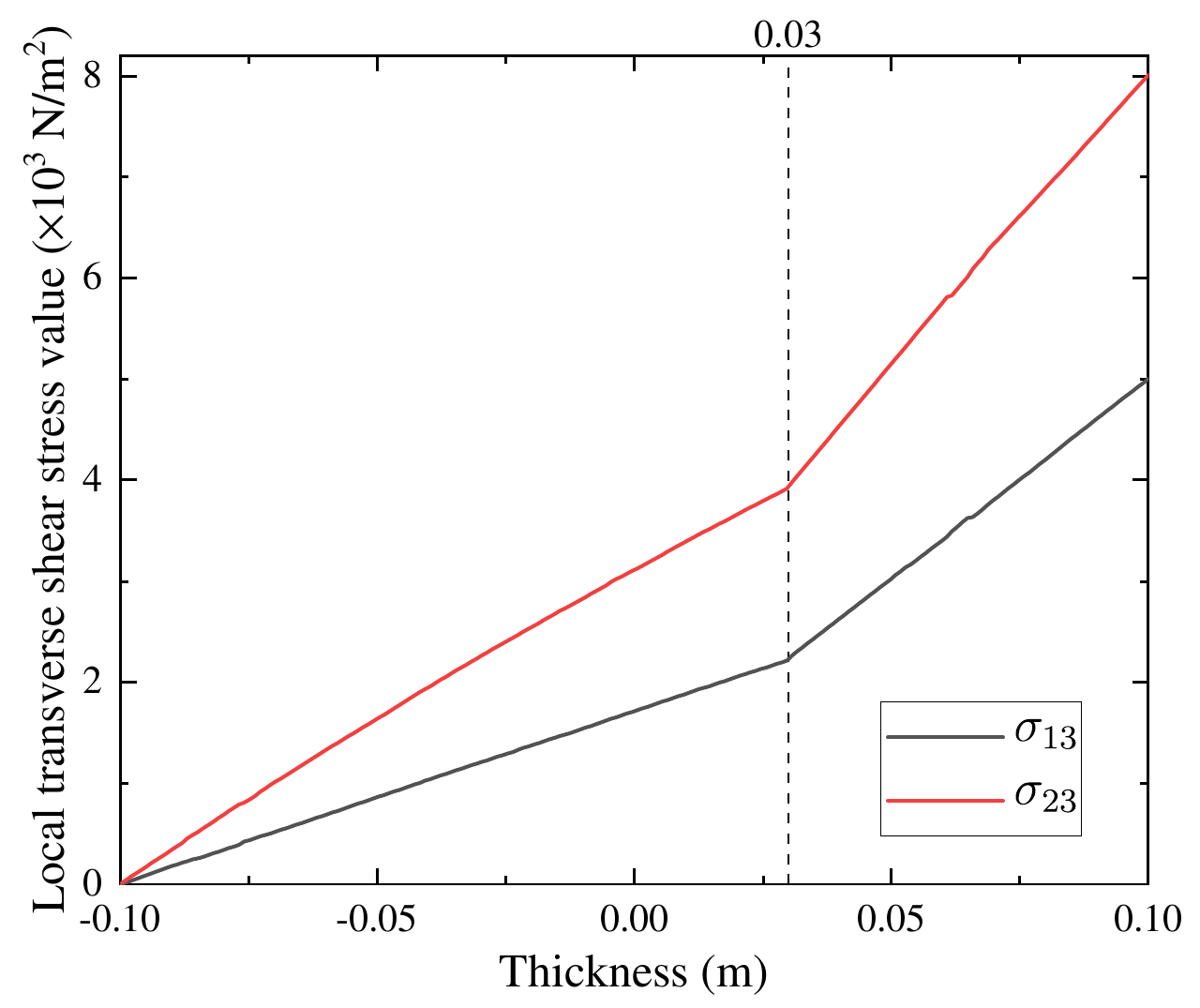}
		\label{transverse_linear_plot2L}
	}
	\subfigure[Variation of $\sigma_{\alpha 3}$ in three-layered spherical shell]{
		\includegraphics[width=8cm]{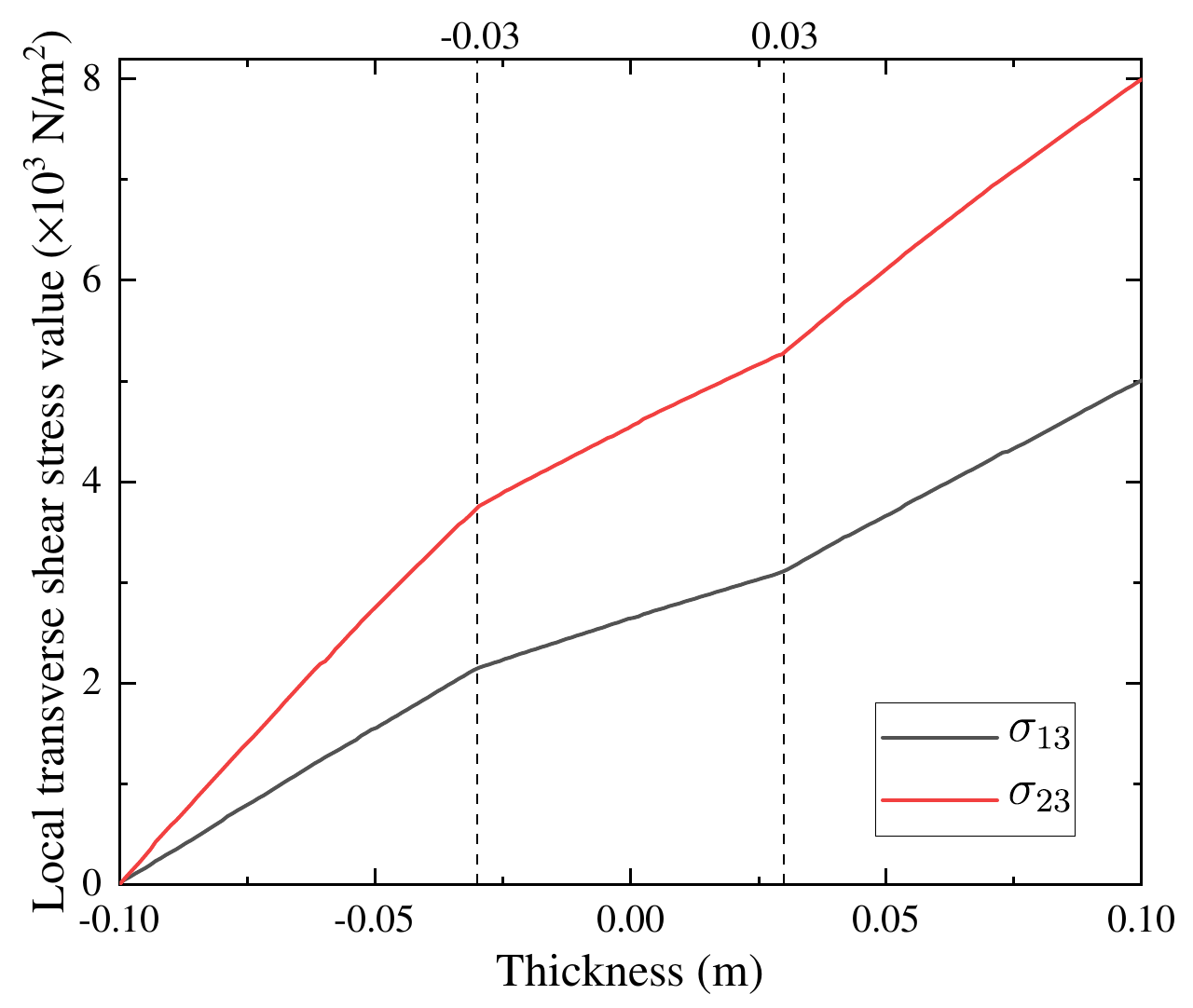}\hspace{0.3cm}
		\label{transverse_linear_plot3L}
	}
	\subfigure[Variation of $\sigma_{\alpha 3}$ in four-layered spherical shell]{
		\includegraphics[width=8cm]{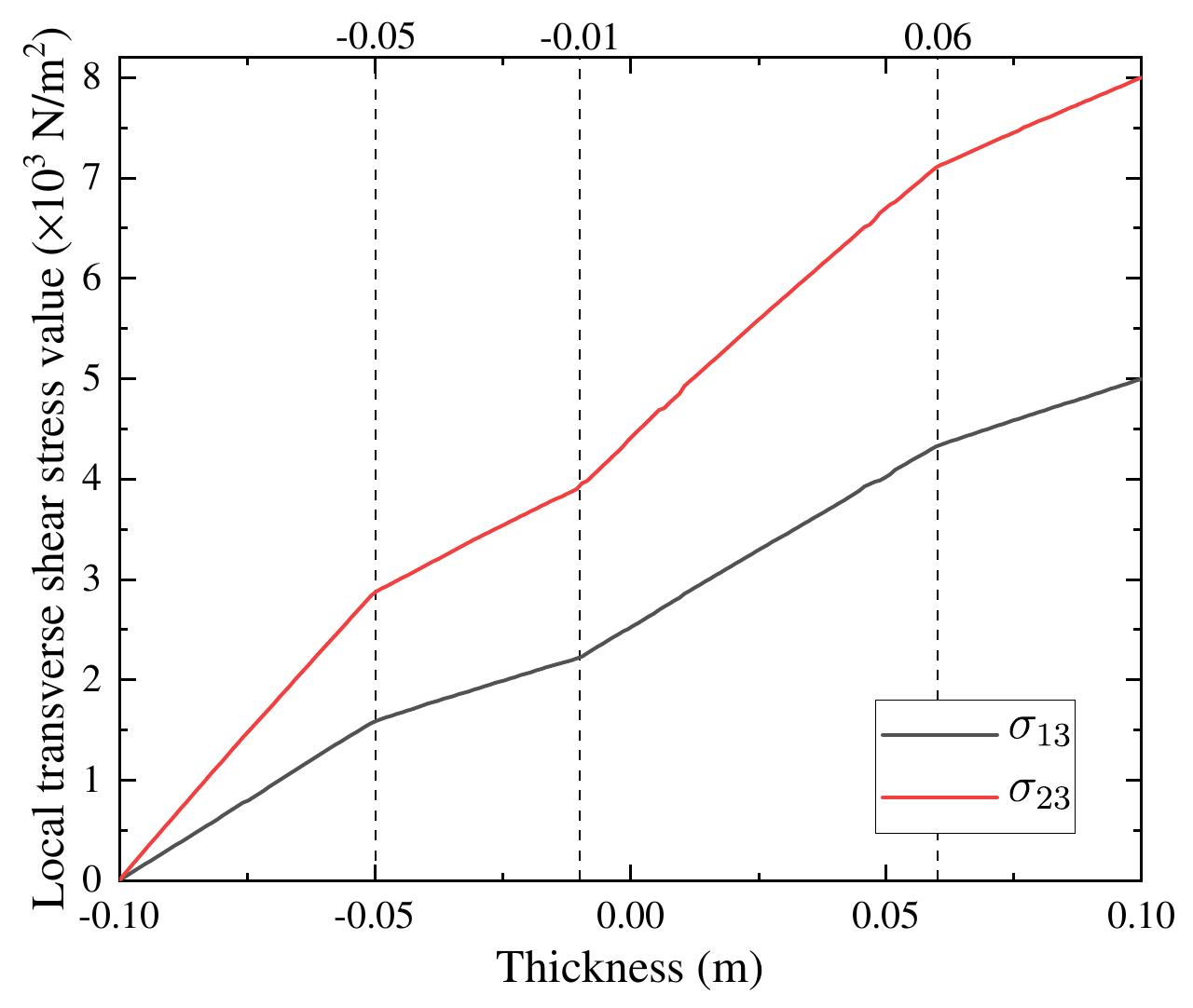}
		\label{transverse_linear_plot4L}
	}
	\caption{Variation of the transverse shear stress components $\sigma_{\alpha 3}$ obtained from fine-mesh FEA along the shell thickness.\label{transverse_linear_plot}}
\end{figure}

The variations of $\sigma_{\alpha 3}$ obtained from direct analysis with extremely fine mesh about the shell thickness coordinate are presented in Fig.~\ref{transverse_linear_plot} for shells with 1 to 4 layers. We can find that, Figs.~\ref{transverse_linear_plot2L}-\ref{transverse_linear_plot4L} all show a piecewise linear distribution along the thickness direction and the turning points of the slopes are exactly located at the junctions of different layers, i.e., the value of the vertical dashed line correspond to the location of the interface, which can be obtained from the thickness ratio $\lambda_{\sss\CL}$ presented in Fig.~\ref{spherical_linear_study}. Based on the theoretical derivation, the distribution of $\sigma_{\alpha 3}$ in a shell with a single layer can be approximated by a straight line simply connecting the top and bottom surfaces to the values of the applied shear forces. Note that, there is a deviation between the actual transverse shear stress and the linear relationship given by AA in Fig.~\ref{transverse_linear_plot1L}, as well as some localized fluctuations observed in Figs.~\ref{transverse_linear_plot2L}-\ref{transverse_linear_plot4L}. These can be considered as the correction effects of higher-order terms, while the trends of $\sigma_{\alpha 3}$ in the MTS have been well captured by the leading-order relationships (Eq.~\eqref{transverse_shear_piecewiselinear}).

\subsubsection{Multi-layered Shells with Applied Surface Shear Loads}\label{Sec_example_shear}
The CST is believed to be unreasonable due to a pre-assumed trivial trasnverse strain state when the investigated MTS undergoes shear deformation. So the common practice is to introduce a SCF $K$ based on the FSDT to compensate for the transverse shear stresses that are uniformly distributed along the thickness, or to resort to the HOTs. However, several issues may occur for the consideration of multi-layered shell structures. First, determining the exact SCF $K$ for laminated/layered shells with complex cross-sections is not an easy task (\cite{gruttmann2017shear}). Second, higher-order kinematic expressions in the HOTs are mainly axiom-based and lack rational support. Here in this section, we will present the deformation responses of different types of shells subject to shear forces, including spherical shells, cylindrical shells, and more generally, ellipsoidal shells. The obtained displacement and stress fields are further compared to those computed based on the 3D solid element FEA, which serve as benchmarks during our validation process. It should also be noted that solutions given by the leading-order equations convergence to the true solution only as the introduced aspect ratio (Eq.~\eqref{epsilon}) approaches zero. Acknowledging that all shell structures inherently exist in 3D space, where thickness, though small, remains a non-negligible quantity relative to the surface dimensions. Consequently, errors inherent in the leading-order model can only be minimized by reducing the aspect ratio or employing higher-order expansions, with truncation errors on the order of $\frac{h}{L}$.

A three-layer spherical shell with a mid-surface radius of $4\,\mathrm{m}$ and a overall thickness of $0.2\,\mathrm{m}$ subject to a given shear force is first considered here. The thickness ratio of each layer is selected to be $\lambda=\left[0.35,0.3,0.35\right]$ from bottom to top, the corresponding Young's moduli and Poisson's ratios are $E=\left[6.825,4,6.825\right]\times10^7\,\mathrm{Pa}$ and $\nu=\left[0.3,0.3,0.3\right]$, respectively. The shear force is imposed on the upper surface of the spherical shell, with the magnitudes of its components along two principal directions being $p_1=5000\,\mathrm{N/m^2}$ and $p_2=8000\,\mathrm{N/m^2}$. Only half of the structure is modeled here, so the symmetric boundary condition is used for the lower boundary (refer to the specific geometry shown in Fig.~\ref{spherical_linear_study}).

\begin{figure}[!htbp]
	\setlength{\abovecaptionskip}{0.3cm}
	\setlength{\belowcaptionskip}{-0.cm}
	\centering
	\subfigure[Displacement (present method, $\mathrm{m}$)]{
		\includegraphics[width=6.5cm]{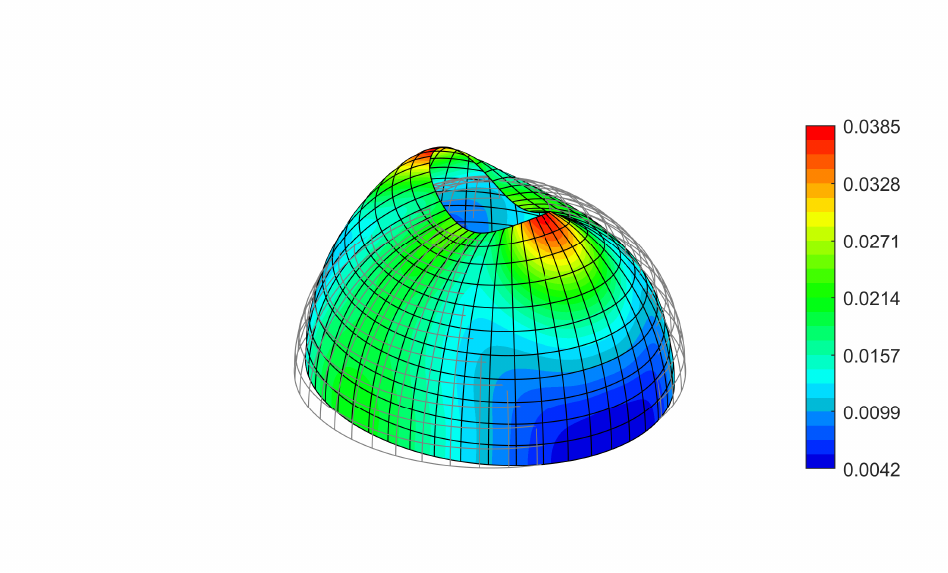}\hspace{1.2cm}
		\label{spherical_disp}
	}
	\subfigure[Displacement (fine-mesh FEA, $\mathrm{m}$)]{
		\includegraphics[width=6.5cm]{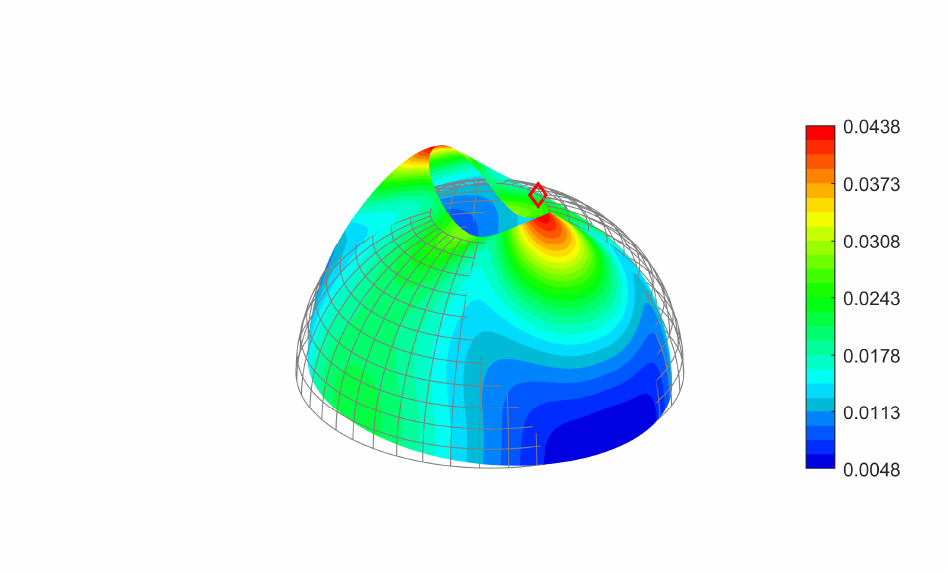}
		\label{spherical_disp_comsol}
	}
	\subfigure[von Mises Stress (present method, $\mathrm{N/m^2}$)]{
		\includegraphics[width=6.5cm]{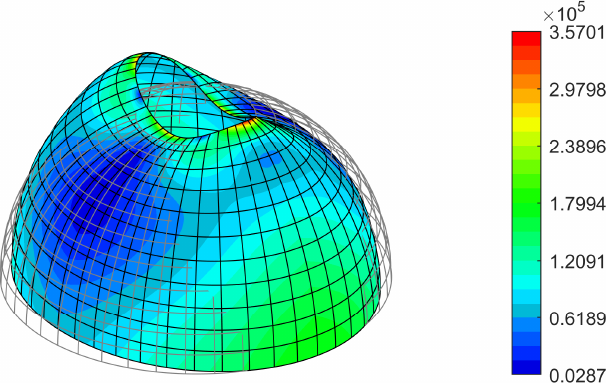}\hspace{1.2cm}
		\label{spherical_vonMises}
	}
	\subfigure[von Mises Stress (fine-mesh FEA, $\mathrm{N/m^2}$)]{
		\includegraphics[width=6.5cm]{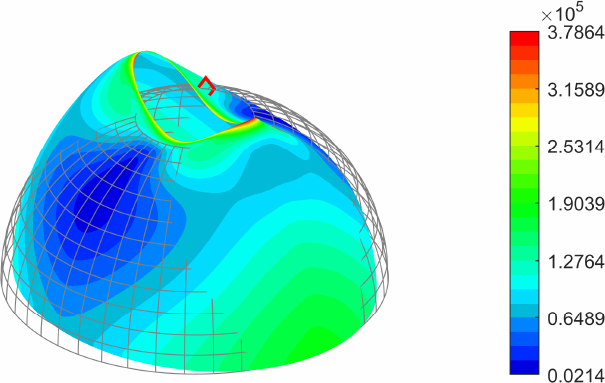}
		\label{spherical_vonMises_comsol}
	}
	\caption{Contour plots of the mid-surface of half of a three-layer spherical shell subject to surface shear forces $p_1=5000\,\mathrm{N/m^2}$, $p_2=8000\,\mathrm{N/m^2}$. Basic parameters are given here: $R=4\,\mathrm{m},\,h=0.2\,\mathrm{m},\,\lambda=\left[0.35,0.3,0.35\right],\,E=\left[6.825,4,6.825\right]\times10^7\,\mathrm{Pa},\,\nu=\left[0.3,0.3,0.3\right]$. The red diamonds mark the positions of the maximum relative errors on the undeformed structure. (a) The magnitude of displacement given by the present framework; (b) displacement field obtained from direct FEA with extremely fine mesh; (c) the von Mises stress distribution given by the present framework; (d) the von Mises stress field obtained from fine-mesh FEA.\label{spherical_shear_contour}}
\end{figure}

Fig.~\ref{spherical_shear_contour} shows the contour plots of one half of the considered spherical shell, including the magnitudes of the displacement and von Mises stress fields and their corresponding results given by the benchmarks (fine-mesh FEA). The maximum relative error of the displacement field is $14.65\%$ at point $\left(1.2361,0,3.8042\right)\mathrm{m}$ and that of the von Mises stress field is $96.45\%$ at point $\left(0.9095,0.8371,3.8042\right)\mathrm{m}$. Note that these two particular positions are located at the free boundary of the MTS, as indicated by the red diamond marks in Figs.~\ref{spherical_disp_comsol} and \ref{spherical_vonMises_comsol}, where the largest deviations in displacement and von Mises stress are concentrated. This is mainly due to the ``BL effects''. A BL is a thin region that contains a large strain energy gradient (\cite{lee2002asymptotic}) and is generally believed to be responsible for failures in predictions of traditional models, which are valid for regions far from the structural boundary. General treatments regarding the BL have been discussed in the works of \cite{howell_kozyreff_ockendon_2008} for linear plates and \cite{green1962boundary} for thin elastic shells. They both successfully derived equations for the BL effects, which finally leading to the change in the boundary conditions on such boundaries without having to solve them explicitly. Therefore, since the BL effects are not of our major concern for the time being, the average relative error is adopted here as a new measure of the model accuracy. Specifically, $80\times80$ data points are selected uniformly, and their corresponding relative errors with the benchmarks are calculated and then averaged.

In this way, the averaged relative error for the displacement is $7.30\%$ and that for the von Mises stress is $5.95\%$, which are both roughly of $\mO\left(h/L\right)$. The present framework thus exhibits a good accuracy in modeling multi-layered spherical shells under shear deformation (except for the BL regions) without the need to introduce additional problem-dependent SCF or intuition-based displacement patterns.

\begin{figure}[!htbp]
	\setlength{\abovecaptionskip}{0.3cm}
	\setlength{\belowcaptionskip}{-0.cm}
	\centering
	\subfigure[Displacement (present method, $\mathrm{m}$)]{
		\includegraphics[width=6.5cm]{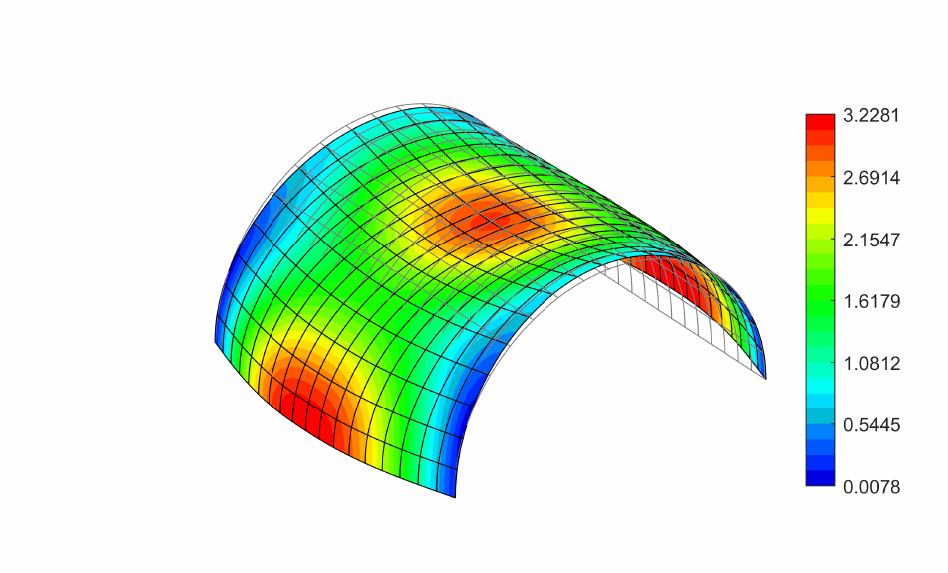}\hspace{1.2cm}
		\label{cylindrical_disp}
	}
	\subfigure[Displacement (fine-mesh FEA, $\mathrm{m}$)]{
		\includegraphics[width=6.5cm]{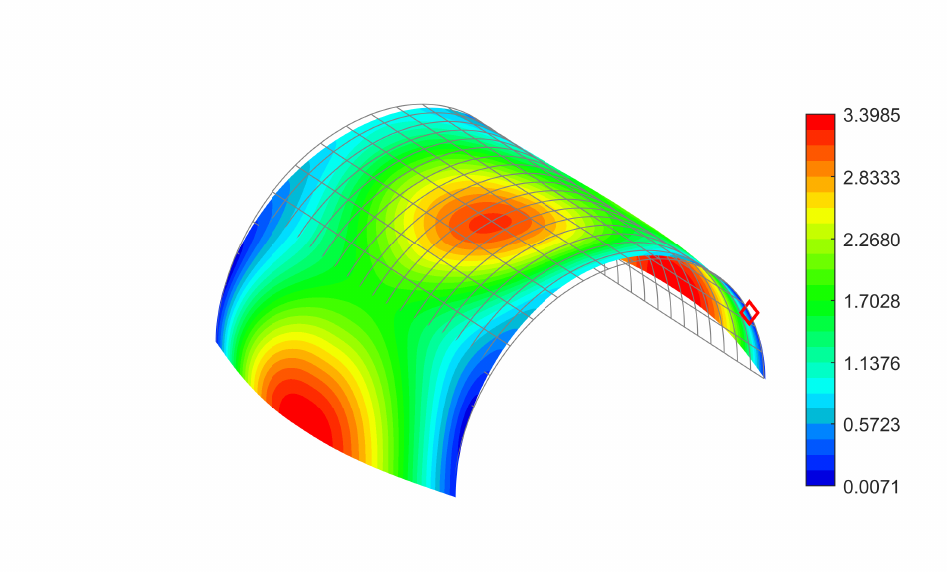}
		\label{cylindrical_disp_error}
	}
	\subfigure[von Mises Stress (present method, $\mathrm{N/m^2}$)]{
		\includegraphics[width=6.5cm]{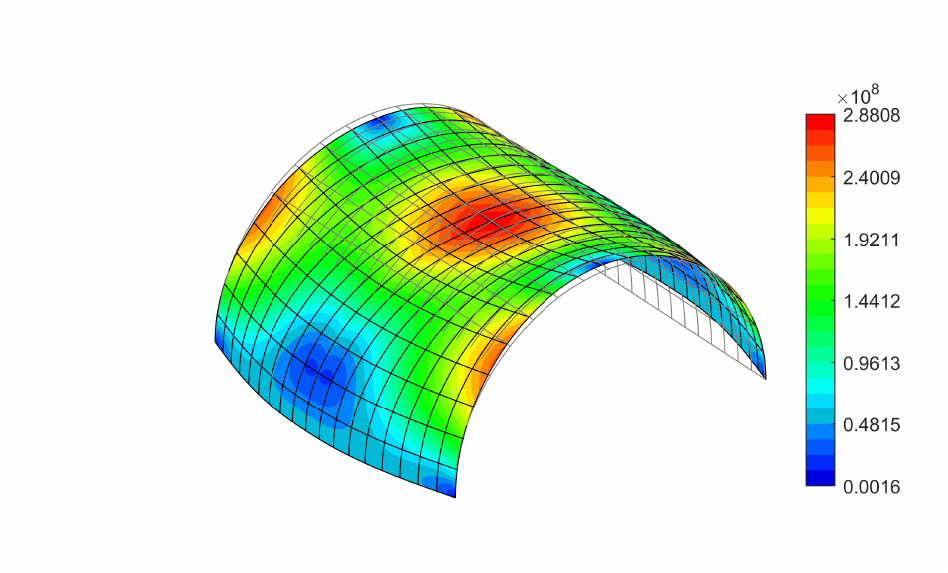}\hspace{1.2cm}
		\label{cylindrical_vonMises}
	}
	\subfigure[von Mises Stress (fine-mesh FEA, $\mathrm{N/m^2}$)]{
		\includegraphics[width=6.5cm]{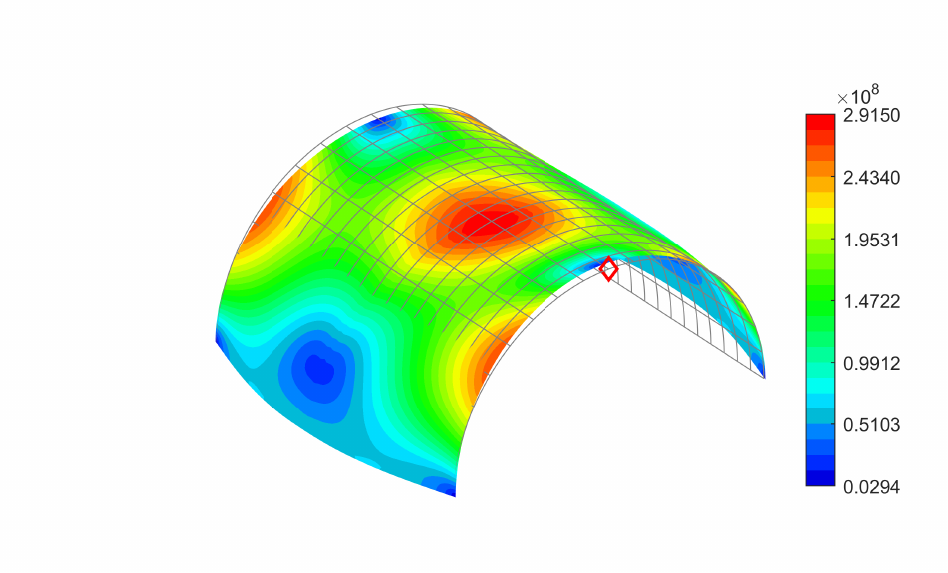}
		\label{cylindrical_vonMises_error}
	}
	\caption{Contour plots of the mid-surface of a three-layer cylindrical shell subject to surface shear forces $\left(p_1,p_2\right)=\left(300,400\right)\mathrm{N/m^2}$. Basic parameters are given here: $R=300\,\mathrm{m},\,L=600\,\mathrm{m},\,h=15\mathrm{m},\,\lambda=\left[0.35,0.3,0.35\right],\,E=\left[72,56,72\right]\,\mathrm{GPa},\,\nu=\left[0.33,0.3,0.33\right]$. Positions of the maximum displacement and stress relative errors are shown in red diamonds. (a) The magnitude of displacement given by the present method; (b) displacement field obtained from direct FEA with extremely fine mesh; (c) the von Mises stress distribution given by the present method; (d) the von Mises stress field obtained from fine-mesh FEA.\label{cylindrical_shear_contour}}
\end{figure}

Stiffened cylindrical shells are widely used in the aerospace industry, with materials such as 6061 and 7075 aluminum alloys commonly employed (\cite{cardarelli2008materials}). In this study, the primary focus is to revisit shell theory through the lens of asymptotic analysis; therefore, structural complexities such as detailed stiffener behavior are temporarily excluded from consideration. Instead, the stiffener-core layer is approximated as an intermediate layer with a reduced effective modulus, allowing the entire structure to be modeled as a three-layer cylindrical shell. The outer layers are assumed to be composed of 7075 aluminum alloy, known for its high strength and hardness. We next consider such a three-layer cylindrical shell with a thickness of $h=15\,\mathrm{m}$, subjected to a shear force $\vp=\left(8,8.2,0\right)\times 10^6\,\mathrm{N/m^2}$ applied to its top surface. The Young’s modulus and Poisson’s ratio for the three layers are selected as $E=[72,56,72]\,\mathrm{GPa}$ and $\nu=[0.33,0.3,0.33]$. All other geometric parameters and boundary conditions are consistent with those shown in Fig.~\ref{shell_obstacle}(a). The maximum relative errors of displacement and von Mises stress occur at points $\left(271.8219,0,126.9364\right)\mathrm{m}$ and $\left(0,0,300\right)\mathrm{m}$, as shown by the red diamonds marked in Figs.~\ref{cylindrical_disp_error} and \ref{cylindrical_vonMises_error}. This demonstrates that the BL effects are also significant in the vicinity of fixed boundaries (or boundaries attached to rigid diaphragms). In this case, the averaged relative errors for the displacement and stress fields are $4.46\%$ and $4.58\%$, respectively. This indicates that the proposed method also excels in capturing the deformation behavior of multi-layered cylindrical shells under shear forces.

\begin{figure}[!h]
	\setlength{\abovecaptionskip}{-1.3cm}
	\setlength{\belowcaptionskip}{-0.cm}
	\centering
	\tikzset{every picture/.style={line width=0.65pt}} %set default line width to 0.65pt
	\begin{tikzpicture}[x=0.75pt,y=0.75pt,yscale=-1,xscale=1]
		%uncomment if require: \path (0,327); %set diagram left start at 0, and has height of 327
		%Image [id:dp29117706136108423] 
		\draw (254.9,141.68) node  {\includegraphics[width=315pt,height=236.25pt]{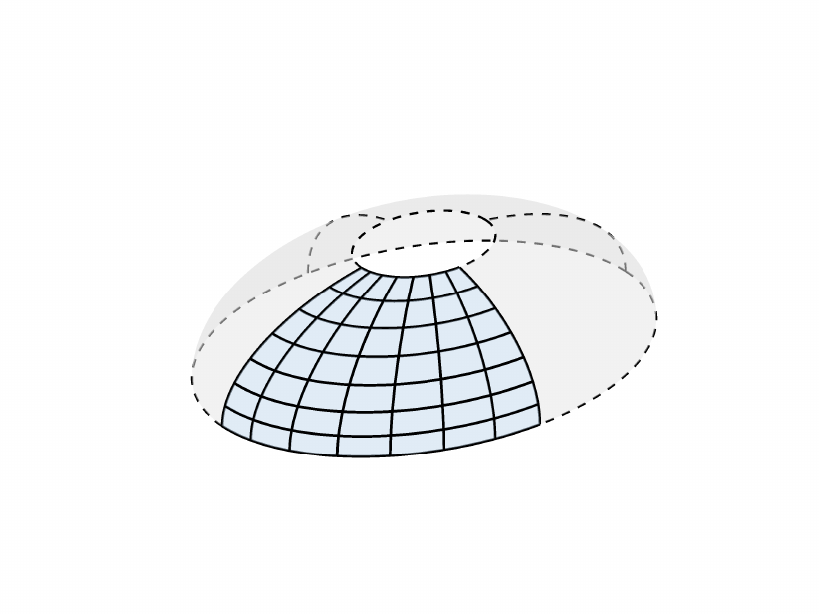}};
		%Straight Lines [id:da7367486822014462] 
		\draw [color={rgb, 255:red, 0; green, 0; blue, 0 }  ,draw opacity=0.5 ]   (261.91,162.87) -- (261.71,60.44) ;
		\draw [shift={(261.7,58.44)}, rotate = 89.88] [fill={rgb, 255:red, 0; green, 0; blue, 0 }  ,fill opacity=0.5 ][line width=0.08]  [draw opacity=0] (9.6,-2.4) -- (0,0) -- (9.6,2.4) -- cycle    ;
		%Straight Lines [id:da5958911929431645] 
		\draw [color={rgb, 255:red, 0; green, 0; blue, 0 }  ,draw opacity=0.5 ]   (261.91,162.87) -- (348.39,218.55) ;
		\draw [shift={(350.07,219.63)}, rotate = 212.78] [fill={rgb, 255:red, 0; green, 0; blue, 0 }  ,fill opacity=0.5 ][line width=0.08]  [draw opacity=0] (9.6,-2.4) -- (0,0) -- (9.6,2.4) -- cycle    ;
		%Straight Lines [id:da42631821974693995] 
		\draw [color={rgb, 255:red, 0; green, 0; blue, 0 }  ,draw opacity=0.5 ]   (261.91,162.87) -- (127.57,214.47) ;
		\draw [shift={(125.7,215.19)}, rotate = 338.99] [fill={rgb, 255:red, 0; green, 0; blue, 0 }  ,fill opacity=0.5 ][line width=0.08]  [draw opacity=0] (9.6,-2.4) -- (0,0) -- (9.6,2.4) -- cycle    ;
		%Straight Lines [id:da4994568231386447] 
		\draw [color={rgb, 255:red, 208; green, 2; blue, 27 }  ,draw opacity=1 ]   (246,195.32) -- (260.04,194.29) ;
		\draw [shift={(263.04,194.07)}, rotate = 175.83] [fill={rgb, 255:red, 208; green, 2; blue, 27 }  ,fill opacity=1 ][line width=0.08]  [draw opacity=0] (5.36,-2.57) -- (0,0) -- (5.36,2.57) -- cycle    ;
		%Straight Lines [id:da06750216819903732] 
		\draw [color={rgb, 255:red, 5; green, 37; blue, 246 }  ,draw opacity=1 ]   (246,195.32) -- (245.76,181.07) ;
		\draw [shift={(245.7,178.07)}, rotate = 89.02] [fill={rgb, 255:red, 5; green, 37; blue, 246 }  ,fill opacity=1 ][line width=0.08]  [draw opacity=0] (5.36,-2.57) -- (0,0) -- (5.36,2.57) -- cycle    ;
		%Straight Lines [id:da3894094313654901] 
		\draw [color={rgb, 255:red, 0; green, 0; blue, 0 }  ,draw opacity=1 ] [dash pattern={on 4.5pt off 4.5pt}]  (365.26,123.41) -- (261.91,162.87) ;
		%Straight Lines [id:da3306351233324729] 
		\draw  [dash pattern={on 4.5pt off 4.5pt}]  (202.81,124.52) -- (261.91,162.87) ;
		%Straight Lines [id:da7878567563033556] 
		\draw  [dash pattern={on 4.5pt off 4.5pt}]  (261.81,110.65) -- (261.91,162.87) ;
		%Straight Lines [id:da2042283647928569] 
		\draw [color={rgb, 255:red, 5; green, 37; blue, 246 }  ,draw opacity=1 ]   (322.22,201.43) -- (321.04,185.51) ;
		\draw [shift={(320.81,182.52)}, rotate = 85.74] [fill={rgb, 255:red, 5; green, 37; blue, 246 }  ,fill opacity=1 ][line width=0.08]  [draw opacity=0] (5.36,-2.57) -- (0,0) -- (5.36,2.57) -- cycle    ;
		%Straight Lines [id:da10605842563053369] 
		\draw [color={rgb, 255:red, 208; green, 2; blue, 27 }  ,draw opacity=1 ]   (322.22,201.43) -- (335.57,196.27) ;
		\draw [shift={(338.37,195.19)}, rotate = 158.86] [fill={rgb, 255:red, 208; green, 2; blue, 27 }  ,fill opacity=1 ][line width=0.08]  [draw opacity=0] (5.36,-2.57) -- (0,0) -- (5.36,2.57) -- cycle    ;
		%Straight Lines [id:da4415975932115477] 
		\draw [color={rgb, 255:red, 208; green, 2; blue, 27 }  ,draw opacity=1 ]   (210.46,164.87) -- (225.41,167.17) ;
		\draw [shift={(228.37,167.63)}, rotate = 188.75] [fill={rgb, 255:red, 208; green, 2; blue, 27 }  ,fill opacity=1 ][line width=0.08]  [draw opacity=0] (5.36,-2.57) -- (0,0) -- (5.36,2.57) -- cycle    ;
		%Straight Lines [id:da20928003880648194] 
		\draw [color={rgb, 255:red, 5; green, 37; blue, 246 }  ,draw opacity=1 ]   (210.46,164.87) -- (213.74,149.45) ;
		\draw [shift={(214.37,146.52)}, rotate = 102.04] [fill={rgb, 255:red, 5; green, 37; blue, 246 }  ,fill opacity=1 ][line width=0.08]  [draw opacity=0] (5.36,-2.57) -- (0,0) -- (5.36,2.57) -- cycle    ;
		%Straight Lines [id:da012906780196415113] 
		\draw [color={rgb, 255:red, 5; green, 37; blue, 246 }  ,draw opacity=1 ]   (158.51,202.72) -- (160.82,187.04) ;
		\draw [shift={(161.26,184.07)}, rotate = 98.39] [fill={rgb, 255:red, 5; green, 37; blue, 246 }  ,fill opacity=1 ][line width=0.08]  [draw opacity=0] (5.36,-2.57) -- (0,0) -- (5.36,2.57) -- cycle    ;
		%Straight Lines [id:da5484407066001828] 
		\draw [color={rgb, 255:red, 208; green, 2; blue, 27 }  ,draw opacity=1 ]   (158.51,202.72) -- (171.07,209.53) ;
		\draw [shift={(173.7,210.96)}, rotate = 208.49] [fill={rgb, 255:red, 208; green, 2; blue, 27 }  ,fill opacity=1 ][line width=0.08]  [draw opacity=0] (5.36,-2.57) -- (0,0) -- (5.36,2.57) -- cycle    ;
		
		% Text Node
		\draw (131.5,215.34) node [anchor=north west][inner sep=0.75pt]  [font=\fontsize{0.71em}{0.85em}\selectfont]  {$x$};
		% Text Node
		\draw (337.28,219.91) node [anchor=north west][inner sep=0.75pt]  [font=\fontsize{0.71em}{0.85em}\selectfont]  {$y$};
		% Text Node
		\draw (263.7,61.84) node [anchor=north west][inner sep=0.75pt]  [font=\fontsize{0.71em}{0.85em}\selectfont]  {$z$};
		% Text Node
		\draw (310.25,145.87) node [anchor=north west][inner sep=0.75pt]  [font=\fontsize{0.71em}{0.85em}\selectfont]  {$a$};
		% Text Node
		\draw (237.28,155.46) node [anchor=north west][inner sep=0.75pt]  [font=\fontsize{0.71em}{0.85em}\selectfont]  {$b$};
		% Text Node
		\draw (251.42,114.71) node [anchor=north west][inner sep=0.75pt]  [font=\fontsize{0.71em}{0.85em}\selectfont]  {$c$};
		% Text Node
		\draw (225.5,226.56) node [anchor=north west][inner sep=0.75pt]  [font=\fontsize{0.93em}{1.12em}\selectfont] [align=left] {{\fontfamily{ptm}\selectfont {\footnotesize sym.}}};
		% Text Node
		\draw (323.34,182.8) node [anchor=north west][inner sep=0.75pt]  [font=\fontsize{0.59em}{0.71em}\selectfont,color={rgb, 255:red, 5; green, 37; blue, 246 }  ,opacity=1 ]  {$p_{2}$};
		% Text Node
		\draw (334.43,197.89) node [anchor=north west][inner sep=0.75pt]  [font=\fontsize{0.59em}{0.71em}\selectfont,color={rgb, 255:red, 208; green, 2; blue, 27 }  ,opacity=1 ]  {$p_{1}$};
	\end{tikzpicture}
	\caption{Illustration of the mid-surface of a three-layered ellipsoidal shell of thickness $h=0.2\,\mathrm{m}$, and the thickness proportion of each layer is $\lambda=\left[0.35,0.3,0.35\right]$. The lengths of three principal semi-axes are $a=8\,\mathrm{m},\,b=6\,\mathrm{m},\,c=4\,\mathrm{m}$. Note that, the NURBS parametric lines on an ellipsoid are not orthogonal, while the ``lines of curvature'' are not intuitive. We hence pick two such mutually orthogonal directions at each point: one aligned with $\vr_{\xi_1}$ and the other perpendicular to it. The components of the applied surface shear force along these two directions are $p_1=5000\,\mathrm{N/m^2},\,p_2=8000\,\mathrm{N/m^2}$. Due to the symmetry of shell geometry and loading conditions, only one eighth of the structure (gridded part) is modeled here.\label{ellipsoidal_illustration}}
\end{figure}

The case of a multi-layered ellipsoidal shell is also included here to test the performance of the present model on non-classical shells. For a sphere (where all points are umbilical) and a cylinder (which has no umbilical points), the principal directions at each point can be uniformly defined, whereas the four umbilical points on an ellipsoidal surface are isolated, making it difficult to pick a local coordinate system that is coordinated with the surroundings. Therefore, as stated in Sec.~\ref{Sec_IGA_implementation}, the local displacement field is finally transformed into the global coordinate system to circumvent the stress singularities associated with an ellipsoidal local coordinate system. The basic parameter settings regarding the ellipsoid are illustrated in Fig.~\ref{ellipsoidal_illustration}, and the material properties assigned to each layer from bottom to top are $E=\left[6.825,4,6.825\right]\times10^7\,\mathrm{Pa},\,\nu=\left[0.3,0.3,0.3\right]$.

\begin{figure}[!htbp]
	\setlength{\abovecaptionskip}{0.3cm}
	\setlength{\belowcaptionskip}{-0.cm}
	\centering
	\subfigure[Displacement (present method, $\mathrm{m}$)]{
		\includegraphics[width=6.5cm]{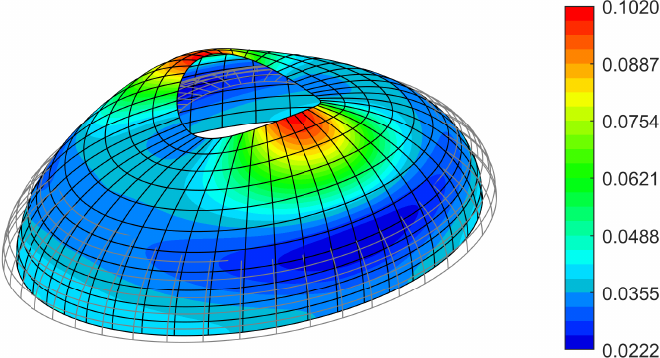}\hspace{1.2cm}
		\label{ellipsoidal_disp}
	}
	\subfigure[Displacement (fine-mesh FEA, $\mathrm{m}$)]{
		\includegraphics[width=6.5cm]{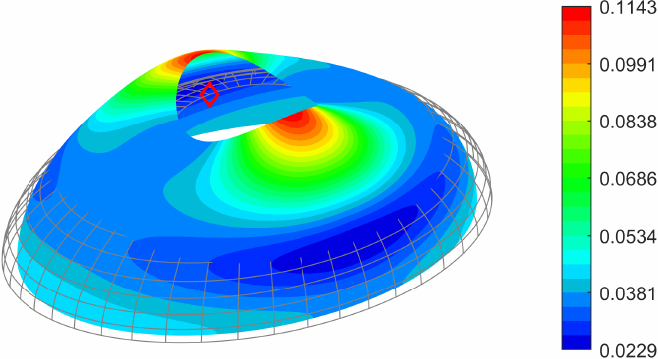}
		\label{ellipsoidal_disp_comsol}
	}
	\subfigure[von Mises Stress (present method, $\mathrm{N/m^2}$)]{
		\includegraphics[width=6.5cm]{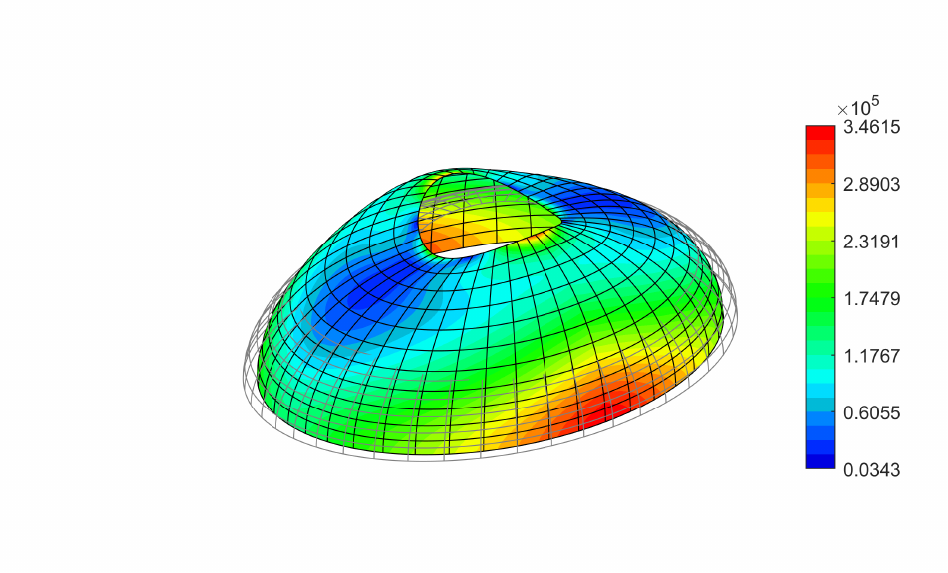}\hspace{1.2cm}
		\label{ellipsoidal_vonMises}
	}
	\subfigure[von Mises Stress (fine-mesh FEA, $\mathrm{N/m^2}$)]{
		\includegraphics[width=6.5cm]{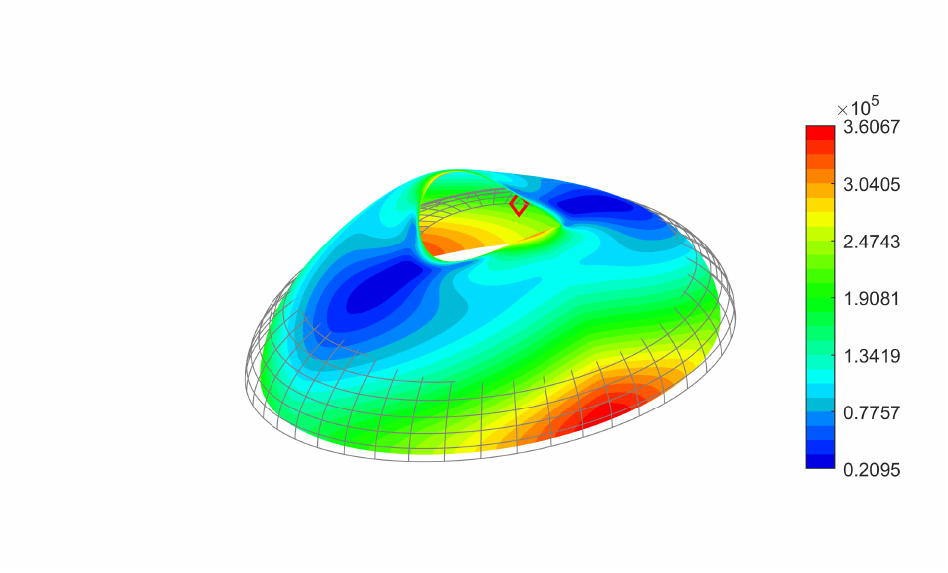}
		\label{ellipsoidal_vonMises_comsol}
	}
	\caption{Contour plots of the mid-surface of a three-layer ellipsoidal shell subject to surface shear forces $\left(p_1,p_2\right)=\left(5000,8000\right)\mathrm{N/m^2}$. The geometry of the shell mid-surface is presented in Fig.~\ref{ellipsoidal_illustration} . Other basic parameters are given here: $h=0.2\,\mathrm{m},\,\lambda=\left[0.35,0.3,0.35\right],\,E=\left[6.825,4,6.825\right]\times10^7\,\mathrm{Pa},\,\nu=\left[0.3,0.3,0.3\right]$. Positions of the maximum displacement and stress relative errors are shown in red diamonds. (a) The magnitude of displacement predicted by the present method; (b) displacement field obtained from direct FEA with extremely fine mesh; (c) the von Mises stress distribution predicted by the present method; (d) the von Mises stress field obtained from fine-mesh FEA.\label{ellipsoidal_shear_contour}}
\end{figure}

The deformation modes of the mid-surface of the investigated ellipsoidal shell are shown in Fig.~\ref{ellipsoidal_shear_contour}. By comparing the results given by the present method (Figs.~\ref{ellipsoidal_disp} and \ref{ellipsoidal_vonMises}) with those given by direct finite element simulation (Figs.~\ref{ellipsoidal_disp_comsol} and \ref{ellipsoidal_vonMises_comsol}), we can find that differences in both displacement and von Mises stress fields are mainly concentrated near the free boundary. And the largest relative errors in these two fields, marked by the red diamonds, are also located on the free edge. We can therefore attribute this to the BL effects, which are inherent in the plate and shell theories.

The averaged relative error for the displacement field is $4.79\%$, while that for the von Mises stress is $4.97\%$. It is worth mentioning that, errors of the examples discussed here are both of order $\epsilon=0.05$ which is exactly the theoretical accuracy of the AA method. It follows that the MTS can be modeled to a high accuracy based solely on the geometric characteristic that the shell thickness is much smaller than its corresponding surface dimensions, without requiring the introduction of the SCF or pre-assumed deformation patterns. This holds true except in the BL regions, where traditional shell theories, such as the Kirchhoff-Love theory, the FSDT, etc., also fail to provide accurate descriptions.

From the numerical examples given above, it can be seen that the leading-order equations derived from AA alone are able to capture the deformation modes of multi-layered shells under shear loading effectively without introducing additional parameters (e.g., the SCF and coefficients associated with higher-order terms in displacement expression, etc.). This suggests the actual role of the applied surface shear force, i.e., it mainly affects the redistribution of in-plane stress components (refer to Eqs.~\eqref{reduced_equilibrium_T1}, \eqref{reduced_equilibrium_T2}, and \eqref{reduced_equilibrium_M}) rather than changing the transverse shear stress components.

\section{Conclusions and Further Discussions}\label{Sec_conclusions}
In the present article, we propose a method within the framework of asymptotic analysis for modeling and interpreting the deformation behavior of multi-layered shells with greater mathematical rigor. With the help of asymptotic methods, we revisit conclusions originally based on kinematic assumptions, establishing a more solid foundation. The novelties presented can be summarized into three main strands.

Firstly, we introduce a classification rule for shell deformation modes based on the order of principal curvatures, as illustrated in Fig.~\ref{Fig_deformation_region_classification}. Specifically, we distinguish between cases where the maximum non-dimensional principal curvature is of $\mO\left(\epsilon\right)$, resembling thin plates; $\mO\left(1\right)$, corresponding to membrane shell theory; or even larger, e.g., $\mO\left(1/\epsilon\right)$), showing a hybrid, combining features of both bending and membrane responses.

Secondly, we derived the specific order relationships for the internal field variables under each classification and demonstrate that curvature, as a geometric factor, changes the their scaling behavior. For example, as the curvature of the shell mid-surface increases from nearly flat to normally curved, the normal displacement transitions from dominating the in-plane components to becoming of the same order. This transition indicates that shell structures are not a direct extension of plate models, and their governing conclusions cannot be indiscriminately inherited. Moreover, properties of the eigenvalue problems associated with free vibration also vary with the scaling and order of the involved variables.

Thirdly, the applied external shear force is shown to primarily affect the redistribution of local in-plane stress components $\sigma_{\alpha\beta}^{(0)}$ (as implied by Eqs.~\eqref{reduced_equilibrium_T1}, \eqref{reduced_equilibrium_T2}, and \eqref{reduced_equilibrium_M}), rather than necessitating a significant consideration of transverse shear stress. In fact, the transverse stress components can be directly recovered from the leading-order equilibrium equations, once the dominant in-plane stress components are determined.

Although the current method has demonstrated certain advantages in achieving a rational understanding of the deformation behavior of shell structures, further discussion is needed on potential areas for future research on this method.

\textbf{\textit{On the Boundary Layer Effects.}} As mentioned in Sec.~\ref{Sec_example_shear}, due to the concentration of strain energy---generally arising from curvature discontinuities, incompatible boundary conditions, and/or irregular loadings---shells often exhibit behaviors near boundaries that differ from those observed far from the boundaries (\cite{lee2002asymptotic}). These boundary phenomena cannot be adequately described by the previously derived equations.Addressing them necessitates a fully three-dimensional model regarding the BL, which consequently leads to modified boundary conditions for the investigated problem.

Specifically, we investigate a thin BL near the edge $\wxi_1=0$, where we have $\wxi_1=\epsilon\bar{\xi}_1$ and $\bar{\xi}_1\sim\mO(1)$. Here the symbol ``--'' denotes the BL-related quantities. For the nontrivial equilibrium of weakly curved shells (Eq.~\eqref{equilibrium_weakly}), the transverse stress components should be rescaled accordingly, i.e., $\wsigma_{\alpha3}=\bsigma_{\alpha3}/\epsilon,\ \wsigma_{33}=\bsigma_{33}/\epsilon^2$. Combined with Eq.~\eqref{stress_scaling}, it is found that all stress components within the BL are of the same order. As a result, the transverse normal stress is two orders of magnitude larger than its interior counterpart, revealing some of the unexpected behaviors of the BL region. However, the challenges posed by BL effects are more than just rescaling, the boundary conditions also call for certain modifications. For instance, five boundary conditions can be obtained from a traction-free edge, $\wT_{11}=\wT_{12}=\wM_{11}=\wM_{12}=\wN_1=0$, of which only four are necessary. Through asymptotic matching of the exterior and interior solutions (\cite{reiss1962theory,howell_kozyreff_ockendon_2008}), or by correcting the interior solution based on the principle of linear superposition (\cite{green1962boundary}), we can finally derive the correction to the aforementioned redundant boundary conditions as follows:
\begin{equation}
	\wT_{11}=\wT_{12}=\wM_{11}=\wN_1-\pd{\wM_{12}}{\wxi_2}=0.
\end{equation}

Additionally, \cite{karamian2002boundary} used the method of exponential solutions to investigate the BL and internal layer properties of developable elastic shells, while \cite{nosier2010boundary} applied displacement-based theories to study the BL of laminated/composite shells. Future research could build upon these approaches to further examine the shell BL effects in more complex scenarios.

\textbf{\textit{On the Applicability Limits.}} When dynamic loading is considered, the influence of inertial forces becomes significant. According to our theory, the product of the maximum principal curvature and the characteristic shell size approaching 1 or 0 corresponds respectively to a complete or constrained eigenvalue problem. Therefore, if the dynamic load amplitude remains moderate and the frequency is low, the shell can still be treated as being in a linear elastic state, with the corresponding leading-order equations taking the form of Eq.~\eqref{equilibrium_weakly}, Eq.~\eqref{equilibrium_normal}, or Eq.~\eqref{equilibrium_significant}, depending on the curvature magnitude. However, when load-induced deformations become large and the materials enter the plastic phase under high strain rates, the applicability of the model extends beyond the scope of this study. In such cases, a renewed asymptotic analysis of the nonlinear geometric and constitutive equations is required, leading to more complex relations. Nevertheless, the original equilibrium equations (as presented by Eq.~\eqref{equilibrium_concise}), expressed in terms of the Cauchy stress, remain valid based on force equilibrium considerations. It should be noted that the present AA, based on local coordinates coinciding with ``lines of curvature'', is primarily focused on effectively capturing the deformation modes and properties of the shell. However, for solving strongly nonlinear problems, the use of invariant tensor notation (\cite{lurie1940general}) within the continuum mechanics framework may offer a viable alternative.

\textbf{\textit{On the Higher-Order Approximations.}} As previously mentioned, asymptotic analysis of the 3D equations proceeds in a stepwise hierarchical manner. At higher orders, improved accuracy in modeling is achieved at the cost of introducing more variables and increasingly complex equations, leading to prohibitive computational expense. Moreover, some studies have questioned the necessity of stepping into higher-order realms in certain situations. \cite{SIMMONDS19922441} noted that for elastic tubes subjected to end conditions such as axisymmetric radial and axial displacements, it is not possible---or asymptotically correct---to recover the response solely by refining the thickness distribution of the kinematic expression, without accounting for BL effects. \cite{yu2002asymptotic}, on the other hand, derived an equivalent single layer (ESL)-like theory directly from the full 3D potential energy using VAM, demonstrating that the resulting accuracy is comparable to that of higher-order or layerwise theories, but with much fewer variables. These examples highlight the need for greater caution when advancing to higher-order analyses, while also revealing the potential power of lower-order models.

From Eqs.~\eqref{constitutive_dimensionless4}-\eqref{constitutive_dimensionless5}, it is evident that the leading-order transverse shear strain asymptotically vanishes, while becoming nonzero at higher orders ($\gamma_{\alpha3}\sim\mO\left(\epsilon^2\right)$ or higher). The corresponding expression for transverse shear stress components is given by Eq.~\eqref{transverse_shear_leading}. Therefore, shear effects become increasingly significant when addressing more complex strain states or cases involving large contrasts in material properties, where higher-order approximations are required. In cases where shear effects are negligible, e.g., when the material moduli across layers differ slightly and the applied shear forces are of moderate order, the contribution of transverse shear stresses to the total energy is limited, and the applied shear forces serve to redistribute the in-plane stresses. Under such circumstances, the leading-order model suffices to accurately capture the shell deformation behavior. The accuracy of AA depends on the truncation error associated with the chosen expansion order in terms of the small parameter $\epsilon$. Consequently, the error of the leading-order model is approximately of order $h/L$. As the ratio of shell thickness to the characteristic length decreases---that is, for thin shells---the error introduced by the asymptotic approximation becomes smaller and strictly holds as $\epsilon\to0$.

\section*{Acknowledgements}
The financial supports from the National Natural Science Foundation of China [12172074, Zhu], and the Fundamental Research Funds for the Central Universities, P.R. China [Zhu] are gratefully acknowledged.

\appendix
\section{Derivation of the Jacobian Determinants Related to Domain Transformation\label{Appendix1}}
Now we derive the specific expressions for several Jacobian determinants in Eqs.~\eqref{internal_virtual_work} and \eqref{external_virtual_work}. For a vector $\vr_\mathrm{e}$ in the physical space, it is equivalent in representation under global and local basis vectors, i.e., $\vr_\mathrm{e} = x_i\boldsymbol{\tve_i} = \wxi_i\frac{\intd\boldsymbol{\wxi_i}}{\|\intd\wxi_i\|}$. Therefore, a volume infinitesimal $\intd\Omega$ in the Cartesian coordinates can be transformed into one in the local orthogonal curvilinear coordinate system by a certain scaling, that is,
\begin{equation}
	\intd\Omega = \left(\intd\boldsymbol{\wxi_1},\intd\boldsymbol{\wxi_2},\intd\boldsymbol{\wxi_3}\right) = \det\left(\left[\pd{\vr_\mathrm{e}}{\xi_1},\pd{\vr_\mathrm{e}}{\xi_2},\pd{\vr_\mathrm{e}}{\xi_3}\right]^\rT\right)\intd\wxi_1\intd\wxi_2\intd\wxi_3,
\end{equation}
where $\left(\bigcdot,\bigcdot,\bigcdot\right)$ denotes the triple product of three vectors. Compared with the second identity in Eq.~\eqref{internal_virtual_work}, the Jacobian determinant measuring physical-to-local volume change is obtained
\begin{equation}\label{jacobian_volume_physical2local}
	\left|J_{\Omega\to\left(\wCP\times\wxi_3\right)}\right| = \det\left(\left[\pd{\vr_\mathrm{e}}{\xi_1},\pd{\vr_\mathrm{e}}{\xi_2},\pd{\vr_\mathrm{e}}{\xi_3}\right]^\rT\right) = h_1h_2h_3,
\end{equation}
here Eq.~\eqref{basis_lame_coefficient} is adopted to derive the second identity of Eq.~\eqref{jacobian_volume_physical2local}.

Similarly, an area infinitesimal $\intd\mathcal{S}_\mathrm{t}$ on the top surface of the TLS can be linked to one in the principal parametric space $\wCP$ by
\begin{equation}
	\left.\intd\mathcal{S}_\mathrm{t}\right|_{\wxi_3 = \frac{1}{2}} = \left(\ve_3,\intd\boldsymbol{\wxi_1},\intd\boldsymbol{\wxi_2}\right)_{\wxi_3 = \frac{1}{2}} = \left(h_1h_2\right)_{\wxi_3 = \frac{1}{2}}\,\intd\wxi_1\intd\wxi_2,
\end{equation}
thus the Jacobian determinant associated with the physical-to-local area change is given by
\begin{equation}\label{jacobian_area_physical2local}
	\left|J_{\partial_\mathrm{t}\Omega\to\wCP}\right| = \left(h_1h_2\right)_{\wxi_3 = \frac{1}{2}}.
\end{equation}

Finally, we consider the change of variables in the last identities of both Eq.~\eqref{internal_virtual_work} and Eq.~\eqref{external_virtual_work}. As illustrated in Fig.~\ref{Fig_midsurface_nurbs}, the area change brought by the regular map $\mathscr{L}$ from the NURBS parametric space $\CP$ to the principal parametric space $\wCP$ is determined by
\begin{equation}
	\intd\wCP = \left(\ve_3,\intd\boldsymbol{\xi_1},\intd\boldsymbol{\xi_2}\right) = \det\left(\left[\pd{\wxi_\alpha}{\xi_\beta}\right]\right)\intd\CP,
\end{equation}
where the determinant of the matrix $\left[\pd{\wxi_\alpha}{\xi_\beta}\right]$ captures the effect of the change of variables between two parametric coordinates, that is,
\begin{equation}\label{jacobian_change_of_variables}
	\left|J_{\wCP\to\CP}\right| = \det\left(\left[\pd{\wxi_\alpha}{\xi_\beta}\right]\right) = 1\bigg/\det\left(\left[\pd{\xi_\beta}{\wxi_\alpha}\right]\right),
\end{equation}
note that the components $\pd{\xi_\beta}{\wxi_\alpha}$ are exactly the aforementioned combination coefficients for the generation of principal directions Eq.~\eqref{mapping_principal_parameter} and can be denoted as $J_{\beta\alpha}$.

With Eqs.~\eqref{jacobian_volume_physical2local},\eqref{jacobian_area_physical2local}, and \eqref{jacobian_change_of_variables}, the two Jacobian determinants involved in the last equations of the virtual work formulations are obtained by
\begin{subequations}\label{jacobian_expression_final}
	\begin{align}
		&\left|J_{\Omega\to\left(\CP\times\xi_3\right)}\right| = h_1h_2h_3/\left|J\right|;\label{jacobian_expression_final1}\\
		&\left|J_{\partial_\mathrm{t}\Omega\to\CP}\right| = \left(h_1h_2\right)_{\wxi_3 = \frac{1}{2}}/\left|J\right|.\label{jacobian_expression_final2}
	\end{align}
\end{subequations}

\bibliographystyle{model5-names}
\biboptions{authoryear}
\bibliography{reference}

\end{document}